\documentclass{article}

\usepackage{amssymb}
\usepackage{amsmath}
\usepackage{algorithm}
\usepackage{bm}

\usepackage{johd}

\title{Embedded Model Bias Quantification with Measurement Noise for Bayesian Model Calibration}

\author{Daniel Andrés Arcones$^{a}$ $^{b}$ $^{*}$, Martin Weiser$^{c}$, Phaedon-Stelios Koutsourelakis $^{b}$, Jörg F. Unger $^{a}$\\
	\small $^{a}$Bundesanstalt für Materialforschung und -prüfung, Berlin, Germany \\
	\small $^{b}$Technical University of Munich, Garching bei München, Germany \\
	\small $^{c}$Zuse Institute Berlin, Berlin, Germany\\\\
	\small $^{*}$Corresponding author: Daniel Andrés Arcones; \tt{daniel.andres-arcones@bam.de} \\
}
\date{23 June 2025}

\begin{document}
	\maketitle
	
	\begin{abstract}
		A key factor in ensuring the accuracy of computer simulations that model physical systems is the proper calibration of their parameters based on real-world observations or experimental data. Inevitably, uncertainties arise, and Bayesian methods provide a robust framework for quantifying and propagating these uncertainties to model predictions. Nevertheless, Bayesian methods paired with inexact models usually produce predictions unable to represent the observed datapoints. Additionally, the quantified uncertainties of these overconfident models cannot be propagated to other Quantities of Interest (QoIs) reliably. A promising solution involves embedding a model inadequacy term in the inference parameters, allowing the quantified model form uncertainty to influence non-observed QoIs. This paper introduces a more interpretable framework for embedding the model inadequacy compared to existing methods. To overcome the limitations of current approaches, we adapt the existing likelihood models to properly account for noise in the measurements and propose two  new formulations designed to address their shortcomings. Moreover, we evaluate the performance of this inadequacy-embedding approach in the presence of discrepancies between measurements and model predictions, including noise and outliers. Particular attention is given to how the uncertainty associated with the model inadequacy term propagates to the QoIs, enabling a more comprehensive statistical analysis of prediction's reliability. Finally, the proposed approach is applied to estimate the uncertainty in the predicted heat flux from a transient thermal simulation using temperature observations.
	\end{abstract}
	\noindent
	
	\paragraph*{Funding Statement}
	This research was supported by the German Research Foundation (DFG) in the special focus program SPP 2388/1 ``Hundert plus – Verlängerung der Lebensdauer komplexer Baustrukturen durch intelligente Digitalisierung'' (Hundred Plus - Extending the service life of complex building structures through intelligent digitalisation) in the subproject C07: Data driven model adaptation for identification of digital twins of bridges.


	\section{Introduction}
	In the last decades, advances in sensor technology have made available large quantities of affordable data that reflect the current state of physical systems and processes. In parallel, simulation models have become ubiquitous in science and engineering for generating predictions on the behaviour of a given system based on physical laws. With the ongoing transition to an Industry 4.0 paradigm and the proliferation of Digital Twins, the challenge of calibrating simulation models based on data from physical sensors is becoming increasingly relevant (\cite{Vaidya2018}). In the specific case of Digital Twins of physical systems that use simulations to predict the behaviour of their real counterpart and make decisions on their behalf, the accuracy and reliability of such predictions are essential (\cite{AndresArcones2023a}). Therefore, their uncertainty must be adequately quantified to achieve such trustworthy predictions.
	
	In most cases, the calibration of the computational models is achieved by estimating a set of parameters that control their behaviour based on available observations from the modeled system. Although more uncertainty sources can be identified (\cite{Walker2003}), this process introduces two main ones: the error attributed to noise or uncontrolled variations in the data and the discrepancy created by the assumptions used to generate the computational model. They can be related to aleatoric and epistemic uncertainty sources, respectively. While noise errors can be satisfactorily estimated from the  observations, the so-called \textit{model discrepancy} or \textit{model inadequacy} presents further challenges. All models are based on assumptions and simplifications that are unavoidable when tackling an infinitely complex reality. Systematic reviews on different approaches for dealing with this model inadequacy exist (\cite{Campbell2006, Bojke2009, Gupta2012, Pernot2017a, Sung2024}), but they all coincide in the need for further research to achieve reliable methods for quantifying the model form uncertainty.
	
	Bayesian approaches are usually implemented to infer the model parameters (\cite{Robert2007, Gelman2013}), which allows to obtain a probability distribution for those parameters based on the observed data.  This distribution will narrow down to the optimal parameter value, which may not be able to represent the observations in the presence of model inadequacy (\cite{Kaipio2006}). One classical solution to this problem is the framework proposed by \cite{Kennedy2001}, which extends the model output with a flexible term that corrects the predictions to better reflect the observations. Since its conception, several extensions have been proposed within the framework (\cite{Bayarri2009,Brynjarsdottir2014,Plumlee2017,Barbillon2024,Leoni2024}). However, one key disadvantage of such implementations is the impossibility of transferring the inferred estimation of the model inadequacy term to other derived Quantities of Interest (QoI) computed  with the same calibrated model (\cite{AndresArcones2023}). In contrast to these \textit{external} correction approaches that associate the uncertainty with the predictions from the model, \textit{internal} corrections arise as an alternative by attributing the uncertainty to its components instead (\cite{Wu2024}). In this way, the inferred uncertainty can be pushed forward to any QoI that uses the same model parameters for its computation.
	
	This family of approaches, also called Parameter Uncertainty Inflation (PUI) methods (\cite{Pernot2017, Pernot2017a}) have been gaining traction in the last years. The most prominent approach is Sargsyan's parameters embedding (\cite{Sargsyan2015}; \cite{Sargsyan2019}), which adds a stochastic dimension to the variables to be inferred and fits it together with the other parameters. This methodology has been successfully implemented in the context of ignition reaction (\cite{Huan2017}). As alternatives to this methodology, \cite{Mortensen2005} proposes the statistical manipulation of the inferred parameter variance, and \cite{Wu2024} suggests inferring the internal model error structure through Kalman filters.
	
	The aforementioned formulations have the advantage of offering a flexible implementation that can be adapted to the requirements of the problem. Despite being the most promising approach for dealing with model error through an internal correction, Sargsyan's proposal requires further refinement for complex inference cases (\cite{Pernot2017}). This paper aims to address the challenges that arise when using the embedded approach for the calibration of simulation models from sensor data, and in particular a) the impact of misspecified noise models, b) the presence of observations that cannot be explained with modifications of the estimated parameters, and c) the uncertainty propagation to other QoIs. A modification  in how the prescribed noise is handled is introduced in the two main likelihood  formulations proposed in \cite{Sargsyan2015} for the embedded problem- the Independent Normal (IN) and Approximate Bayesian Computation (ABC) likelihoods. Additionally, two new formulations based on the statistical convergence of the residuals distribution aimed to alleviate the flaws of IN and ABC are proposed in this paper: the Global Moment-matching (GMM) and Relative Global Moment-matching (RGMM) likelihoods. Further insight is provided for the choice in the control parameters of the likelihoods and their potential shortcomings. Finally, an analysis on the propagation of the uncertainties through QoI is developed.
	The QoIs obtained from the computational model are observed by \textit{virtual} sensors (\cite{AndresArcones2023a}). In contrast with \textit{real} sensors that are installed in the physical system, virtual sensors observe values provided by the computational model. They are not limited to observable quantities as the real sensors, but can represent any QoI. Quantifying the uncertainty in the output of virtual sensors is key for an assessment of the reliability of the predictions of QoIs.
	
	The advantages and disadvantages of each formulation will be first illustrated on a  one-dimensional,  linear example with different variations that reflect the possible errors in the observations. Afterwards, a more complex transient thermal model simulating the heat transfer through a reinforced concrete structure is investigated. The objective of the more complex case is to test the propagation of the inferred uncertainty to a non-related QoI. The structure of the rest of the paper is as follows: Section \ref{sec:methodology} presents the embedded approach, the likelihood functions and the proposed extensions, Section \ref{sec:applications} includes the simple, linear model with its variations (Section \ref{sec:linear}) and the complex transient thermal one (Section \ref{sec:thermal}), and conclusions and possible extensions are provided in Section \ref{sec:conclusion}. The methods here implemented are included in the python package \texttt{probeye} \cite{BAMResearch2024}.
	
	\section{Methodology}
	\label{sec:methodology}
	\subsection{Model form uncertainty framework}
	Let $f$ be a computable, deterministic  function $f:\Theta\times\mathbb{R}^{n_x}\to\mathbb{R}^{n_z}$ for a $n_x$-dimensional input $x\in\mathbb{R}^{n_x}$, parametrized through a set of $n_\theta$ parameters $\bm{\theta}\in\Theta\subseteq\mathbb{R}^{n_\theta}$, which have been restricted to real values for simplicity. This function $f$ models the real response $z:\mathbb{R}^{n_x}\to\mathbb{R}^{n_z}$, defined as a real-valued map that generates the $n_z$-dimensional response of a real system. In that case, an additional error term $\varepsilon_\text{model}=z-f$ rooted on the inability of $f$ to exactly reproduce $z$ is introduced. This model inadequacy term $\varepsilon_\text{model}$ will be referred to as \textit{model form uncertainty}. If the same response is measured through sensor observations $y$, a noise $\varepsilon_\text{noise}$ will introduce an additional model inadequacy with $z$. The relation between observations and computational model predictions can be expressed as
	\begin{equation}
		y = z(x) + \varepsilon_\text{noise} =  f(\theta,x) + \varepsilon_\text{model} + \varepsilon_\text{noise},
		\label{eq:noise_model}
	\end{equation}
	modeling the model inadequacy terms additively. This formulation was introduced in the seminal paper of \cite{Kennedy2001}, where they state the need for including the model error in the calibration of computational models. Alternative formulations are possible, such as a multiplicative relation between model inadequacy terms and model predictions.
	
	The objective of the so-called \emph{inverse problem} is to estimate the values for the parameters $\bm{\theta^*}$ such that $f$ best approximates $z$ as
	\begin{equation}
		\bm{\theta^*}=\arg\min_{\bm{\theta}\in\Theta}\left\|z-f(\theta,\cdot)\right\|,
	\end{equation}
	\noindent
	where $\|\cdot\|$ is a distance to be defined later. As $z$ is not generally known, it is commonly substituted by the set of $n_y$ sensor observations $\bm{y}$ for known input parameters $\bm{x}$, where bold notation will be used to denote vector quantities. The inverse problem then transforms to
	\begin{equation}
		\bm{\theta^*}=\arg\min_{\bm{\theta}\in\Theta}\left\|\bm{y}-f(\theta,\bm{x})\right\|.
	\end{equation}
	Bayesian approaches are a popular choice to solve the inverse problem (\cite{Kaipio2006}). They provide a posterior probability density $\pi(\bm{\theta}|\bm{y})$ on  the parameters  $\bm{\theta}$ given the observations $\bm{y}$ and the  computational model $\bm{f(\theta,x)}$. The application of Bayes' theorem yields
	\begin{equation}
		\pi(\bm{\theta}|\bm{y})=\frac{\pi(\bm{\theta})\pi(\bm{y}|\bm{\theta})}{\pi(\bm{y})}
		\label{eq:forward_model}
	\end{equation}
	where $\pi(\bm{\theta}|\bm{y})$ is the \textit{posterior} probability distribution, $\pi(\bm{\theta})$ is the \textit{prior} probability distribution that encompasses the previous knowledge on the latent parameters, $\pi(\bm{y}|\bm{\theta})$ is the \textit{likelihood} of the observations $\bm{y}$ having been generated by $\bm{\theta}$ and $\pi(\bm{y})$ is the \textit{marginal} probability distribution of the observations. We assume the prior distribution $\pi(\bm{\theta})$ to be given, such that the remaining key challenge is the choice and evaluation of the likelihood function $\mathcal{L}(\bm{\theta}) = \pi(\bm{y}|\bm{\theta})$ for the given set of observations $\bm{y}$. Bayesian inference approaches are based on the evaluation of the relationship from Equation \ref{eq:forward_model} to obtain the posterior distribution of the latent parameters $\bm{\theta}$.
	
	Once the posterior distribution of the latent parameters is obtained, it can be pushed forward to generate predictions $\bm{f}_P(\bm{\theta,x}|\bm{y})$ using the same forward model $f$ or they can be used in a different model $g(\bm{\theta}):\Theta\to\mathbb{R}$, for example, that computes a QoI using the same model parameters but evaluating a different quantity. In the first case $\bm{f}_P(\bm{\theta,x}|\bm{y})$ can usually be compared with $\bm{y}$ and indirectly $\bm{z}(\bm{x})$ through $\varepsilon_\text{noise}$. 
	
	The calculation of $\mathcal{L}(\bm{\theta})$ requires the evaluation of the computational model $f$ under the assumption that it generates the real system's response $z$. However, that is not generally the case and the model error must be considered. \cite{Kennedy2001} propose to include the model form uncertainty $\varepsilon_\text{model}$ in the inference procedure formulated as a Gaussian Process added to the predicted response as in Equation \ref{eq:noise_model}. Despite its potential applications, this model inadequacy term lacks physical meaning and cannot be employed outside of the use case with which it is inferred. Therefore, the uncertainty quantified by $\varepsilon_\text{model}$ cannot be propagated to other QoI through the model $g$. Notice that this model inadequacy term does not aim to identify the deficiencies in the model and correct them, but to introduce a term that envelops their effects such that the response is corrected.
	
	More importantly, adding the model inadequacy term to the predicted outputs does not address one of the main challenges of using classical Bayesian inference approaches with imperfect models: the posterior distributions of the latent parameters collapse to Dirac-delta distributions and that produce model responses that do not correspond with the system that generated the observation. This is commonly known as the problem of misspecification. The Bernstein-von Mises theorem states that under a set of conditions of continuity, differentiability and non-singularity, the posterior distribution $\pi(\bm{\theta}|\bm{y})$ converges in total variation (TV) distance with the true generating process $\pi_{\bm{\theta}_0}$ to a multivariate normal distribution centered at the maximum likelihood estimator (MLE) $\hat{\bm{\theta}}$ and covariance matrix $n_y^{-1}\mathcal{I}(\bm{\theta}_0)^{-1}$. Here, $n_y$ is the number of observations in $\bm{y}$ and $\mathcal{I}(\bm{\theta}_0)$ is Fisher's information matrix at the true values $\bm{\theta}_0$ of the latent parameters (\cite{Vaart2000}). Formally, this can be formulated as
	\begin{equation}
		\left\|\pi(\bm{\theta}|\bm{y})-\mathcal{N}(\hat{\bm{\theta}},n_y^{-1}\mathcal{I}(\bm{\theta}_0)^{-1})\right\|_{TV}\xrightarrow{\pi_{\bm{\theta}_0}}0.
	\end{equation}
	This result is crucial for constructing confidence intervals for the parameters and predictive responses, linking Bayesian and frequentist statistics. For large values of $n_y$, the posterior distribution becomes concentrated around the MLE $\hat{\bm{\theta}}$. However, \cite{Kleijn2012} demonstrated that the confidence intervals derived from Bernstein-von Mises theorem application only reflect the real credibility of the predictions in the case of perfect models that can reproduce the observations exactly.
	
	In the case of imperfect models, where discrepancies exist between the model and the data, the posterior distribution still converges to a multivariate normal distribution that concentrates around the MLE $\hat{\bm{\theta}}$ for large $n_y$. However, $\hat{\bm{\theta}}$ may not generate the observations due to the model inadequacy even for large $n_y$, and therefore the associated confidence intervals for the predictive response cannot be interpreted as credible intervals with respect to the true system, as demonstrated by \cite{Kleijn2012}. This implies that the predicted distributions do not adequately capture the variability of the true system, making them unsuitable for quantifying system uncertainty.
	
	When the model inadequacy vector $\delta_\text{model}$, that models $\varepsilon_\text{model}$ at the observation points, is implemented as a correction to the model response, the combined model $\bm{f}_P+\delta_\text{model}$  can reproduce exactly the observations $\bm{y}$. As a result, the Bernstein-von Mises theorem holds within the domain of the parameters updated during this process. However, the model $\bm{f}_P$ alone cannot reliably quantify uncertainty without correcting the $\varepsilon_\text{model}$, as it is limited to the observations' domain. Similar to the scenario where no model inadequacy term is used, the model produces overly concentrated posterior distributions for both the parameters and the response, which may fail to represent the observations accurately. As the computation of QoIs through $g(\bm{\theta})$ usually depends only on the parameters and not on the corrected model response $\bm{f}_P+\varepsilon_\text{model}$, the uncertainty in the QoI will not be representative of the credibility of the model.
	
	A promising solution to these challenges is the use of an embedded formulation where the inadequacy is added to the model through the latent parameters. The objective is to augment those parameters with an additional stochastic variable that introduces random variations in the pre-existing model parameters. It is the introduction of this variability what prevents the latent parameters from presenting a concentrated posterior for large $n_y$. Following the same philosophy as Kennedy and O'Hagan's (KOH) framework, no corrections are introduced in the structure of the physical model which allows the approach to be used non-intrusively with parametrized black-box models. This embedding is presented through Sections \ref{sec:embedding} to \ref{sub:hierarchical}, the likelihood definition and evaluation in Sections \ref{sub:likelihood} and \ref{sub:selection}, and the calculation of predicted variables through $\bm{f}_P$ and QoIs through $g$ are presented in Section \ref{sub:QoI}.
	
	\subsection{Embedding of Model Form Uncertainty} 
	\label{sec:embedding} 
	The objective of the embedding is that, despite the unavoidable collapse of the posterior distributions of the latent parameters to Dirac-deltas, the distribution of the response should preserve the variance present in the observations. This is achieved by treating the latent parameters $\bm{\theta}$ as random variables with known distributions. The objective is now to infer the values of the parameters of the distribution. We will denote these random varialbes as $\tilde{\bm{\theta}}$. This idea is similar to the one employed in hierarchical bayesian approaches, where the prior distribution of $\bm{\theta}$ is unknown. This turns the originally deterministic model into a stochastic one. With this approach, the extended model's response $f\left(\tilde{\bm{\theta}}\right)$ does not necessarily collapse to a Dirac-delta distribution, as a given sample of the inferred parameters still generates a non-concentrated distribution of the random variable $\tilde{\bm{\theta}}$.
	
	There are different ways to model the random vector $\tilde{\bm{\theta}}$. One well-known approach, employed in  \cite{Sargsyan2015} and \cite{Sargsyan2019}, uses Polynomial Chaos Expansions (PCE) to represent the stochastic, latent parameters. Specifically, the parameter vector $\tilde{\bm{\theta}}$ is expressed as a series expansion as
	\begin{equation}
		\tilde{\bm{\theta}} \sim \sum_j \alpha_j \Psi_j(\bm{\xi}),
	\end{equation}
	\noindent
	where $\alpha_j \in \mathbb{R}^{n_\theta}$ are the expansion's coefficients (included in the latent parameter space $\Theta$), $\Psi_j: \mathbb{R}^{n_\xi} \to \mathbb{R}$ are orthogonal polynomials, and $\bm{\xi}$ is a set of stochastic variables (also called the stochastic germ) with $\bm{\xi}\in\mathbb{R}^n_\xi$ where $n_\xi$ is the number of input random variables considered. This PCE-based approach allows for a flexible representation of the distribution of $\tilde{\bm{\theta}}$.
	
	In this paper, we propose a more interpretable alternative. Instead of representing $\tilde{\bm{\theta}}$ as a polynomial expansion, we adopt an explicit representation where the stochasticity is introduced through an additive model inadequacy term. This approach is in line with the proposal of \cite{Oliver2015} for model error characterization and the work of \cite{Strong2014} on internal model discrepancies decomposition. Specifically, we model the random vector $\tilde{\bm{\theta}}$ as
	\begin{equation}
		\tilde{\bm{\theta}} = \bm{\theta}^m + \delta(\bm{\theta}^b),
		\label{eq:embedded_bias}
	\end{equation}
	\noindent
	where $\bm{\theta}^m$ represents the deterministic part, and $\delta(\bm{\theta}^b)$ is a parameterized zero-mean random vector that captures the stochastic deviation from $\bm{\theta}^m$. The parameters $\bm{\theta}^b$ control the magnitude of the stochasticity, and the explicit form of $\delta(\bm{\theta}^b)$ is chosen based on prior assumptions about the model inadequacy structure.
	
	For simplicity, and because the exact shape of the distribution is often unknown, we assume that the model inadequacy $\delta$ is a random variable that depends on $\bm{\theta}^b$, following a normal distribution. Specifically, we assume  
	\begin{equation}
		\delta \left(\bm{\theta}^b\right) \sim \mathcal{N}(0, \mathrm{diag}(\bm{\theta}^b)),
	\end{equation}
	\noindent
	where $\bm{\theta}^b \in \mathbb{R}^{n_\theta}$ controls the variance of each parameter in $\tilde{\bm{\theta}}$. This formulation offers a clear interpretation of the model inadequacy associated with each parameter, making it easy to to identify the sources of uncertainty and to calibrate the model accordingly. If the parameters are constrained to be positive or known to follow a different distribution, this assumption may need to be revisited. Nevertheless, it is often possible to transform other distributions to a normal through isoprobabilistic transformations, e.g. the Nataf transform.
	
	Unlike the PCE approach, which implicitly incorporates model inadequacy through a series expansion and introduces many latent parameters for the PCE coefficients, our explicit model inadequacy representation offers a more direct and interpretable description of uncertainty. This is particularly valuable for model calibration, where identifying the sources of model inadequacy is crucial. Additionally, since the calculation of QoIs may only require some of the calibrated parameters, defining the embedded model inadequacy $\bm{\theta}^b$ independently for each parameter ensures that only the uncertainty associated with the relevant parameter is transferred. Separating the model inadequacy representation from the inference process also enables a more transparent treatment of uncertainty, reducing the identifiability issues that can arise when simultaneously fitting both the inadequacy and model parameters.
	
	For the remainder of this paper, we will use the formulation in Equation \ref{eq:embedded_bias}, assuming that $\delta(\bm{\theta}^b) \sim \mathcal{N}(0, \mathrm{diag}(\bm{\theta}^b))$. Non-additive inadequacy terms or other probability distributions are possible, but they do not affect the generality of the method presented here. Variance structures with correlation can analogously be considered for $\delta$. To generate a stochastic response from the computational model, the probability distribution of $\tilde{\bm{\theta}}$ must be pushed through the model. This presents computational challenges, as the forward model must be evaluated for each sample of $\tilde{\bm{\theta}}$. To address this, we implement a PCE approximation of the model's response while keeping our identifiable embedding, which is detailed in Section \ref{sec:pce}.
	Finally, because the model outputs are now stochastic, traditional inference methods that rely on deterministic outputs are no longer applicable. In Section \ref{sub:likelihood}, we introduce likelihood formulations inspired by Approximate Bayesian Computation (ABC) methods, which leverage summary statistics to infer the latent parameters of the model. These methods provide a robust framework for handling stochastic models, enabling a direct comparison between the stochastic model response and the available data. Instead of replicating individual observations exactly, the approach focuses on fitting their statistical moments, resulting in more reliable predictions and avoiding overfitted posterior distributions.  
	
	In Section \ref{sec:applications}, we demonstrate how introducing the stochastic extension significantly improves the probability of the observations being generated by the posterior-predicted model. This improvement is measured through the Mahalanobis distance between the predicted and observed data and is shown to outperform standard Bayesian approaches.
	
	\subsection{Forward model evaluation and Polynomial Chaos Expansion (PCE)}
	\label{sec:pce}
	The evaluation of the stochastic response of the forward model requires the propagation of the uncertainty introduced by the embedding. A closed-form representation of the response is generally not possible, therefore sampling-based methods are required. A very popular approach for uncertainty propagation in recent years has been the use of generalized Polynomial Chaos Expansions (PCE) (\cite{Xiu2002}) to approximate the stochastic response once the uncertainty has been propagated. This approach consists of approximating the response $f$ by a linear combination of a basis of orthonormal polynomials $\left\lbrace\Psi_j \right\rbrace_{j=0}^D $ truncated at degree $d$. 
	
	Let $\Psi_j:\mathbb{R}^{n_\xi}\to\mathbb{R}$ be the number of input random variables, where $\xi\in\mathbb{R}^{n_\xi}$ is the stochastic germ that determines the realization of a given random variable and $n_\xi = n_b$. These stochastic germs must follow an independent distribution inherited from $\theta$. For example, if $\theta$ is normal, $\xi$ must also follow a normal distribution. The assumption of independent input random variables, which can be enforced through the use of isoprobabilistic transformations such as Nataf or Rosenblatt transforms (\cite{Jakeman2019}). The number of polynomials $D$ is computed as $D=\binom{d+n_b}{d}$, where $n_b$ is the number of input random variables to be considered. 
	
	Then, for each output $f_i(\theta)=f(\theta, x_i)$, which refers to the model response evaluated at the $i$-th point $x_i$ of the domain $x$ (i.e., $f_i$ is $f$ evaluated at $x_i$ with fixed $\theta$), its approximation via PCE $\tilde{f}_i(\theta)$ is 
	\begin{equation}
		f_i(\theta)\approx\tilde{f}_i(\theta)=\sum_{j=0}^{D}\alpha_{ij}\Psi_{j}(\xi).
	\end{equation}
	\noindent
	As it can be observed, $\theta$ generates a single approximation for each $\tilde{f}_i(\theta)$, which are random variables that can be sampled by evaluating the PCE for different values of $\xi$. To express this more concisely, the approximation can be written in vector form $\mathbf{f}(\theta)=\left[f_1(\theta), f_2(\theta),...,f_N(\theta) \right]$ as
	\begin{equation}
		\mathbf{f}(\theta) \approx \tilde{\mathbf{f}}(\theta) = \sum_{j=0}^{D} \bm{\alpha}_j \Psi_j(\xi),
	\end{equation}
	\noindent
	where $\alpha_{ij}$ are the PCE coefficients associated with each polynomial $\Psi_j$ that fit the approximation to $f_i$  (\cite{Sudret2021}). The choice of the base of orthonormal polynomials $\Psi$ depends on the distribution of the input random variables to be propagated. Askey's scheme (\cite{Xiu2002}) assigns popular distributions with their corresponding polynomial basis that ensures orthonormality and numerical stability. 
	
	Computing the $\alpha_{ij}$ coefficients requires solving the problem of minimizing the distance $\mathcal{E}$ between $\mathbf{f}(\theta)$ and $\tilde{\mathbf{f}}(\theta)$ as in
	\begin{equation}
		\mathcal{E}=\left\|\mathbf{f}(\theta,x)-\tilde{\mathbf{f}}(\theta,x)\right\|_2=\sqrt{\mathbb{E}\left[\mathbf{f}(\theta,x)-\tilde{\mathbf{f}}(\theta,x) \right]^2}.
	\end{equation}
	\noindent
	Here, the $\alpha$ coefficients depend on $x$ (the spatial domain), and the norm $\|\cdot\|_2$ is defined as the $L_2$ norm, with the inner product as:
	\begin{equation}
		\langle f, g \rangle = \int f(\xi) g(\xi) \pi_\xi(\xi) \, d\xi,
	\end{equation}
	\noindent
	where $\pi_\xi(\xi)$ is the probability density function associated with $\xi$. The PCE coefficients that minimize $\mathcal{E}$ are given by
	\begin{equation}
		\alpha_{ij} = \frac{1}{\left\langle \Psi_j, \Psi_j\right\rangle }\int_\xi f_i(\theta)\Psi_j(\xi)\pi_\xi(\xi)\,d\xi,
	\end{equation}
	\noindent
	where $\pi_\xi(\xi)$ is the probability density associated with the random variable $\xi$. The term "stochastic germ" here refers to the random input variable, but "outcome" or "sample" could also be used. This integral can be computed using a Gauss quadrature scheme with weights $w_1, w_2, ..., w_P$ and nodes $\xi_1, \xi_2, ..., \xi_P$, where $P=p^{n_b}$, the number of quadrature points. The expansion coefficients are then computed as
	\begin{equation}
		\alpha_{ij} = \frac{1}{\left\langle \Psi_j, \Psi_j\right\rangle }\sum_{k=1}^{P}w_kf_i(\theta(\xi_k))\Psi_j(\xi_k)\pi_\xi(\xi_k),
	\end{equation}
	\noindent
	where the factor $\pi_\xi(\xi_k)$ is included in the quadrature weights $w_k$.
	
	The full process to evaluate a sample through the forward model is summarized in Algorithm \ref{alg:pce}. The Python package \texttt{chaospy} (\cite{Feinberg2015}) has been used for the PCE implementation in this paper. The result of such evaluation is the PCE of the forward model's response, which is stochastic and requires post-processing. The polynomial formulation provides direct access to the statistical moments of the response. Let $\mu^h$ be the vector of means of the response $\tilde{\mathbf{f}}(\theta, x)$. Then, each of its entries can be obtained as
	\begin{equation}
		\mu_i^h = \mathbb{E}\left[\tilde{f}_i(\theta)\right] = \alpha_{i0},
		\label{eq:pce_mu}
	\end{equation}
	\noindent
	which holds under the assumption that $\Psi_0 = 1$. Analogously, the vector of variances $\sigma^h$ is computed as
	\begin{equation}
		\left(\sigma_i^h\right)^2 = \text{Var}\left[\tilde{f}_i(\theta)\right] = \sum_{j=1}^D \alpha_{ij}^2,
		\label{eq:pce_sigma}
	\end{equation}
	\noindent
	because the variance is the sum of squares of the non-constant coefficients.
	
	\begin{algorithm}[!t]
		\caption{Forward model evaluation with Polynomial Chaos Expansion (PCE) and pseudo-spectral projection}
		\label{alg:pce}
		Given a sample $\theta$:\\
		\textit{Step 1}. Build a joint distribution $\mathcal{J}$ of the stochastic latent parameters $\theta^b$. \label{step01}\\
		\textit{Step 2}. If  $\mathcal{J}$ is not composed of independent variables, perform an isoprobabilistic transformation to make them independent, e.g., Nataf or Rosenblatt transform. \label{step02}\\
		\textit{Step 3}. Compute the weights $w_1, w_2, ..., w_D$ and nodes $\xi_1, \xi_2, ..., \xi_D$ for the Gauss quadrature scheme of degree $p$ given $\mathcal{J}$. \label{step03}\\
		\textit{Step 4}. Generate the orthonormal polynomial basis $\Psi$ of degree $d$ using Askey's scheme. \label{step04}\\
		\textit{Step 5}. Evaluate the forward model at the nodes $\xi$ for the sampled $\theta$.\\
		\textit{Step 6}. For each entry in $\mathbf{f}$, compute the PCE coefficients $\alpha_j$ using Gauss integration with weights $w$, nodes $\xi$, and the corresponding model evaluations.
	\end{algorithm}
	
	\subsection{Embedded model form uncertainty as a hierarchical Bayesian problem}
	\label{sub:hierarchical}
	The inverse problem with an embedded model form uncertainty can be interpreted within a hierarchical Bayesian framework. First, a set of \textit{hyperpriors} is prescribed for $\bm{\theta}^m$ and $\bm{\theta}^b$, from which they are sampled. Then, a prior distribution for $\bm{\theta}$ is defined through Equation \ref{eq:embedded_bias}. Finally, the forward model is evaluated based on $\bm{\theta}$. This structure follows the sequential definition of latent variables that typically appears in hierarchical Bayesian models (\cite{Robert2007}). This formulation can also be extended to use Approximate Bayesian Computation (ABC) (\cite{Turner2013}), which has been applied to inferring the model inadequacy in industrial electric motor simulations (\cite{John2021, John2021a}). However, the classical hierarchical Bayesian framework fundamentally differs from the embedded approach in several key ways, both in implementation and computational efficiency.
	
	This can be observed by comparing the posterior distribution for each approach. In the original embedded approach, the posterior distribution is expressed as:
	\begin{equation}
		\pi(\bm{\theta}^m,\bm{\theta}^b|\bm{y})\propto\pi(\bm{y}|\bm{\theta}^m,\bm{\theta}^b)\pi(\bm{\theta}^m,\bm{\theta}^b).
	\end{equation}
	\noindent
	Here, the stochastic vector of embedded variables $\tilde{\bm{\theta}}=\bm{\theta}^m+\bm{\delta}(\tilde{\bm{\theta}}^b)$ is not sampled nor treated as latent variables in the standard sense. Instead, the forward model $\bm{f}(\tilde{\bm{\theta}}^m,\tilde{\bm{\theta}}^b,\bm{x})$ produces a stochastic response directly, characterized by a distribution. This is achieved by pushing the uncertainty through the forward model using a PCE, which efficiently represents the random nature of $\delta(\tilde{\bm{\theta}}^b)$ without requiring direct sampling of $\tilde{\bm{\theta}}$.
	
	In contrast, the posterior distribution in the hierarchical Bayes framework is expressed as
	\begin{equation}
		\pi(\bm{\theta}^m,\bm{\theta}^b|\bm{y})\propto\pi(\bm{y}|\tilde{\bm{\theta}})\pi(\tilde{\bm{\theta}}|\bm{\theta}^m,\bm{\theta}^b)\pi(\bm{\theta}^m,\bm{\theta}^b).
	\end{equation}
	\noindent
	In practice, this approach necessitates access to and sampling from the conditional distribution $\pi(\tilde{\bm{\theta}}|\bm{\theta}^m,\bm{\theta}^b)$, often requiring iterative methods like Gibbs sampling (\cite{Robert2004,Robert2007}), which increases computational cost. Often a sample $\tilde{\bm{\theta}}_s$ is drawn from the conditional distribution $\pi(\tilde{\bm{\theta}}|\bm{\theta}^m,\bm{\theta}^b)$. The forward model $\bm{f}$ is evaluated for $\tilde{\bm{\theta}}_s$ as $\bm{f}(\tilde{\bm{\theta}}_s,\bm{x})$, and the response is deterministic. 
	
\begin{figure}[!h]
		\centering
		{\includegraphics[width=\textwidth]{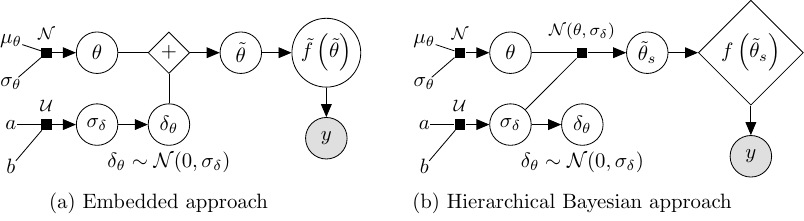}}
		{\caption{Bayesian graph for the inference of the parameters involved in the embedded formulation of the model form uncertainty. (a) Embedded approach. (b) Classical hierarchical Bayesian approach with Gibbs' sampling. Following usual notation (\cite{Dietz2010, Obermeyer2019}), circled values with white background represent latent variables, circled shaded values represent observations, rhomboids represent deterministic operations, and black squares represent drawing a sample from the indicated distribution. In this case, $\theta\sim\mathcal{N}(\mu_\theta,\sigma_\theta)$ and $\sigma_\delta\sim\mathcal{U}(a,b)$. Notice that the main difference is that $\tilde{\theta}$ is sampled in (b) but not in (a). This leads to the response $\tilde{f}(\tilde{\theta})$ being a stochastic variable in (a) obtained from pushing the distribution of $\tilde{\theta}$ through $f$ by building the PCE; but in (b) a deterministic response $f(\tilde{\theta}_s)$ is obtained for the sampled $\tilde{\theta}_s$}\label{fig:hierarchical_bayes}}
	\end{figure}

	We highlight the key difference between embedded and most hierarchical formulations: while hierarchical approaches factorize the parameters, sampling them and inputting these samples into the model to generate a deterministic response, the embedded formulation directly propagates the distributions of the parameters, resulting in a stochastic response. These distinctions are illustrated in Figure \ref{fig:hierarchical_bayes}, where the inadequacy in a single parameter $\tilde{\theta}$ is modeled such that $\tilde{\theta} = \theta^m + \delta(\theta^b)$, where $\delta \sim \mathcal{N}(0,\sigma_\delta)$. Equivalently, $\tilde{\theta} \sim \mathcal{N}(\theta^m, \sigma_\delta)$ for this simple case. This comparison highlights that both the embedded and hierarchical formulations share similar modeling principles but differ fundamentally in how the forward model response is is computed. The embedded approach handles the forward model stochastically, leveraging Polynomial Chaos Expansion (PCE) to propagate uncertainty efficiently, while the hierarchical Bayesian approach results in a deterministic forward model after sampling $\tilde{\theta}$.
	
	In simple cases such as the one presented in Figure \ref{fig:hierarchical_bayes}, the advantages of using an embedding approach over a hierarchical one may not be obvious. However, as the number of parameters and levels of dependency increases, the hierarchical approach necessitates further nested iteration loops for sampling conditional distributions, which often results in slower convergence rates. In contrast, the embedding approach reduces this issue by sampling all variables only once, albeit at the cost of introducing more complex polynomial chaos expansion (PCE) structures. Nonetheless, the curse of dimensionality associated with PCE can be mitigated using sparse integration schemes (\cite{Constantine2012}).
	
	A significant consequence of the stochastic nature of the response in the embedded approach is that classical likelihood models commonly employed in hierarchical Bayesian models are not directly applicable. Instead, PCE facilitates rapid computation of moments, which can be utilized for moment-matching likelihood computations inspired by approximate Bayesian computation (ABC) methods. By explicitly fitting the statistical moments of the predictions, this approach offers additional insights and enhances control over the update procedure compared to the hierarchical approach.
	
	Hierarchical methods may be preferable in scenarios where efficiently sampling the conditional distributions of latent parameters is feasible. This is particularly true for cases with a limited number of parameters or complex stochastic structures requiring higher-degree polynomials in the PCE for accurate approximation. In these situations, despite the expected slower convergence rates of hierarchical Bayesian methods using iterative algorithms, they may still outperform the embedding approach due to the additional forward model evaluations required.
	
	While both methodologies have practical applications, an exhaustive comparison between them is beyond the scope of this paper. \cite{Pernot2017a} compared, among others, an ABC implementation of the embedded approach and a hierarchical one, concluding that the former is a better fit for the estimation of parameters under model form uncertainty. For improved results, the hierarchical formulation required a local modification of the parameters, effectively increasing their number. For a more comprehensive discussion on hierarchical Bayesian approaches and their computational challenges, refer to \cite{Robert2004, Robert2007, Turner2013, Li2024}.
	
	\subsection{Likelihood formulation}
	\label{sub:likelihood}
	Due to the uncertainty propagation, the output $\bm{f}(\bm{\theta})$ of evaluating the forward model with a given sample $\bm{\theta}$ or its PCE approximation $\bm{\tilde{f}}(\bm{\theta})$ is a stochastic random variable that follows a probability distribution described by the PCE approximation. While a modified approach that utilizes the statistical moments of the distribution is feasible, directly evaluating a classical Gaussian likelihood function based on the residuals between the predictions, $\bm{\tilde{f}}(\bm{\theta})$, and the observations, $\bm{y}$, is not possible because their domains are not directly comparable, $\bm{\tilde{f}}(\bm{\theta})$ being a random variable and $\bm{y}$ being deterministic observations. Approximate Bayesian Computation (ABC) approaches provide a methodology to obtain the likelihood $\pi_\text{ABC}(\bm{y}|\bm{\theta})$ from the statistical moments of $\bm{\tilde{f}}(\bm{\theta})$, which are readily available from the PCE, enabling the formulation of a likelihood function.
	
	Throughout this paper, an ensemble-based Monte Carlo-Markov Chain (MCMC) sampler using a stretch move (\cite{Goodman2010}) from the package \texttt{emcee} (\cite{ForemanMackey2013}) is used. A threshold on the \emph{effective sample size} ($ESS$) for each of the latent parameter chains as described in Appendix \ref{ap:ess_threshold} is set as stopping criteria for the MCMC algorithm.
	
	\subsubsection{ABC-Likelihood without noise}
	Classical ABC approaches usually include two assumptions: (1) measurements are free of noise, $\varepsilon_\text{noise}\overset{!}{=}0$, and (2) the model response can fully represent the system, $\varepsilon_\text{model}\overset{!}{=}0$. Therefore, the model response is expected to fully represent the system's real behaviour as $\bm{z}=\bm{f}(\bm{\theta})$. Those approaches define a residual function $\rho(\bm{y},\bm{\tilde{f}}(\bm{\theta})):\mathbb{R}^{n_y}\times\mathbb{R}^{n_z}\to\mathbb{R}$ between observations $\bm{y}$ and predictions $\bm{\tilde{f}}(\bm{\theta})$, which usually defines a distance, and aim at reducing it below a tolerance value $\epsilon$ by the following steps (\cite{Sisson2018}):
	\begin{enumerate}
		\item Sample $\bm{\theta}$ from $p(\bm{\theta})$.
		\item Evaluate the forward model $\bm{z}=\bm{f}(\bm{\theta})\sim p(\bm{z}|\bm{\theta})$.
		\item Accept or reject $\bm{\theta}$ if $\rho(\bm{y},\bm{f}(\bm{\theta}))<\epsilon$.
	\end{enumerate}
	If the model is approximated, such as the forward model built in Section \ref{sec:pce}, its evaluation would correspond to $\bm{\tilde{z}}=\bm{\tilde{f}}(\bm{\theta})\sim p(\bm{\tilde{z}}|\bm{\theta})$, where the model form uncertainty is expected to be fully represented in the stochastic response of the computational model, preserving assumption (2). When this tolerance $\epsilon\to 0$, the predictions are expected to reproduce exactly the observations, and therefore the samples of $\theta$ would belong to the associated posterior distribution. As $\bm{y}$ is deterministic and $\bm{\tilde{f}}(\bm{\theta})$ stochastic, the residual function $\rho$ is evaluated on its statistical moments $t$ such that
	\begin{enumerate}
		\setcounter{enumi}{2}
		\item[3a.] Accept or reject $\bm{\theta}$ if $\rho(t(\bm{y}),t(\bm{\tilde{f}}(\bm{\theta})))<\epsilon$.
	\end{enumerate}
	A possibility is to use the mean and standard deviation as summary statistics under the assumption of them being statistically \textit{sufficient}. The predictions $\bm{\tilde{f}}(\bm{\theta})$ are approximated at each output dimension by $t(\bm{\tilde{f}}(\bm{\theta}))=[\bm{\mu}^h,\bm{\sigma}^h]$. The output observations $\bm{y}$ are approximated by its statistical moments as $t(\bm{y})=[\bm{\mu},\bm{\sigma}]$, where $\bm{\mu}\approx\bm{y}$ and $\bm{\sigma}\approx \gamma|\bm{\mu}^h-\bm{y}|$ with $\gamma$ an additional factor to be specified. The approximation $\bm{\sigma}\approx \gamma|\bm{\mu}^h-\bm{y}|$ is necessary due to only one observation being available. A reasonable assumption would be that on, average, the observations $\bm{y}$ are located a given distance from the mean $\bm{\mu}^h$ controlled by their standard deviations $\bm{\sigma}$, leading to the previous approximation. The choice of summary statistics is not unique and introduces a potential model inadequacy rooted in the approximation that is not present in the formulation (\cite{Fearnhead2012}).
	
	The acceptance criterion defines the \textit{indicator function} or \textit{kernel} that governs the sampling procedure. Following the criteria from Step 3, the uniform kernel $I$ indicates if a sample is accepted and can be defined as
	\begin{equation}
		I(\bm{\theta},\bm{y})=\begin{cases}
			1 & \text{if}~\rho(t(\bm{y}),t(\bm{\tilde{f}}(\bm{\theta})))<\epsilon\\
			0 & \text{else}
		\end{cases}.
	\end{equation}
	\noindent
	Applying Bayes' theorem with the chosen kernel, it is possible to express the approximate posterior distribution of the parameters as
	\begin{equation}
		\pi_\text{ABC}(\bm{\theta}|\bm{y})\propto\pi(\bm{\theta})\int_{\mathbb{R}^{n_z}}I(\bm{\theta},\bm{y})\pi(\bm{z}|\bm{\theta})\text{d}\bm{z}
	\end{equation}
	\noindent
	where $\pi(\bm{\theta})$ is the prior distribution for the latent parameters $\bm{\theta}$ and $\pi(\bm{z}|\bm{\theta})$ the probability of the model output for a given $\bm{\theta}$. For a deterministic model $f$, the probability $\pi(\bm{z}|\bm{\theta})$ follows a Dirac's delta distribution located at the value for $\bm{f}(\bm{\theta})$. Analogously, the summary statistics $t(\bm{\tilde{z}})$ of the approximated output $\bm{\tilde{z}}=\bm{\tilde{f}}(\bm{\theta})$ are deterministic as well, therefore $\pi(t(\bm{\tilde{z}})|\bm{\theta})$ follows the same Dirac's delta distribution when they are used as a lower-dimensional representation of the model output. Under this consideration, it can be observed that the kernel acts as the likelihood function that is commonly defined in the classical Bayesian approaches. In particular, the likelihood defined by the aforementioned uniform kernel assigns probability 1 to those samples that fulfil $\rho(t(\bm{y}),t(\bm{\tilde{z}}))<\epsilon$ and 0 elsewhere.
	
	Following the same reasoning, it is possible to implement other likelihood functions that reciprocally induce different kernels that govern the acceptance criteria. Such is the case for the \textit{moment-matching} likelihood used in \cite{Sargsyan2015} and \cite{Sargsyan2019}, which adopts an exponential structure:
	\begin{equation}
		\mathcal{L}^{\text{ABC}}(\bm{\theta})=\left( 2\pi\epsilon^2\right)^{-\frac{1}{2}}\prod_{i=1}^{n_y}\exp\left(-\frac{(\mu_i^h-y_i)^2+\left( \sigma_i^h-\gamma|\mu_i^h-y_i|\right) ^2}{2\epsilon^2} \right) 
		\label{eq:likelihood_sargsyan}
	\end{equation}
	\noindent
	hence
	\begin{equation}
		\pi_\text{ABC}(\bm{\theta}|\bm{y})\propto\pi(\bm{\theta})\mathcal{L}^\text{ABC}(\bm{\theta}).
		\label{eq:post_abc}
	\end{equation}
	\noindent
	This likelihood assigns higher probabilities to those samples that minimize the squared distance between the summary statistics $[\bm{\mu}^h,\bm{\sigma}^h]$ and the observations, defined as 
	\begin{equation}
		\rho(t(\bm{y}),t(\bm{\tilde{f}}(\bm{\theta})))=\begin{bmatrix}
			\bm{\mu}^h-\bm{y} \\
			\bm{\sigma}^h-\gamma|\bm{\mu}^h-\bm{y}| 
		\end{bmatrix}^T\begin{bmatrix}
			\bm{\mu}^h-\bm{y} \\
			\bm{\sigma}^h-\gamma|\bm{\mu}^h-\bm{y}| 
		\end{bmatrix}.
	\end{equation}
	\noindent
	While this distance has proven convergence (\cite{Prangle2017}), it is primarily influenced by the statistical moment with the largest variance . Notably, the value of $\bm{\mu}^h$ is mainly affected by variations in the original latent parameters, whereas $\bm{\sigma}^h$ is contingent upon the likelihood parameters associated with the embedded model inadequacy. The moment-matching likelihood is justified by the assumption of normality in the differences between the predicted summary statistics and those of the observations. The goal is to achieve an exact match between the predicted and observed statistical moments, which occurs as $\epsilon$ approaches 0, assuming there is no model inadequacy —that is, when the model can perfectly replicate the real system. When the measurement noise $\varepsilon_\text{noise}$ is either zero or can be completely absorbed by the model form uncertainty, the posterior predictive distribution generated using this likelihood converges to the true distribution, provided that the summary statistics are sufficient.
	
	The choice of $\gamma$ is also not trivial and depends on the assumptions imposed on the residuum's distribution. It can be argued that, supposing a normal independent probability distribution for the residuum, $\gamma=\sqrt{\frac{\pi}{2}}$ should be chosen over $\gamma=1$ if the desired expected predictive distribution is supposed to have, on average, the magnitude of the residuum as the standard deviation. For a variable $u$ that follows a zero-mean normal distribution $u\sim\mathcal{N}(0,\sigma)$, its absolute value $|u|$ follows a half-normal distribution (\cite{Leone1961, Gelman2013}), and thus
	\begin{equation}
		\mathbb{E}[|u|]=\sigma\sqrt{\frac{2}{\pi}}.
		\label{eq:mean_seminormal}
	\end{equation}
	Therefore, if $\bm{\sigma}^h$ on average should be equal to the residuum $\bm{\mu}^h-\bm{y}$ at every point and not to the expectation of $|\bm{\mu}^h-\bm{y}|$, a correction of $\gamma=\sqrt{\frac{\pi}{2}}$ is required. In such case, the residuum obtained from the observations is, on average, one standard deviation $\bm{\sigma}^h$ from the mean. \cite{Sargsyan2019} proposes taking $\gamma=1$ as a general case instead, specifying then $\bm{\sigma}\approx |\bm{\mu}^h-\bm{y}|$, observing that alternative values of $\gamma$ should be informed by knowledge on the system (\cite{Sargsyan2019}). However, for the likelihood of Equation \ref{eq:likelihood_sargsyan}, that would imply underestimating the predictive variance, hence we will employ $\gamma=\sqrt{\frac{\pi}{2}}$ when applying this ABC likelihood. Further discussion on the impact of modifying $\gamma$ and $\epsilon$ for this likelihood can be found in \cite{Sargsyan2015}. Equation \ref{eq:post_abc} can be implemented in an MCMC framework to estimate the posterior distribution (\cite{Marjoram2003, Sisson2010}). 
	
	\subsubsection{ABC-Likelihood with noise}
	\label{sub:abc}
	The likelihood formulated in Equation \ref{eq:likelihood_sargsyan} does not explicitly include the measurement noise that appears in the observations. It supposes that the moments can be matched exactly up to a constant $\epsilon$, which in reality is generally not true. In cases where measurement noise is present, it is not possible to reduce $\epsilon$ to zero, which invalidates the moment-matching condition and introduces an error in the posterior. Therefore, the explicit inclusion of noise is required to generate a reliable posterior based on real sensor observations.
	
	Three main alternatives have been proposed by \cite{Schaelte2020} to include noise in ABC approaches: (1) controlling the noise through $\epsilon$, (2) simulating samples from the noisy model $f(\theta)+\varepsilon_\text{noise}$ and (3) including the noise model in the sample acceptance criteria, and therefore the likelihood function. Fitting the tolerance to the expected noise is not trivial, as the variation in the summary statistics does not necessarily coincide with the prescribed measurement noise, and it limits the application to uniform noise models (\cite{Daly2017}). Alternatively, simulating samples leads to low acceptance rates and largely increases the number of required model evaluations per sample. 
	Finally, the modification of the acceptance criteria has been proven to correctly converge to the posterior of the summary statistics with noise (\cite{Wilkinson2013}). Based on this property, a modification to include the noise in the moment-matching likelihood function is proposed here. 
	
	Following \cite{Wilkinson2013}, the acceptance step of the general ABC outline is modified to include the noise error:
	\begin{enumerate}
		\setcounter{enumi}{2}
		\item[3b.] Accept $\bm{\theta}$ with probability $\frac{\pi_\varepsilon(\rho(t(\bm{y}),t(\bm{\tilde{z}})))}{c}$.
	\end{enumerate}
	where $\pi_\varepsilon(\rho(t(\bm{y}),t(\bm{\tilde{z}})))$ represents the probability error model for the summary statistics, where $\rho(t(\bm{y}),t(\bm{\tilde{z}}))$ defines a residual function between them, typically a signed distance,  and $c$ is a normalization constant. Applying the MCMC framework and using $\pi_\varepsilon(\rho(t(\bm{y}),t(\bm{\tilde{z}})))$ as likelihood, the posterior distribution of the latent parameters can be approximated as
	\begin{equation}
		\pi_\text{ABC}(\bm{\theta}, \bm{z}|\bm{y})\approx\pi_\text{ABC}(\bm{\theta}, t(\bm{z})|t(\bm{y}))\propto\pi_\varepsilon(\rho(t(\bm{y}),t(\bm{\tilde{z}})))\pi(\bm{\theta}).
		\label{eq:ABC_likelihood_bayes}
	\end{equation}
	\noindent
	The proof of convergence is presented in Appendix \ref{ap:convergence_noise} under consideration that $t(\bm{\tilde{z}})=t\left(\bm{\tilde{f}}(\bm{\theta})\right)$ is deterministic. 
	
	The chosen summary statistics are the same as for the original moment-matching likelihood: the mean and standard deviation of the model. The error $\varepsilon_\text{noise}=\bm{y}-\bm{\tilde{z}}$ follows a prescribed homogeneous Gaussian noise model $\mathcal{N}(0,\sigma_N\bm{I})$. Therefore, the noise associated with the means has variance $\sigma_N^2$. On the other hand, the matching of predicted and estimated standard deviations is considered exactly up to the tolerance $\epsilon$. The predicted standard deviation now not only considers the variation in the output $\bm{\sigma}^h$ but also includes the noise vector $\bm{\sigma}_N$. Therefore, the second moment comparison with noise is $\bm{\sigma}=\bm{\sigma}^h+\bm{\sigma}_N\approx \gamma|\bm{\mu}^h-\bm{y}|$. With this new formulation, the variance in the residuals $\bm{u}=|\bm{\mu}^h-\bm{y}|$ will be explained by the prescribed measurement error first if possible, and the predicted variance would intuitively only be increased to cover for the remaining variance. For the aforementioned homogeneous Gaussian error, the noisy moment-matching likelihood function is
	\begin{equation}
		\mathcal{L}^{\text{ABC}}(\bm{\theta})=\left( 2\pi\epsilon^2\sigma_N^2\right)^{-\frac{n_y}{2}} \prod_{i=1}^{n_y}\exp\left(-\frac{(\mu_i^h-y_i^h)^2}{2\sigma_N^2}-\frac{\left( \sigma_i^h+\sigma_N-\gamma|\mu_i^h-y_i^h|\right) ^2}{2\epsilon^2} \right). 
		\label{eq:noisy_moment_matching}
	\end{equation}
	\noindent 
	This modified likelihood presents several advantages in the case of measurements with noise. First and most prominently, the predicted and sampled means are no longer matched exactly, as their fitness is evaluated considering the known measurement noise structure. This avoids over-fitting to the samples and provides more informative posterior distributions that consider a larger part of the available information. Additionally, the variance from measurement noise is no longer absorbed by the stochastic part of the model prediction, which avoids overestimating the model form uncertainty.
	
	Nevertheless, both moment-matching likelihoods are sensitive to badly represented models. As proven in \ref{ap:behaviour_noise}, if the prescribed noise is larger than the residuals, the inference procedure will converge to a posterior distribution that not necessarily represents the real system. This occurs when the sampling algorithm favours samples with a larger residual in order to exactly match the standard deviation of the prediction with the prescribed noise model. However, this situation can be avoided by a careful selection of the prescribed noise model, as is generally the case in practice. Alternatively, the noise parameter can be inferred as a latent parameter, which is possible only for the noisy moment-matching likelihood function.
	
	\subsubsection{Independent normal likelihood}
	\label{sec:independent_normal}
	
	An alternative to moment-matching likelihoods is to adapt the classical Gaussian Likelihood to take into consideration the predictive variance. Based on the assumption that the residuals follow an independent normal distribution with mean zero and standard deviation of the predicted one, \cite{Sargsyan2015} propose
	\begin{equation}
		\mathcal{L}^{\text{IN}}(\bm{\theta})=(2\pi)^{-\frac{n_y}{2}}\prod_{i=1}^{n_y}(\sigma_i^h)^{-1}\exp{\left(-\frac{\left(y_i-\mu_i^h\right)^2}{2(\sigma_i^h)^2}\right)}.
	\end{equation}
	If noise is accounted for in the variance computation, the likelihood must be modified accordingly, resulting in
	\begin{equation}
		\mathcal{L}^{\text{IN}}(\bm{\theta})=(2\pi)^{-\frac{n_y}{2}}\prod_{i=1}^{n_y}\left(\sqrt{(\sigma_i^h)^2+\sigma_N^2}\right)^{-1}\exp{\left(-\frac{\left(y_i-\mu_i^h\right)^2}{2\left((\sigma_i^h)^2+\sigma_N^2\right)}\right)}.
		\label{eq:in_noise}
	\end{equation}
	\subsubsection{Global Moment-matching (GMM) Likelihood with noise}
	The \textit{Global Moment-matching Likelihood} aims to avoid the fitting of the predictive moments to each $y_i$ individually. This is achieved by considering the predictive distributions of  $f^h(\bm{\theta}, \boldsymbol{x}_i)$ centered at zero as the components of a gaussian mixture model $F(\theta)$. Then, a likelihood function is formulated by treating the observed residuals $y_i-\mu^h_i$ as potential samples from $F(\theta)$. This approach utilizes the sampling distributions of the moments of a normal distribution to define $\pi_\varepsilon$ in Equation \ref{eq:ABC_likelihood_bayes}, as shown in Appendix \ref{ap:convergence_noise}.
	
	First, we observe that at each single set of coordinates $\boldsymbol{x}_i$ for all $i=1,2,...,n_y$ observation points, we can define a random variable $f^h(\bm{\theta},\boldsymbol{x}_i)$ with known moments $\mu^h(\bm{\theta},\boldsymbol{x}_i)$ and $\sigma^h(\bm{\theta},\boldsymbol{x}_i)^2+\sigma_N^2$ that describes the predictions of the model at $\boldsymbol{x}_i$ for a given sample of the latent parameters $\bm{\theta}$. We center every observation by the predictive mean as $f^h(\bm{\theta},\boldsymbol{x}_i)-\mu^h(\bm{\theta},\boldsymbol{x}_i)\sim\mathcal{N}\left(0,\sqrt{\sigma^h(\bm{\theta},\boldsymbol{x}_i)^2+\sigma_N^2}\right)$. Then, we define $F(\bm{\theta})$ as a gaussian mixture model with equal weights following \cite{Murphy2012} as 
	\begin{equation}
		F(\bm{\theta})=\frac{1}{n_y}\sum_{i=1}^{n_y} \left(f^h(\bm{\theta},\boldsymbol{x}_i)-\mu^h(\bm{\theta},\boldsymbol{x}_i)\right)\sim\frac{1}{n_y}\sum_{i=1}^{n_y}\mathcal{N}\left(0,\sqrt{\sigma^h(\bm{\theta},\boldsymbol{x}_i)^2+\sigma_N^2}\right).
	\end{equation}
	The moments of $F(\bm{\theta})$ can be directly computed as 
	\begin{align}
		t_1(F(\bm{\theta}))=\mu_F=\text{E}\left[F(\bm{\theta})\right] &= 0\\
		t_2(F(\bm{\theta}))=\sigma_F^2=\text{Var}\left(F(\bm{\theta})\right)&=\overline{\left(\sigma^h(\bm{\theta},\boldsymbol{x}_i)^2\right)}+\sigma_N^2,\label{eq:gmm_var_f}
	\end{align}
	where, following usual notation, $E\lbrack\cdot\rbrack$ denotes expectation of $\cdot$, $\text{Var}(\cdot)$ denotes variance of $\cdot$ and $\bar{\cdot}$ denotes the average of $\cdot$ over all $i$. We choose $F(\bm{\theta})$ to represent the population. We observe that these are estimates of the population, as its true moments are not available only from samples. Its corresponding sampling distributions are obtained by applying the Cochran theorem \citep{Cochran1934}. In particular, the distribution of the sampled mean and variance from $F(\bm{\theta})$ follow
	\begin{align}
		\mu_U-\mu_F&\sim\mathcal{N}\left(0,\frac{\sigma_F}{\sqrt{n_y}}\right)\label{eq:gmm_mean}\\
		\frac{n_y\sigma_U^2}{\sigma_F^2}&\sim\chi_{n_y-1}^2,\label{eq:gmm_var}
	\end{align}
	where $\mu_U$ and $\sigma_U$ are the moments of the samples and $\chi_{n-1}^2$ is a chi-square distribution of $n-1$ degrees of freedom, which must be obtained from the residuals $u_i=y_i-\mu^h(\bm{\theta},\boldsymbol{x}_i)$ with $i=1,2,...,n_y$, which are realizations of the random variable $U\in\mathbb{R}$. Applying the derivations for ABC likelihood with noise from Appendix~\ref{ap:convergence_noise} to Equations \ref{eq:gmm_mean} and \ref{eq:gmm_var}, the probability distribution of the error based on the samples can then be formulated as
	\begin{equation}
		\pi_\varepsilon\left(\rho(t(U(\bm{\theta})),t\left(F(\bm{\theta})\right))\right)=\pi_{t_1}\left(t_1(U(\bm{\theta}))-t_1\left(F(\bm{\theta})\right)\right)\pi_{t_2}\left(\frac{t_2\left(U(\bm{\theta})\right)}{t_2(F(\bm{\theta}))}\right).
		\label{eq:composed_noise_gmm}
	\end{equation}
	In this case, the first moment $t_1$ will be the mean and $t_2$ will be the variances of the set of samples of $x$. We will use Equations \ref{eq:gmm_mean} and \ref{eq:gmm_var} as error models for $\pi_{t_1}$ and $\pi_{t_2}$, respectively; requiring modeling of the error in $\pi_{t_1}$ as a difference and in $\pi_{t_2}$ as a quotient. 
	The mean $t_1(U(\bm{\theta}))=\mu_U$ and variance $t_2(U(\bm{\theta}))=\sigma_U^2$ are directly estimated from the samples as
	\begin{align}
		\hat{t}_1(U(\bm{\theta}))=\hat{\mu}_U&=\overline{\boldsymbol{u}}\\
		\hat{t}_2(U(\bm{\theta}))=\hat{s}_U^2&=\frac{\sum\limits_{i=0}^{n_y}u_i}{n_y-1},\label{eq:gmm_var_u}
	\end{align}
	where $\overline{\bm{u}}$ represents the mean of the residuals vector $\bm{u}$. The dependency of $\boldsymbol{u}$ and its statistical moments of $\bm{\theta}$ will not be explicitly stated for notation purposes. Assuming that $\sigma_F$ is a good approximation for the population, the sampled variance $\sigma_U^2$ is a sufficient statistic for $\sigma_F^2$. Therefore, the requirements from \cite{Wilkinson2013} for a valid ABC probability distribution with noise hold for Equation \ref{eq:composed_noise_gmm} and the population with variance $\sigma_F^2$. 
	
	Although the assumption of independence among samples is typically not valid for computational models with correlated outputs, the correlation structure of the residuals is often unknown a priori. While computing and incorporating correlation effects is beyond the scope of this paper, one potential strategy would involve including the covariance structure in the calculation of $\sigma_F^2$.
	
	The likelihood function for the mean would directly be obtained by evaluating the error model of Equation \ref{eq:composed_noise_gmm} as
	\begin{equation}
		\mathcal{L}_1^{\text{GMM}}(\bm{\theta})=\pi_{t_1}\left(\hat{t}_1(U(\bm{\theta}))-t_1\left(F(\bm{\theta})\right)\right)=\pi_{t_1}\left(\bar{\bm{u}}\right)= \left(\frac{2\pi}{n_y} \sigma_F^2\right) ^{-\frac{1}{2}}\exp\left(-\frac{n_y\bar{\bm{u}}^2}{2\sigma_F^2} \right),
		\label{eq:gmm-l1}
	\end{equation} 
	\noindent
	which is equivalent to the independent normal likelihood from section \ref{sec:independent_normal}.This likelihood formulation can be interpreted as the probability of the estimated first moment of the samples $\bm{u}$ assigned by the error model obtained from the sample distribution of Equation \ref{eq:gmm_mean}.
	We proceed analogously to build the likelihood function for the second moment. The density function of a $\chi^2_{n-1}$ distribution can be expressed as
	\begin{equation}
		f(x,n-1)=\frac{x^{\frac{n-1}{2}-1}\exp(-\frac{1}{2}x)}{2^\frac{n-1}{2}\Gamma\left(\frac{n-1}{2} \right) }.
	\end{equation}
	\noindent
	Hence, substituting $x=n_y\hat{s}_U^2/\sigma_F^2$ in the $\chi^2_{n_y-1}$ density function, it follows
	\begin{equation}
		\mathcal{L}_2^{\text{GMM}}(\bm{\theta})=\pi_{t_2}\left(\frac{\hat{t}_2(U(\bm{\theta}))}{t_2\left(F(\bm{\theta})\right)}\right)=\pi_{t_2}\left(\frac{\hat{s}_U^2}{\sigma_F^2}\right)=\frac{1}{2^\frac{n_y-1}{2}\Gamma\left(\frac{n_y-1}{2} \right) }\exp\left(-\frac{n_y\hat{s}_U^2}{2\sigma_F^2}\right)\left( \frac{n_y\hat{s}_U^2}{\sigma_F^2}\right)  ^{\frac{n_y-1}{2}-1}
		\label{eq:gmm-l2}
	\end{equation}
	\noindent
	where $\hat{s}_U^2$ and $\sigma_F^2$ are obtained from Equations \ref{eq:gmm_var_u} and \ref{eq:gmm_var_f}. 
	The global moment-matching likelihood will then be
	\begin{equation}
		\pi_\varepsilon\left(\rho(t(U(\bm{\theta})),t(F(\bm{\theta})))\right)=\pi_\varepsilon^{\text{GMM}}(\bm{\theta})=\mathcal{L}_1^{\text{GMM}}(\bm{\theta})\mathcal{L}_2^{\text{GMM}}(\bm{\theta})
		\label{eq:gmm_l12}
	\end{equation}
	and in logarithmic form,
	\begin{equation}
		\begin{split}
			\log\pi_\varepsilon^{\text{GMM}}(\bm{\theta})&=-\frac{1}{2}\log\left( \frac{2\pi}{n_y} \sigma_F^2 \right) -\frac{n_y\bar{u}^2}{2\sigma_F^2}\\
			&-\frac{n_y-1}{2}\log 2 - \log\left(\Gamma\left( \frac{n_y-1}{2}\right)  \right)  -\frac{n_y\hat{s}_U^2}{2\sigma_F^2}+\left(\frac{n_y-1}{2}-1\right)\log\left(\frac{n_y\hat{s}_U^2}{\sigma_F^2}\right).
		\end{split}
		\label{eq:gmm-log}
	\end{equation}
	
	\subsubsection{Relative Global Moment-matching (RGMM) Likelihood with noise}
	\label{sub:rgmm}
	Building the GMM likelihood directly form the residual of the observations may lead to outliers from points where the predicted variance is high. Therefore, in cases with very heteroscedastic noise or model form uncertainties, the previous likelihood will lead to biased estimations of the embedded parameters, which are mainly controlled by such effects. To compensate this, we propose a \textit{relative} version of the GMM likelihood, that normalizes the centered predictions by the predicted variance. In effect, the Gaussian Mixture Model with the relative residuals is directly
	\begin{equation}
		F_r(\bm{\theta})=\frac{1}{n_y}\sum_{i=1}^{n_y} \left(\frac{f^h(\bm{\theta},\boldsymbol{x}_i)-\mu^h(\bm{\theta},\boldsymbol{x}_i)}{\sqrt{\sigma^h(\bm{\theta},\boldsymbol{x}_i)^2+\sigma_N^2}}\right)\sim\frac{1}{n_y}\sum_{i=1}^{n_y}\mathcal{N}\left(0,1\right).
	\end{equation}
	Then, the moments of $F_r(\bm{\theta})$ are directly $\mu_{F_r}=0$ and $\sigma_{F_r}^2=1$. We apply once more Cochran's theorem for the sampled moments of $F_r(\bm{\theta})$ as
	\begin{align}
		\mu_{U_r}-\mu_{F_r}&\sim\mathcal{N}\left(0,\frac{\sigma_{F_r}}{\sqrt{n_y}}\right)\label{eq:rgmm_mean}\\
		\frac{n_y\sigma_{U_r}^2}{\sigma_{F_r}^2}&\sim\chi_{n_y-1}^2,\label{eq:rgmm_var}
	\end{align}
	where $\mu_{U_r}$ and $\sigma^2_{U_r}$ are the moments normalized by the predictive variances. Defining the relative residuals as
	\begin{equation}
		{u_r}_i=\frac{y_i-\mu^h(\bm{\theta},\bm{x}_i)}{\sqrt{\sigma^h(\bm{\theta},\boldsymbol{x}_i)^2+\sigma_N^2}}~~~~\text{with }i=1,2,...,n_y,
	\end{equation}
	these sample moments are calculated as 
	\begin{align}
		\hat{t}_1(U_r(\bm{\theta}))=\hat{\mu}_{U_r}&=\bar{\boldsymbol{u}}_r\\
		\hat{t}_2(U_r(\bm{\theta}))=\hat{s}_{U_r}^2&=\frac{\sum\limits_{i=0}^{n_y}{u_r}_i}{n_y-1}.\label{eq:rgmm_var_u}
	\end{align}
	Following the same reasoning as for the GMM likelihood, we obtain the components of the corresponding RGMM likelihood function $\pi_\varepsilon^{RGMM}(\bm{\theta})=\mathcal{L}_1^{RGMM}(\bm{\theta})\mathcal{L}_2^{RGMM}(\bm{\theta}),$ where
	\begin{equation}
		\mathcal{L}_1^{\text{RGMM}}(\bm{\theta})= \left(\frac{2\pi}{n_y} \sigma_{F_r}^2\right) ^{-\frac{1}{2}}\exp\left(-\frac{n_y\bar{\bm{u}}_r^2}{2\sigma_{F_r}^2} \right),
		\label{eq:rgmm-l1}
	\end{equation} 
	and
	\begin{equation}
		\mathcal{L}_2^{\text{RGMM}}(\bm{\theta})=\frac{1}{2^\frac{n_y-1}{2}\Gamma\left(\frac{n_y-1}{2} \right) }\exp\left(-\frac{n_y\hat{s}_{U_r}^2}{2\sigma_{F_r}^2}\right)\left( \frac{n_y\hat{s}_{U_r}^2}{\sigma_{F_r}^2}\right)  ^{\frac{n_y-1}{2}-1}.
		\label{eq:rgmm-l2}
	\end{equation}
	The likelihood in logarithmic form is
	\begin{equation}
		\begin{split}
			\log\pi_\varepsilon^{\text{RGMM}}(\bm{\theta})&=-\frac{1}{2}\log\left( \frac{2\pi}{n_y} \sigma_{F_r}^2 \right) -\frac{n_y\bar{u}_r^2}{2\sigma_{F_r}^2}\\
			&-\frac{n_y-1}{2}\log 2 - \log\left(\Gamma\left( \frac{n_y-1}{2}\right)  \right)  -\frac{n_y\hat{s}_{U_r}^2}{2\sigma_{F_r}^2}+\left(\frac{n_y-1}{2}-1\right)\log\left(\frac{n_y\hat{s}_{U_r}^2}{\sigma_{F_r}^2}\right).
		\end{split}
		\label{eq:rgmm-log}
	\end{equation}
	
	
	\subsection{Summary on likelihood formulations}
	\label{sub:selection}
	
	In this article, four likelihood functions are compared: ABC-Likelihood with noise (ABC), Global Moment-matching (GMM) likelihood, Relative Global Moment-matching (RGMM) and Independent Normal (IN) likelihood with noise. GMM and RGMM are new proposals developed in this work, while ABC and IN have been adapted from \cite{Sargsyan2019} to explicitly include noise. As a common feature, the four of them aim to match the mean and variance of the predicted output distribution with the data. A summary of the likelihood models is presented in Table \ref{tab:likelihood_models}.
	
	\begin{table}[!h]
		{\caption{Moment-matching likelihoods summary}}{
				\renewcommand{\arraystretch}{1.5}
				\begin{tabular}{|p{3.7cm}|p{2cm}|p{4cm}|p{2.9cm}|}
					\hline
					\multicolumn{1}{|c|}{\textbf{Likelihood model}}   & \multicolumn{1}{c|}{\textbf{Formulation}} & \multicolumn{1}{c|}{\textbf{Mean matching}}   & \multicolumn{1}{c|}{\textbf{Variance matching}}\\
					\hline
					Approximate Bayesian Computation (ABC) & Equation \ref{eq:noisy_moment_matching} & $-\frac{(\mu_i^h-y_i^h)^2}{2\sigma_N^2}$ & $-\frac{\left( \sigma_i^h+\sigma_N-\gamma|\mu_i^h-y_i^h|\right) ^2}{2\epsilon^2}$\\\hline
					Independent Normal & Equation \ref{eq:in_noise} & $e^{-\frac{\left(y_i-\mu_i^h\right)^2}{2\left((\sigma_i^h)^2+\sigma_N^2\right)}}$ & $\left(\sqrt{(\sigma_i^h)^2+\sigma_N^2}\right)^{-1}$\\\hline
					Global Moment-matching (GMM)  & Equations \ref{eq:gmm-l1} and \ref{eq:gmm-l2} &  $\mu_{U}-\mu_{F}\sim\mathcal{N}\left(0,\frac{\sigma_{F}}{\sqrt{n_y}}\right)$ &
					$\frac{n_y\hat{s}_{U}^2}{\sigma_{F}^2}\sim\chi_{n_y-1}^2$\\\hline
					Relative Global Moment-matching (RGMM) &Equations  \ref{eq:rgmm-l1} and \ref{eq:rgmm-l2} & $\mu_{U_r}-\mu_{F_r}\sim\mathcal{N}\left(0,\frac{\sigma_{F_r}}{\sqrt{n_y}}\right)$ &
					$\frac{n_y\hat{s}_{U_r}^2}{\sigma_{F_r}^2}\sim\chi_{n_y-1}^2$\\
					\hline
				\end{tabular}
			\label{tab:likelihood_models}
		}
	\end{table}
	
	\subsection{Pushed-forward prediction of Quantities of Interest}
	\label{sub:QoI}
	The solution of the inverse problem provides the joint posterior distribution $\pi(\bm{\theta}^m,\bm{\theta}^b|\bm{y})$. The predicted response $\bm{f}_P(\bm{\theta}^m,\bm{\theta}^b,\bm{x}|\bm{y})$ is to be computed. To do so, the posterior distribution of the latent parameters should be propagated through the forward model. The response generated by $\bm{f}$ is stochastic, which increases the complexity of the propagation. A common approach (\cite{Huan2017, Sargsyan2019}) is to use an estimator $\bm{\hat{\theta}}$ of the latent parameters, typically a maximum-a-posteriori (MAP) one. The predicted response is then 
	\begin{equation}
		\bm{f}_P(\bm{\hat{\theta}}^m,\bm{\hat{\theta}}^b,\bm{x}|\bm{y})=\bm{f}\left(\bm{\hat{\theta}}^m,\bm{\hat{\theta}}^b, \bm{x}\right).
		\label{eq:predicted_response}
	\end{equation}
	Due to the embedding, $\bm{f}_P$ generates a stochastic response that usually does not have a tractable form. Therefore, either the polynomial chaos expansion specified in Section \ref{sec:pce} or Monte-Carlo methods can be used to obtain the moments of the approximated predicted response $\bm{\mu}^h_P$ and  $\bm{\sigma}^h_P$. One of the main characteristics of $\bm{f}_P$ as shown by \cite{Sargsyan2019} is that the variance of the predicted response does not reduce to zero for large sample sets, preserving the variance due to the dataset and the model inadequacy.
	
	A special case is the propagation of the uncertainty obtained for the inferred parameters through the embedding and solution of the inverse problem to other Quantities of Interest (QoI) that did not take part in the inference procedure. The function $\bm{g}$ that represents the QoI based on the inferred parameters is defined analogously to $\bm{f}$ following Equation \ref{eq:forward_model}. Then, the posterior distribution must be pushed through $\bm{g}$ in the same way as for $\bm{f}_P$. The predicted QoI will include the uncertainty expressed through the inferred embedded latent variables.
	
	Nevertheless, using an estimator for $\bm{\theta}$ presents several drawbacks. First, $\bm{\hat{\theta}}$ are pointwise estimators despite generating a predictive distribution $\bm{\hat{f}}_P$. The information they provide is limited, losing the structure and potential correlations that are present in the joint posterior distribution of the latent parameters. Pointwise estimators fail as well in multimodal distributions to represent other local maxima apart from the optimum, which may be relevant for the QoI. Additionally, the distribution obtained from $\bm{f}_P(\bm{\hat{\theta}}^m,\bm{\hat{\theta}}^b,\bm{x}|\bm{y})$ only preserves the uncertainty provided by the model form uncertainty parameters $\bm{\theta}^b$, disregarding the variance obtained through the inference procedure, which leads to potentially overconfident predictions.
	
	As an alternative, in this article we propose a full propagation of the posterior distributions of the latent parameters as 
	\begin{equation}
		\bm{f}_P(\bm{\theta}^m,\bm{\theta}^b,\bm{x}|\bm{y})=\bm{f}\left(\bm{\theta}^m,\bm{\theta}^b, \bm{x}\right),
		\label{eq:pushed_predicted_response}
	\end{equation}
	or $\bm{g}(\bm{\theta}^m,\bm{\theta}^b,\bm{x}|\bm{y})$ for QoIs other than the predictions. By propagating the joint posterior distribution, the full information obtained during the inference procedure is preserved. This allows not only obtaining a predictive posterior distribution $\bm{f}_P$ or $\bm{g}$, but also perform inference on its parameters, providing additional insight on its reliability. For example, it is possible to obtain the pushed-posterior distribution of $\bm{\mu}^h_P$, where propagating the estimators would have only provided a point value. Two main challenges arise with this approach: the increasing number of evaluations of $\bm{f}$ or $\bm{g}$ and the composition of stochastic variables. First, the additional evaluations come from the need of sampling the joint posterior distribution of the latent parameters, which implies evaluating $\bm{f}$ or $\bm{g}$ for each of the samples. These samples of $\bm{\theta}$ define random variables as inputs, therefore $\bm{f}$ and $\bm{g}$ necessarily generate a stochastic response. To reduce the number of evaluations of $\bm{f}$ and $\bm{g}$, a PCE approximation of the response or the use of surrogate models are available. If the QoI is the value of a given realization of $\bm{f}_P$ or $\bm{g}$, the posterior distribution must account for the variability in the sampled $\bm{\theta}$ and the variance from the predicted distribution itself. In this paper, this situation will be solved by sampling from the pushed-posterior distributions  $\bm{f}_P(\bm{\theta}_i^m,\bm{\theta}_i^b,\bm{x}|\bm{y})$ or $\bm{g}(\bm{\theta}_i^m,\bm{\theta}_i^b,\bm{x}|\bm{y})$ for every $(\bm{\theta}_i^m,\bm{\theta}_i^b)$ in $n_P$ samples of $\bm{\theta}$ and analyzing the resulting dataset.
	
	\section{Applications and discussion}
	\label{sec:applications}
	\subsection{Application case: simple example}
	\label{sec:linear}
\begin{figure}[!h]
		\centering
		{\includegraphics[width=0.6\textwidth]{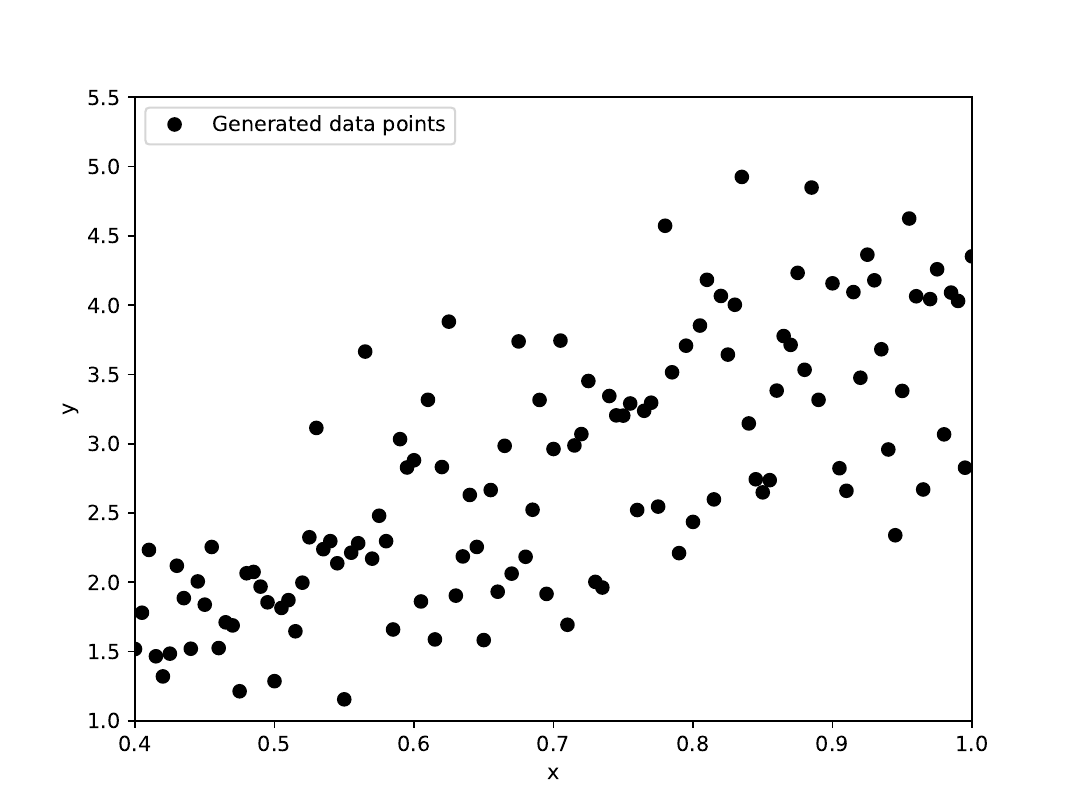}}{\caption{Generated dataset of the simple example for 120 data points}\label{fig:simple_dataset}}
		
	\end{figure}

	This example aims to test the behaviour of the different likelihood formulations under known conditions. The same model is used to generate the dataset and to compute the predictions:
	\begin{equation}
		y = \theta x+\varepsilon_N
		\label{eq:simple_model}
	\end{equation}
	where $x\in[0,1]$ is the input variable, $y$ is the output variable, $\theta$ is the slope parameter to be inferred and $\varepsilon_N$ is a white noise perturbation. The generator aims to provide a sample of observations generated by the computational model given $\theta\sim\mathcal{N}(4.0,1.0)$. The observations $\bm{y}$ are then generated by evaluating Equation \ref{eq:simple_model} with a random value of $\theta$ from the aforementioned distribution for each entry of the input vector $\bm{x}$ and adding a random perturbation of $\varepsilon_N\sim\mathcal{N}(0,\sigma_\text{true}^2)$ with $\sigma_\text{true}=0.01$. If not specified for a particular analysis, $\bm{x}$ is composed of 120 equidistant samples covering the range $[0.4,1.0]$. An example of a generated dataset is presented in Figure \ref{fig:simple_dataset}.
	
	The extended variable $\theta^*$ with the embedding is defined as
	\begin{equation}
		\theta^*= t + \delta_b, ~~~~\delta_b\sim\mathcal{N}(0,\sigma_b^2)
	\end{equation}
	where $t$ would correspond to the original latent variable and $\sigma_b$ is the new additional latent variable associated with the model inadequacy. This is equivalent to imposing $\theta^*\sim\mathcal{N}(t,\sigma_b^2)$ and using $\theta^*$ instead of $\theta$ in Equation \ref{eq:simple_model}. As the forward model is a linear function and $\theta^*$ follows a normal distribution, a closed-form expression of $\bm{f}$ is available in this case and can be formulated as
	\begin{equation}
		\bm{f}(t,\sigma_b,\bm{x})=\theta^*\bm{x}+\bm{\varepsilon}_N\Rightarrow f(t,\sigma_b,x_i)\sim\mathcal{N}(tx_i,(\sigma_b^2x_i^2+\sigma_N^2))~~~\forall i=1,2,...,n_y.
	\end{equation}
	The existence of a closed-form solution provides analytical expressions for the statistical moments of $f(t,\sigma_b,x_i)$ for every $x_i$ in $\bm{x}$ and sampled $t$ and $\sigma_b$. Nevertheless, a PCE approximation of the forward model response is built in the interest of testing the whole methodology developed in Section \ref{sec:methodology}. As the response is a linear function and the input is normal, a first degree PCE with Hermite polynomials and two Gauss points is exact and the approximation error tends to zero. The statistical moments of the computed response $\bm{\mu}^h$ and $\bm{\sigma}^h$, which are necessary to evaluate the likelihood functions in the inference procedure, are then obtained from Equations \ref{eq:pce_mu} and \ref{eq:pce_sigma}. 
	
	To obtain the posterior probability $\pi(t,\sigma_b|\bm{y})$, an MCMC algorithm with the tested likelihood models is applied to the generated dataset $\bm{y}$. The algorithm is run until convergence with a threshold of the estimated sample size $\widehat{ESS}=837$ or a maximum of 10000 MCMC samples. The $\widehat{ESS}$ is calculated following Appendix \ref{ap:ess_threshold}. The arbitrarily chosen maximum number of samples was sufficiently large such that it was never reached.
	
\begin{figure}[!h]
		\centering
		{
			\includegraphics[width=0.9\textwidth]{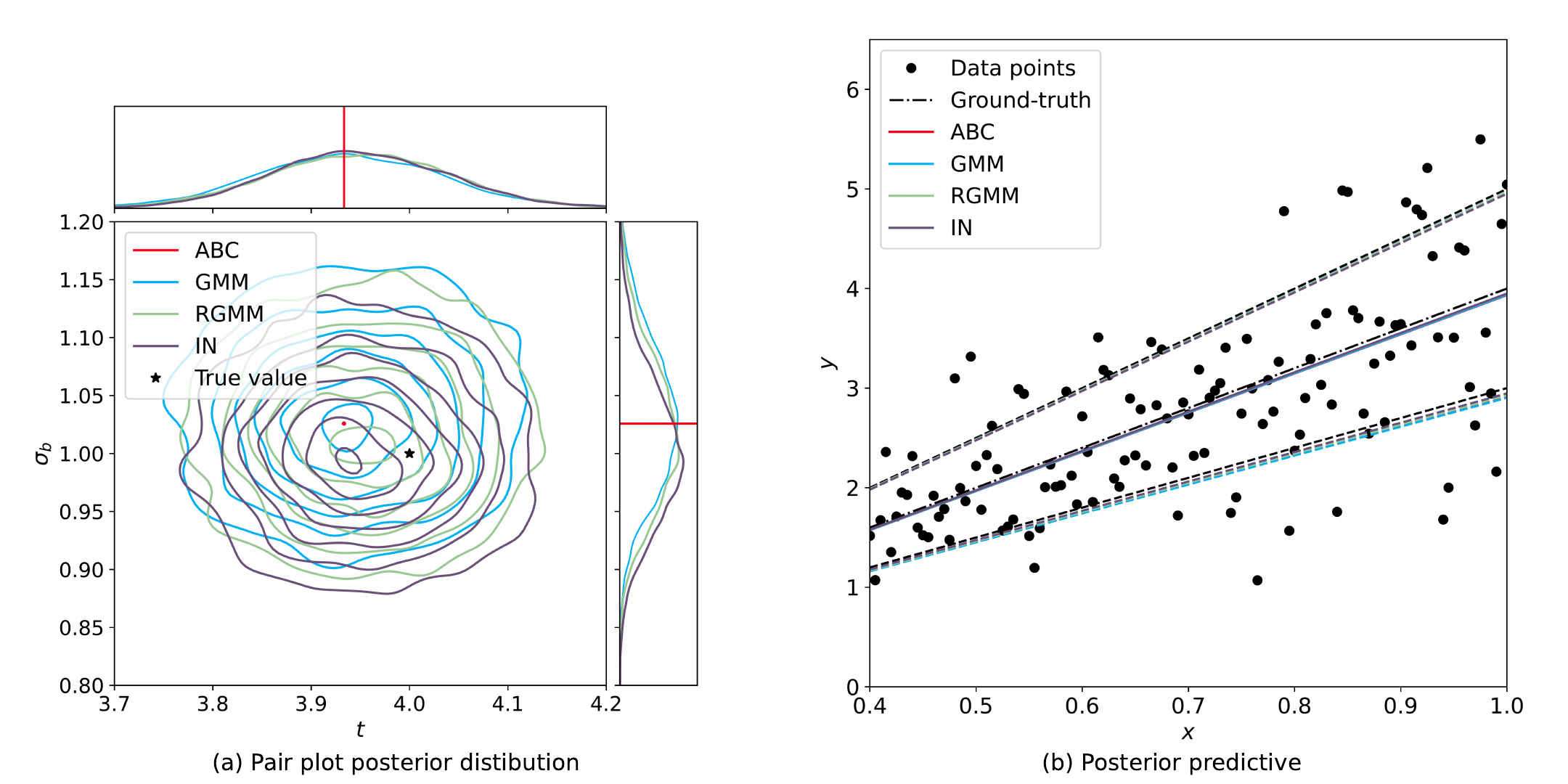}
		}{
			\caption{Comparison for the converged solution of linear example. (a) Pair plot of the posterior distribution for $t$ and $\sigma_b$. (b) Predictive distribution from the mean of the posterior distributions of $t$ and $\sigma_b$. Dashed lines indicate the interval of $\mu^h\pm\sigma^h$}
			\label{fig:simple_comparatives}
		}
	\end{figure}

	The prior distributions for the parameters are $\pi(t)\sim\mathcal{N}(4.5,0.5^2)$ and $\pi(\sigma_b)\sim\mathcal{LN}(-1,0.5)$. The prior for $\sigma_b$ is chosen as a log-normal distribution to ensure the positivity. The chosen parameters lead to $\sigma_b\in\left[e^{-1.5},e^{-0.5}\right]$ in the range of one standard deviation. 	
	The pair plots for the posterior $\pi(t,\sigma_b|\bm{y})$ from applying the MCMC algorithm with each likelihood formulation and the posterior predictive for its mean are shown in Figure \ref{fig:simple_comparatives}. Each likelihood leads to converged values close to the true generating parameters $t=4.0$ $\sigma_b=1.0$. The results are presented in Table \ref{tab:simple_results}.
	
	\begin{table}[h]
\centering
\caption{Posterior results for the linear model and different likelihoods}
\label{tab:simple_results}
\begin{tabular}{l|ccc|ccc}
 & \multicolumn{3}{c}{$t$} & \multicolumn{3}{c}{$\sigma_b$} \\
Likelihood & Mean & Std & $\widehat{ESS}$ & Mean & Std & $\widehat{ESS}$ \\
ABC & 3.93 & 0.00 & 866 & 1.03 & 0.00 & 845 \\
GMM & 3.95 & 0.13 & 901 & 0.89 & 0.15 & 868 \\
RGMM & 3.97 & 0.13 & 896 & 1.03 & 0.14 & 855 \\
IN & 3.95 & 0.09 & 1117 & 1.00 & 0.06 & 866 \\
\end{tabular}
\end{table}

	The different likelihood models provide posterior distributions centered around the same values of the latent parameters. The distribution provided by the ABC likelihood presents significantly less variance than the others. This is due to the fitting of $\sigma_b$ being exact up to $\epsilon$ as a difference, while the other likelihoods have larger acceptance probabilities for the values of $\sigma_b$. Alternative values for $\epsilon$ would have been less restrictive in the fitting of the moments, generating a flatter likelihood distribution that would have ledd to a wider posterior. There is no significant difference in the convergence speed for this particular dataset. The difference between the means of the parameters and the true generating ones is rooted in the specific realization of the dataset. However, it must be noted that ABC is the only likelihood that does not cover the true generating parameters for any reasonable confidence interval, as it aims to fit exactly the moments of particular sample dataset.
	
\begin{figure}[!h]
		\centering
		{
			\includegraphics[width=\textwidth]{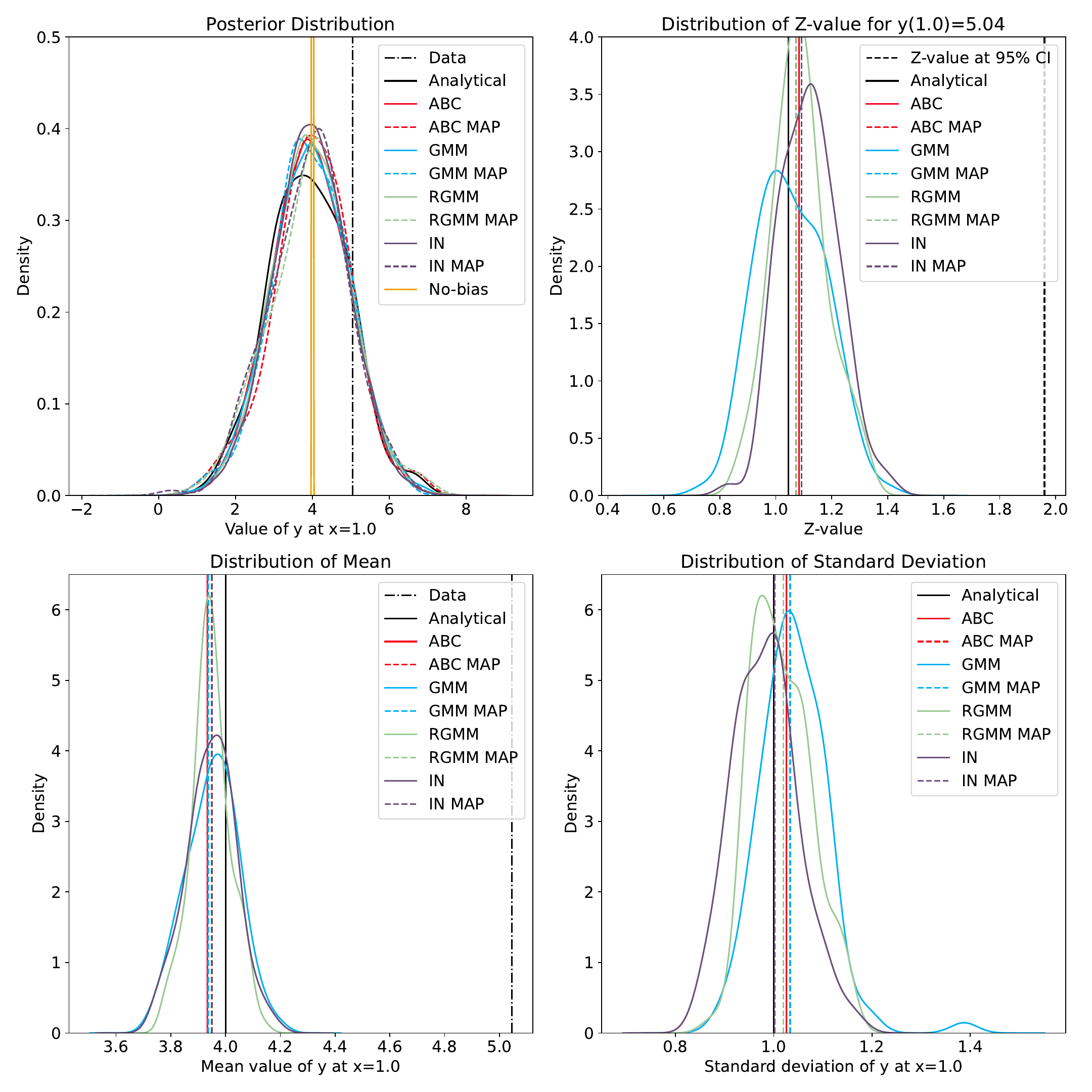}
		}{
			\caption{Analysis of Quantities of Interest by propagating the posterior distribution of $t$ and $\sigma_b$ for the simple example}
			\label{fig:simple_qoi}
		}
	\end{figure}

	Four QoI are evaluated for $x=1.0$: the output $y$, the mean of the predictive distribution $\mu_P^h$, the standard deviation $\sigma_P^h$ and the z-value of $y(1.0)$. The QoI generally can have physical meaning or simply be indicators of performance. In this case, due to the simplicity of the application, they represent quantities of statistical importance that can be used for an analysis on the fitness of the predictions. In a classical inference problem, this quantities would be deterministic, which would not allow for an analysis of the reliability of the QoIs themselves. The output $y$ is directly the pushed-forward prediction $\bm{f}_P=\bm{f}\left(\bm{t},\bm{\sigma_b}, \bm{x}\right)$, as described in Section \ref{sub:QoI}. The other QoIs are calculated for each sample of the posterior distribution of the parameters. In particular, the z-value $|Z|=\left|\frac{\mu_P^h-y(1.0)}{\sigma_P^h}\right|$ can be used for testing the hypothesis $H_1:$ the value $y(1.0)$ has not been generated by a normal distribution $\mathcal{N}(\mu_P^h,\sigma_P^h)$ against the null hypothesis $H_0$. The critical threshold of $|Z|$ for a confidence level of 95\% is 1.96, allowing to reject $H_0$ if the value of $|Z|$ is significantly larger. For comparison, the equivalent QoIs obtained with the MAP estimator of $t$ and $\sigma_b$ are calculated as well. The results are plotted in Figure \ref{fig:simple_qoi}.
	
	The obtained propagated distribution is comparable for all likelihoods, either using an estimator of the posterior or sampling from the distribution. The distributions for $\mu_P^h$ and $\sigma_P^h$ reflect the same conclusions from the pair plot representation. This is due to the linear nature of the system, which propagates directly the posterior distribution of the parameters, as $t$ controls $\mu_P^h$ and $\sigma_b$, $\sigma_P^h$. The z-values are below 1.96 for all likelihoods, failing to reject $H_0$, meaning it is plausible that the data point $y(1.0)=5.04$ was generated by the corresponding posterior distribution. In comparison, if no inadequacy term had been introduced and the posterior only considered the prescribed noise as source of variance, the corresponding z-value would have been 104, which allows to reject $H_0$, inferring that the data point could not have been generated by the prediction. It is noted that if only the estimators where propagated, it would not be possible to generate distributions for three of the QoI, and the values indicated in Figure \ref{fig:simple_qoi} as dashed lines would be taken as the predicted ones.
	
	\subsubsection{Dataset dependency analysis}
\begin{figure}[!h]
		\centering
		{\includegraphics[width=\textwidth]{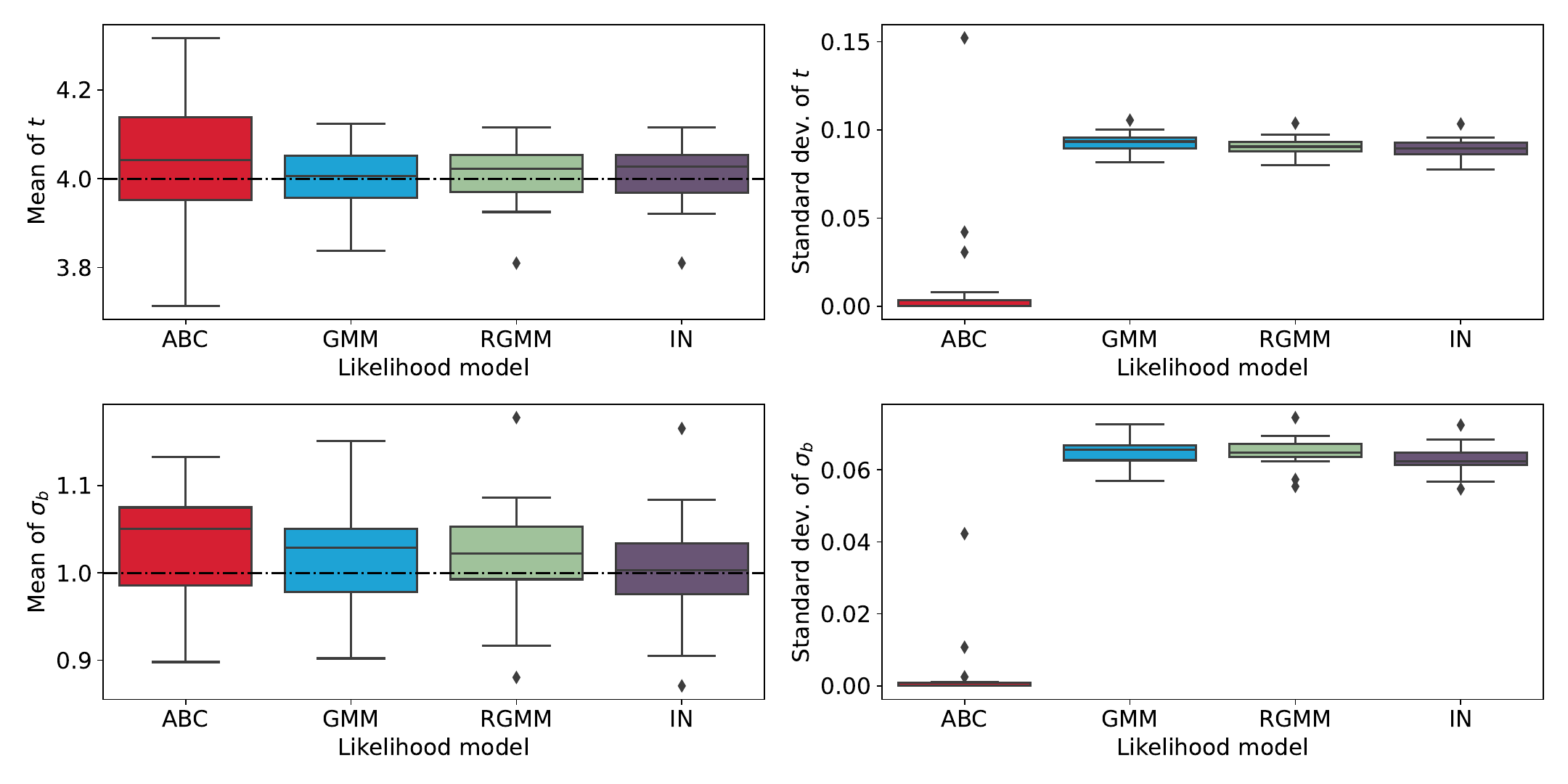}}{	\caption{Box plots for the latent parameters of the embedded model with different likelihoods. Outliers are indicated as rhomboids. Mean and standard deviation of the posterior distribution after 1000 MCMC evaluations with 10 walkers and 200 burn-in steps for 20 different realizations of the dataset. Mean (top left) and standard deviation (top right) for the slope variable $t$ and mean (bottom left) and standard deviation (bottom right) of the model inadequacy scale parameter $\sigma_b$. True generating values indicated as a dot-dashed line in the means} 
			\label{fig:linear_box_plot_repetitions}}
	\end{figure}

	While the obtained results provide insight into the properties of the different likelihoods for this case, it is necessary to compare them with several datasets generated analogously. The described procedure was repeated for 20 datasets with different random seeds for the data generation. The summary of results is presented in Figure \ref{fig:linear_box_plot_repetitions}.
	
\begin{figure}[!h]
		\centering
		{\includegraphics[width=\textwidth]{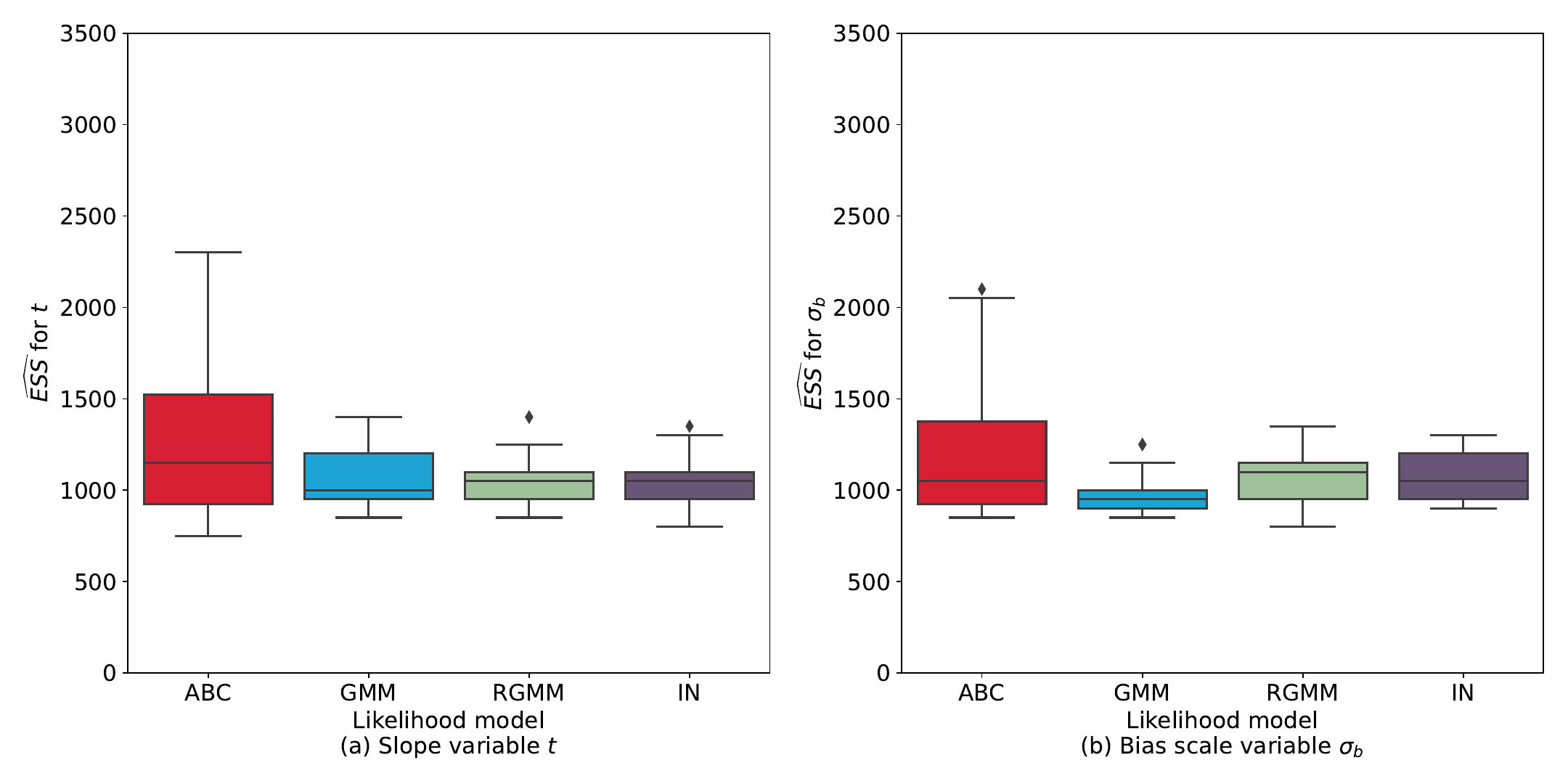}}{	\caption{Boxplot of required number of MCMC samples until reaching the threshold $\widehat{ESS}=837$ in both variables for the simple example}
			\label{fig:convergence_repetition}}
	\end{figure}

	As it can be observed, the mean value of $t$ using ABC varies the most from dataset to dataset, with the others being mostly equivalent. Coincidentally, ABC presents the smallest variance in the posterior of $t$, followed by IN, GMM and RGMM. Due to the limited dataset, ABC tends to overfit to the available points, whose statistics may present a discrepance from the generating model ones, as ABC aims to fit the moments exactly. When comparing the model form uncertainty results related to $\sigma_b$, the means for RGMM and IN are comparable, . The only difference comes from GMM, for which the limited dataset does not hold the assumptions of the predictions sharing the same distribution, which is required for this likelihood.
	An analysis on the speed of convergence is presented in Figure \ref{fig:convergence_repetition}. The ABC likelihood presents a  larger variance in the number of MCMC samples to converge, with a significant amount of outliers for the $\widehat{ESS}$ of $\sigma_b$. Due to the exact matching of the statistical moments, the ABC likelihood may require longer chains in the MCMC algorithm. It must be noted that this can be controlled by the value of the tolerance parameter $\epsilon$. The other likelihood models present comparable convergence behaviour with each other.
	
	\subsubsection{Noise influence analysis}
\begin{figure}[!h]
		{
			\includegraphics[width=\textwidth]{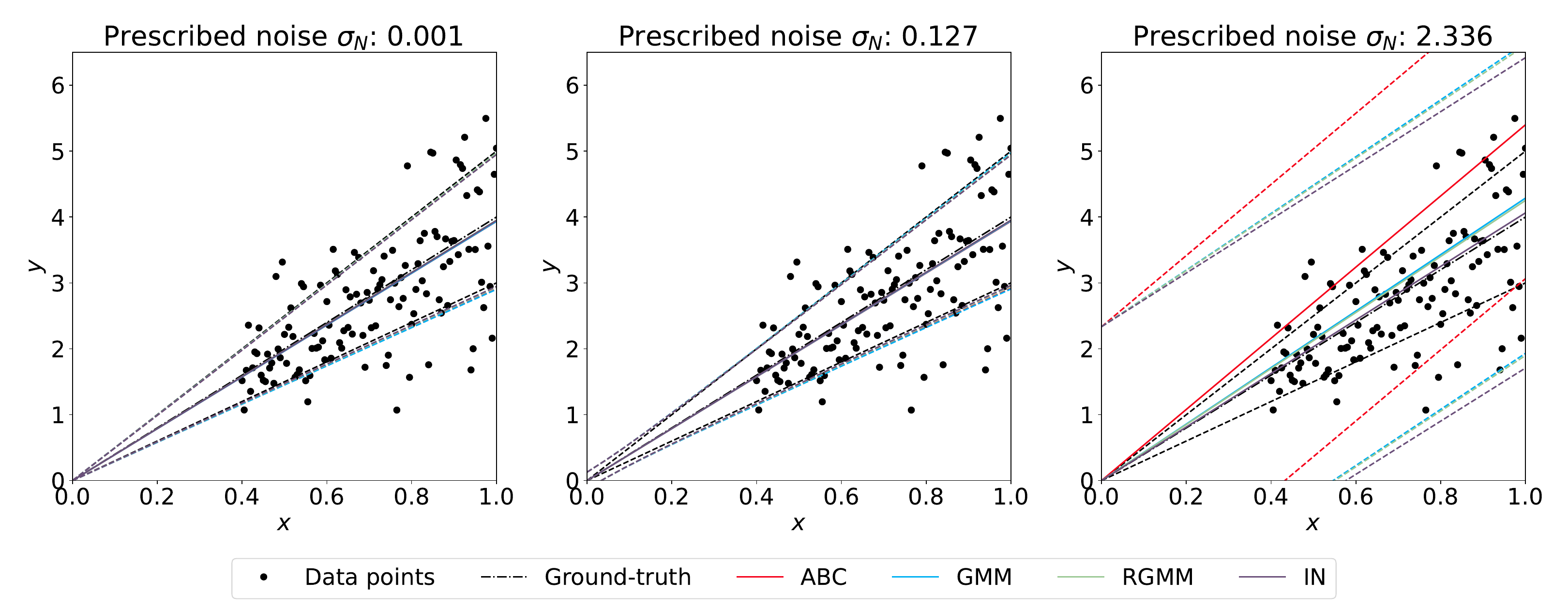}}{
			\caption{Posterior prediction comparison for noise value $\sigma_N$ of 0.001, $e^{-2}$ and $e^{0.85}$, chosen using a logarithmic rule. Dashed lines indicate the interval of $\mu^h\pm\sigma^h$}
			\label{fig:linear_noise_posterior}}
	\end{figure}

\begin{figure}[!h]
		{
			\includegraphics[width=\textwidth]{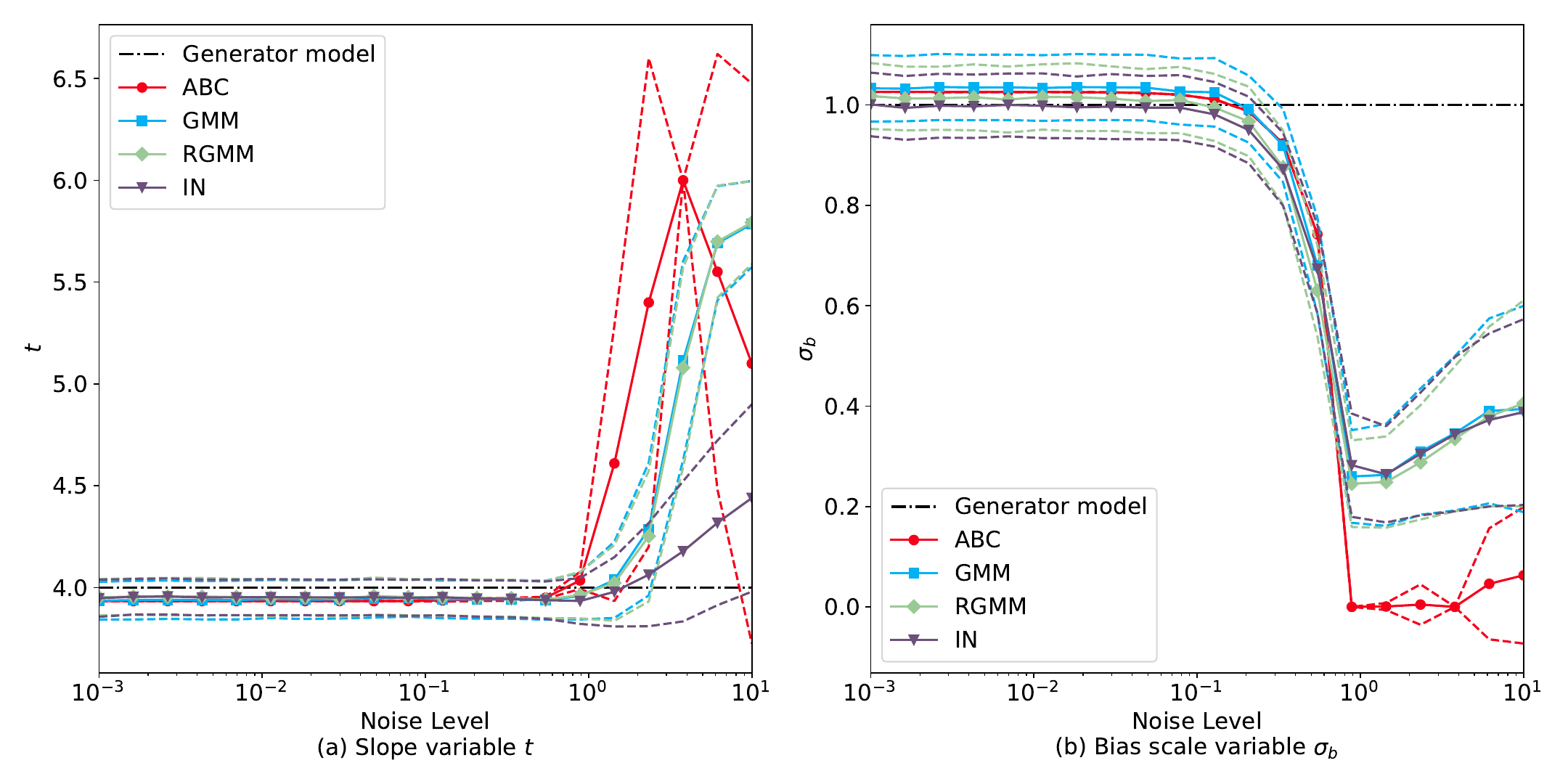}}{
			\caption{Influence of prescribed noise value $\sigma_N$ for the posterior of (a) slope variable $t$ and (b) inadequacy scale variable $\sigma_b$. Dashed lines indicate the interval of $\pm\sigma$}\label{fig:linear_noise_influence}}
	\end{figure}

	It has been proven in Section \ref{sec:methodology} and Appendix \ref{ap:behaviour_noise} that the ABC likelihood is very sensitive to the prescribed additive noise $\sigma_N$, and in particular to its misspecification. We repeat the basic experiment, but with 20 different values for the prescribed $\sigma_N$ between 0.001 and 10.0 and same prior distribution. The posterior predictive results are presented in Figure \ref{fig:linear_noise_posterior} and the compared influence of the noise in Figure \ref{fig:linear_noise_influence}.
	
	As expected, the ABC likelihood is the most sensitive to noise misspecification, particularly for values of $\sigma_N$ exceeding 1.0. When $\sigma_N < 0.01$, the prescribed noise underestimates the true generative noise, and the resulting variability is absorbed by the model form uncertainty. However, due to the scale difference, this effect is not visible in the results. For $0.01 < \sigma_N$, the noise is overestimated for at least some data points, reducing the contribution of $\sigma_i^h$ to the predictive variance at those locations. Notably, for $\sigma_N > 0.4$, the prescribed noise surpasses the variance of the data generator at certain points, leading to significant shifts in the inferred parameters. Once $\sigma_N > 1.0$, all data samples exhibit overestimated variance. This overestimation strongly affects the latent parameter $t$ in ABC, which attempts to match both the mean and variance precisely. Although other likelihoods also degrade in performance as noise increases, their sensitivity is less pronounced than that of ABC. Additionally, for large $\sigma_N$, ABC exhibits irregular behavior due to the emergence of local minima in the likelihood, causing some chains to converge prematurely and introducing numerical instability. In contrast, GMM, RGMM, and especially IN are considerably more robust to noise misspecification. While GMM and RGMM yield estimates of $t$ that deviate more from the true value ($t = 4.0$) than IN, their confidence intervals remain narrow. For IN, the interval widens with increasing noise. It is also worth noting that due to heteroscedasticity in the discrepancy between predictions and observations—stemming from randomness in $\theta$—$\sigma_b$ remains nonzero to account for prediction errors not captured by the homoscedastic noise $\sigma_N$. These effects are particularly pronounced for unrealistically large values of prescribed noise.
	
	\subsubsection{Offset influence analysis}
\begin{figure}[!h]
		\centering
		{
			\includegraphics[width=\textwidth]{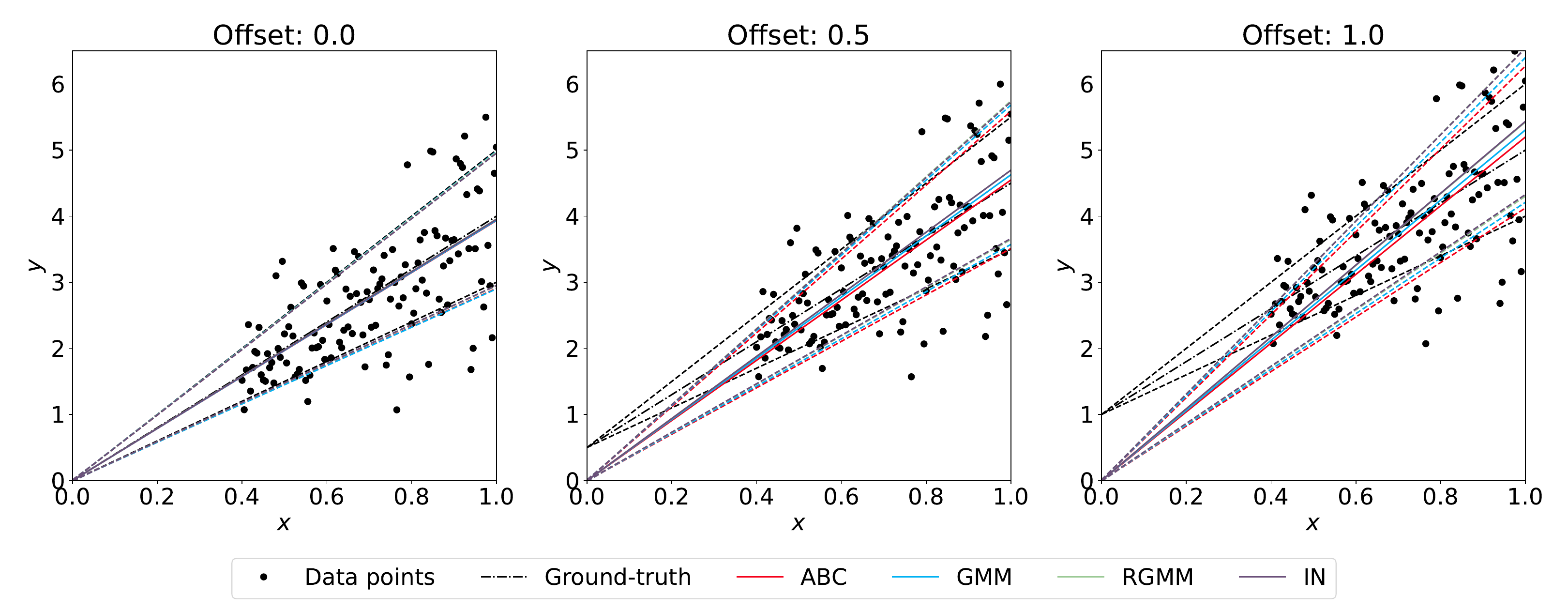}
		}{\caption{Posterior prediction comparison for offset values of 0.0, 0.5 and 1.0}
			\label{fig:linear_offset_posterior}}
	\end{figure}

\begin{figure}[!h]
		\centering
		{
			\includegraphics[width=\textwidth]{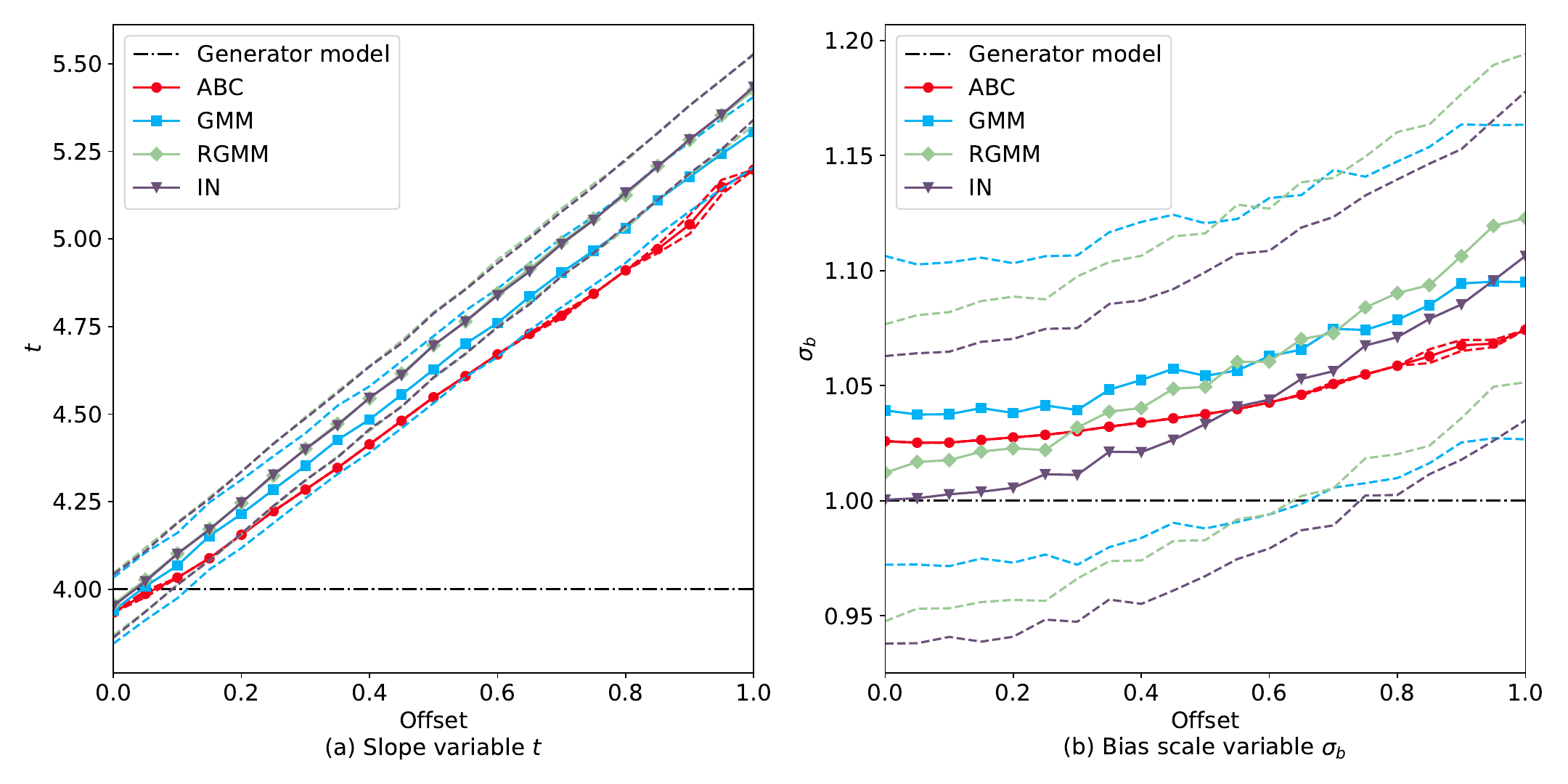}
		}{\caption{Influence of prescribed offset $\Delta y$ for the posterior of (a) slope variable $t$ and (b) model inadequacy scale variable $\sigma_b$. Dashed lines indicate the interval of $\pm\sigma$}
			\label{fig:linear_offset_influence}}
	\end{figure}

	Another influential factor in the choice of the likelihood formulation is the existence of data points that cannot be replicated by the response function. This can be simulated by including an additive offset $\Delta y$ to the generator as 
	\begin{equation}
		y = \theta x + \Delta y +\varepsilon_N
		\label{eq:offset_model}
	\end{equation}
	while keeping the original computational model. This case is relevant, for example, for non-stationary problems that depend on initial conditions which may introduce such an offset. An offset in the range between 0.0 and 1.0 is tested. The posterior predictive results are presented in Figure \ref{fig:linear_offset_posterior} and the influence comparison in Figure \ref{fig:linear_offset_influence}.
	Increasing offset values lead to higher values for $t$ to try to fit the observed points. At the same time, this leads to the need for larger variances to cover the dataset. This effect is more pronounced in RGMM and IN as the offset impacts in particular values with smaller $\sigma_i^h$ that need larger changes in $t$ and $\sigma_b$ to cover the new dataset, and coincidentally are those with the largest weight in those likelihood models. Comparing the values for the mean of $t$ with offsets of 0.0 and 1.0, it can be observed that IN and RGMM evolve less favorably than GMM with the addition of such an offset, as it was expected as the marginalizing effect is stronger in such likelihood models. For smaller datasets where the individual points have a larger weight, we expect this tendency to be more accute, leading to improvements of RGMM over IN as well driven by the addition of the second moment fitting. 
	
	\subsubsection{Outlier influence analysis}
\begin{figure}[!h]
		{
			\includegraphics[width=\textwidth]{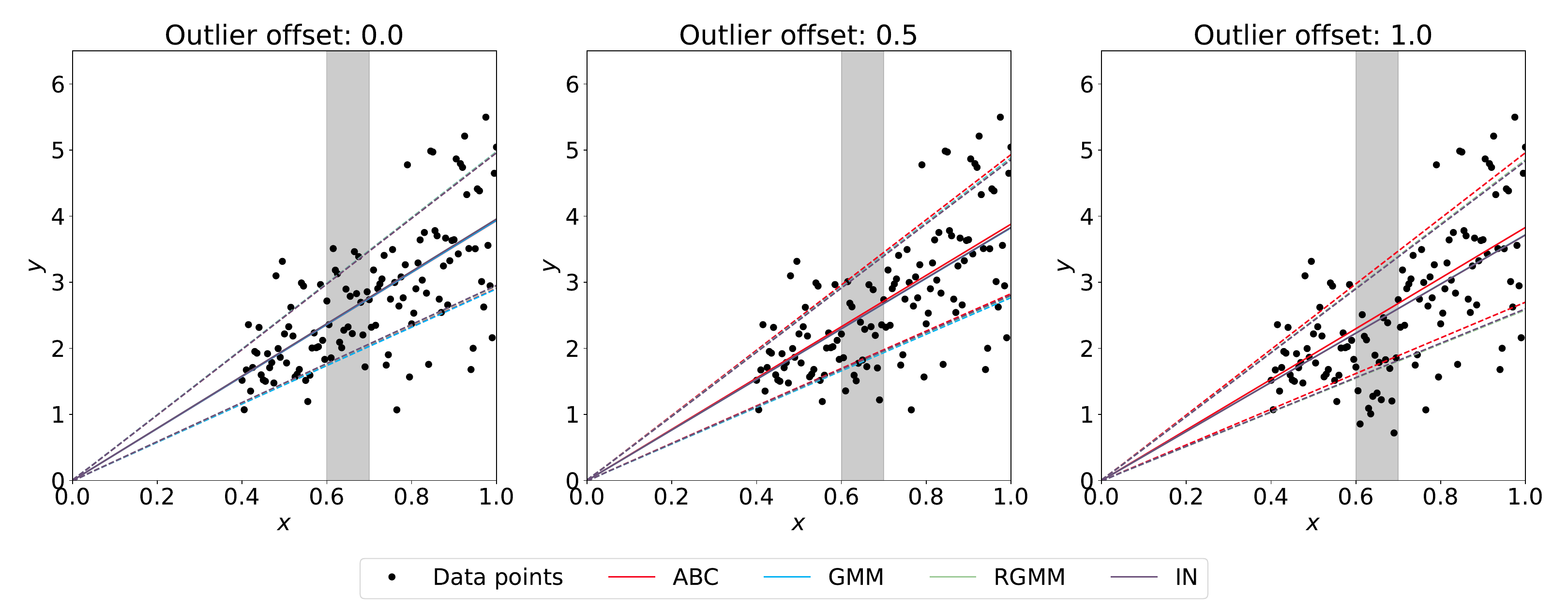}
		}{\caption{Posterior prediction comparison for outlier's offset values of 0.0, -0.5 and -1.0. The zones with outliers are shaded in grey. Dashed lines indicate the interval of $\mu^h\pm\sigma^h$}
			\label{fig:linear_outlier_posterior}}
	\end{figure}

\begin{figure}[!h]
		{
			\includegraphics[width=\textwidth]{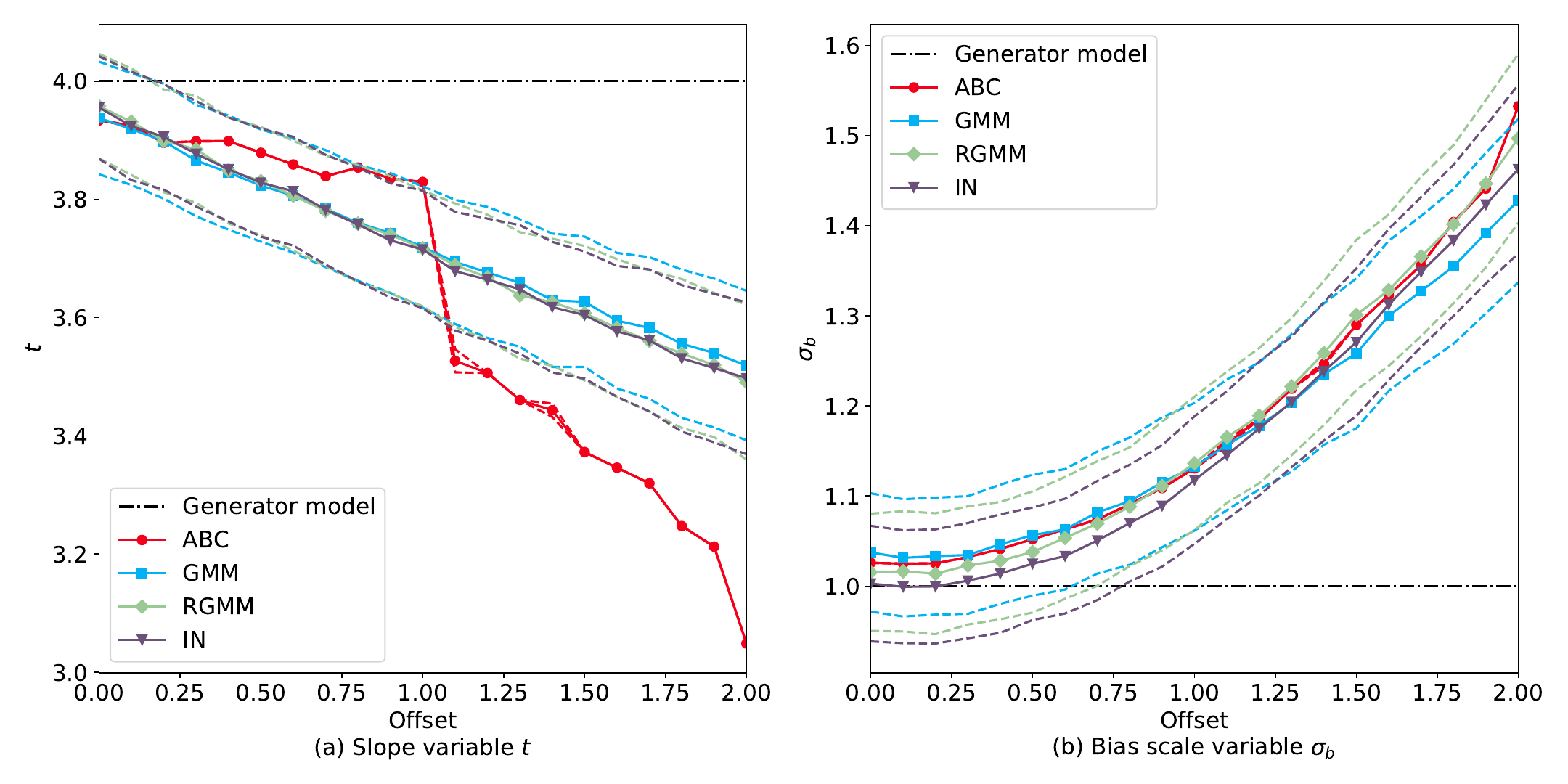}
		}{\caption{Influence of outlier data magnitude for the posterior of (a) slope variable $t$ and (b) model inadequacy scale variable $\sigma_b$. Dashed lines indicate the interval of $\pm\sigma$}
			\label{fig:linear_outlier_influence}}
	\end{figure}

	An alternative situation in which an offset is present is if it only affects a given subset of data points, creating a group of outliers that influence the inference. In this case, Equation \ref{eq:offset_model} is modified to only affect a subset of samples such that
	\begin{equation}
		y =
		\begin{cases} 
			\theta x +\varepsilon_N & \text{if } x \in[0.4,0.6)\cup(0.7,1.0] \\
			\theta x - \Delta y +\varepsilon_N & \text{if } x \in[0.6,0.7]
		\end{cases}.
	\end{equation}
	The offset ranges from 0.0 to 2.0 in 21 samples. The posterior predictive results are presented in Figure \ref{fig:linear_outlier_posterior} and the comparison in \ref{fig:linear_outlier_influence}.
	
	
	In this case, ABC is the most affected in comparison with the other likelihood models. The outliers can be represented by the predicted response, but the fitting of every point with equal weight provokes larger deviations in ABC. Nevertheless, if the outliers are located at data points with low predicted variance, it is expected that RGMM and IN show a larger influence.	
	
	
	\subsection{Application case: transient thermal simulation of reinforced concrete}
	\label{sec:thermal}
\begin{figure}[!h]
		\centering
		{\includegraphics[width=0.8\textwidth]{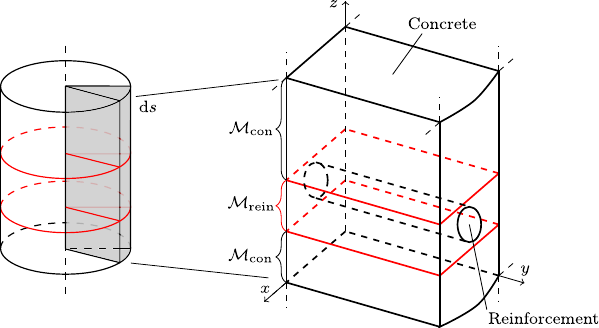}}{\caption{Schematic diagram of the heat example case for a differential slice d$s$}\label{fig:concrete_cube}}
	\end{figure}

\begin{figure}[!h]
		\centering
		{
			\includegraphics[width=\textwidth]{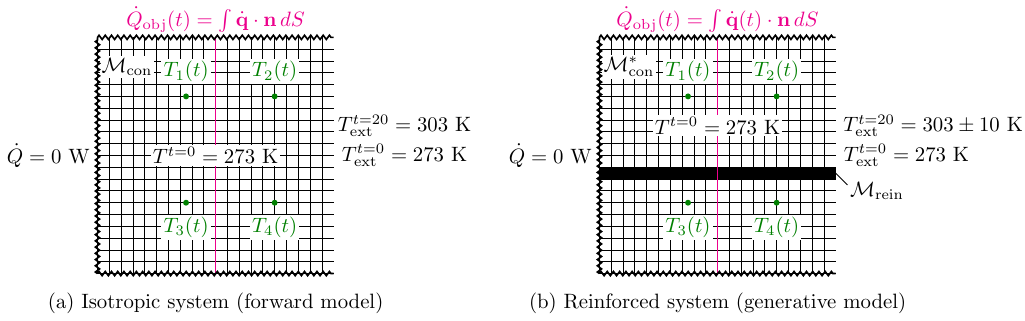}
		}{
			\caption{Diagrams of the systems modeled in the thermal example. (a) System with isotropic material properties $\mathcal{M}_\text{con}$ and (b) system with a band with modified material properties $\mathcal{M}_\text{rein}$ that represent a reinforcement bar in that region. The temperature sensors $T_i(t)$ for $i={1,2,3,4}$ act as \textit{real} sensors and are used for updating the material parameters of $\mathcal{M}_\text{con}$. The \textit{virtual} sensor $\dot{Q}_\text{obj}(t)$ predicts the heat through the midline of the system obtained by integrating its normal heat-flow obtained from the gradient of the temperature field. The Quantity of Interest $Q_\text{obj}(t)$ is the cumulative heat through the middle line at a given $t$}
			\label{fig:thermal}
		}
	\end{figure}

	The second application case consists of a transient thermal 2D simulation of the section of a reinforced concrete cylinder with constant external temperature. Due to symmetry and under the assumption of an infinitely long cylinder, such a system can be modeled as a 2D square section with adiabatic boundary conditions at the top, bottom and left-hand sides and constant temperature at the right-hand side. The Figure \ref{fig:concrete_cube} shows a schematic representation of the case. The adiabatic conditions represent the periodic nature of the reinforcement for the sides and an isolation layer for the back. The system is reduced to from a 3D cube to the 2D section assuming constant properties in X direction. The diagrams in Figure \ref{fig:thermal} represent the generative and forward models for this application case. The objective is to obtain the total cumulative heat $Q_\text{obj}(t)$ that passes through the middle line of the section at a given time. 
	
	The transient thermal system follows the heat equation for temperature $T$
	\begin{equation}
		\rho c_p\frac{\partial T}{\partial t}-\nabla\cdot(k\nabla T)=\dot{q}_V
		\label{eq:heat_eq}
	\end{equation}
	where $\rho$ is the density of the material, $c_p$ its heat capacity, $k$ its conductivity and $\dot{q}_V$ is the volumetric heat flux in the system. There exists a linear relationship between $k$ and $c_p$ for a given $\rho$ which is the diffusivity $\alpha=k/(\rho c_p)$, which requires correcting the heat component as $\dot{q}_V/(\rho c_p)$. The heat flux vector field at a given time $t$ is defined as 
	\begin{equation}
		\dot{\mathbf{q}(t)}=-k\nabla T,
	\end{equation}
	where $k$ is calculated for a certain $\rho$, $c_p$ and $\alpha$, and the heat flux through the midline is obtained by integrating the normal heat flux field over the facet $S$ as
	\begin{equation}
		\dot{Q}_{\text{obj}}(t) = \int \dot{\mathbf{q}}(t) \cdot \mathbf{n}\, dS.
		\label{eq:heat}
	\end{equation}
	The cumulative heat at time $t_n$ is obtained by integrating $\dot{Q}_{\text{obj}}(t)$ over time, which in this case will be done using a trapezoidal rule as
	\begin{equation}
		Q_{\text{obj}}(t_n) = \int_{t_0}^{t_n}\dot{Q}_{\text{obj}}(t)\, dt \approx \sum_{i=0}^{n-1} \frac{\Delta t_i}{2} \left[ \dot{Q}_{\text{obj}}(t_i) + \dot{Q}_{\text{obj}}(t_{i+1}) \right],
		\label{eq:cumulative_heat}
	\end{equation}
	where $t_i$ are the discretized time points and $\Delta t_i$ is the interval between two consecutive times $t_i$ and $t_{i+1}$.
	
	Altogether, the system's geometry, material properties $\rho$, $c_p$ and $k$ and its thermal and initial boundary conditions are required to calculate $\dot{Q}_{\text{obj}}(t)$. In this case, the squared slab has a length of 0.4 m in each direction, with adiabatic boundary conditions in three sides ($\dot{Q}=0$ W) and fixed external temperature $T_\text{ext}=303$ K. The initial temperature at all points of the system is $T(t=0)=273$ K and we assume a known density for the concrete of 2300 kg/m$^3$. The diffusivity value $\alpha$ determines the thermal behaviour and must be estimated from the observations. 
	
	For the parameter estimation, two material models are implemented: an isotropic one $\mathcal{M}_\text{con}$ for the forward model and a biphasic one $\mathcal{M}_\text{con}+\mathcal{M}_\text{rein}$ representing the real reinforced system for the generative model. $\mathcal{M}_\text{con}$ is based on the assumption that the whole system shares the same material properties. Therefore, $\mathcal{M}_\text{con}=\lbrace\rho=2300\text{ kg/m}^3, c_p=900 \text{ J/(kg$\cdot$ K)}, \alpha\rbrace$ with $\alpha$ to be inferred. The values of $\rho$ and $c_p$ have been chosen to reflect concrete properties at $T=293$ K according to EN 1992-1-2:2004 (\cite{EN2004}). A reference value for the conductivity from these conditions is $k=2.0$ W/(m$\cdot$K). This simplified model responds to the situation where the location of the reinforcement bars is not known beforehand or their effect on the simulation is disregarded.
	
	The biphasic model $\mathcal{M}^*_\text{con}+\mathcal{M}_\text{rein}$ is used to generate the ``real'' observations required for the parameter estimation. The system is divided into two materials, $\mathcal{M}^*_\text{con}$ for the concrete and $\mathcal{M}_\text{rein}$ for the reinforcement. The properties for the concrete are taken from EN 1992-1-2:2004 (\cite{EN2004}) as $\mathcal{M}^*_\text{con}=\lbrace\rho=2300\text{ kg/m}^3, c_p=900 \text{ J/(kg$\cdot$ K)}, \alpha=9.66\text{e-}7\text{ m$^2$/s}\rbrace$. To obtain $\mathcal{M}_\text{rein}$, a mix of material properties from the steel of the reinforcement and the concrete part must be considered. The region denoted for $\mathcal{M}_\text{rein}$ has a height of 0.02 m located between Y=0.16 m and Y=0.18 m, which, assuming the same depth, leads to a total surface perpendicular to the reinforcement of 0.0004 m$^2$.  Using a standard bar diameter of 12 mm, the fractional volumes for concrete and steel are
	\begin{equation}
		f_\text{steel}=\frac{V_\text{steel}}{V_\text{total}}=\frac{A_\text{steel}\cdot L}{A_\text{total}\cdot L}=\frac{0.012^2\cdot0.25\cdot\pi}{0.0004}=0.283
	\end{equation}
	and
	\begin{equation}
		f_\text{concrete}=1-f_\text{steel}=0.717.
	\end{equation}
	The material parameters $M_i$ of $\mathcal{M}_\text{rein}$ are then calculated following a Voigt model as
	\begin{equation}
		M_i=f_\text{steel}M_\text{steel}+f_\text{concrete}M_\text{concrete},
	\end{equation}
	which for $\mathcal{M}_\text{steel}=\lbrace\rho=7850\text{ kg/m}^3, c_p=440 \text{ J/(kg$\cdot$ K)}, \alpha=1.56\text{e-}5\text{ m$^2$/s}\rbrace$ (obtained from EN 1993-1-2:2005 (\cite{EN2005}) with $k=54.0$ W/(m$\cdot$K)) leads to $\mathcal{M}_\text{rein}=\lbrace\rho=3871\text{ kg/m}^3, c_p=770 \text{ J/(kg$\cdot$ K)}, \alpha=5.61\text{e-}6\text{ m$^2$/s}\rbrace$ with $k=16.7$ W/(m$\cdot$K). The material properties for the generative model are summarized in Table \ref{tab:thermal_material_properties}.
	There is a possible argument to use a Reuss model for some of the material properties, in particular the conductivity $k$. The mixed materials with such a model would be obtained as
	\begin{equation}
		\frac{1}{M_i}=\frac{f_\text{steel}}{M_\text{steel}}+\frac{f_\text{concrete}}{M_\text{concrete}}.
	\end{equation}
	However,the use of Voigt's model provides a better approximation than Reuss' model for all the studied properties when the heatflow is parallel to the reinforcement (\cite{BenAmoz1970}). More complex formulations are possible, but would only be justified if the direction of the reinforcement were unknown and non-homogeneous.
	
	\begin{table}[h!]
		{\caption{Material properties used for the generative model of the thermal application case}}{\centering
			\begin{tabular}{|c|c|c|c|c|c|}
				\hline
				\textbf{Material} &\textbf{Density} &\textbf{Heat Capacity}  & \textbf{Conductivity}& \textbf{Diffusivity} &\textbf{Volumetric}\\
				\textbf{Model} & $\rho$ [kg/m$^3$] & $c_p$ [J/(kg$\cdot$K)] &$k$ [W/(m$\cdot$K)] & $\alpha$ [m$^2$/s]&\textbf{Fraction} $f$\\
				\hline
				$\mathcal{M}^*_\text{con}$ & 2300 & 900 & 2.0 & $9.66 \text{e-}7$ & 0.717 \\
				\hline
				$\mathcal{M}_\text{steel}$ & 7850 & 440 & 54.0 & $1.56 \text{e-}5$ & 0.283 \\
				\hline
				$\mathcal{M}_\text{rein}$ & 3871 & 770 & 16.7 & $5.61 \text{e-}6$ & - \\
				\hline
			\end{tabular}
		}
		\label{tab:thermal_material_properties}
	\end{table}
	
	In a real application, it would not be possible to obtain direct measurements of the heat $\dot{Q}_\text{objective}$ to infer the value of $\alpha$ in $\mathcal{M}_\text{con}$. It is more common to have available temperature readings from sensors installed inside or at the surface of the concrete specimen. Four temperature sensors $T_i$ for $i=\lbrace1,2,3,4\rbrace$ are placed into the system as in Figure \ref{fig:thermal} at the coordinates from Table \ref{tab:thermal_sensors}. The sensors in the generative model will provide temperature measurements that at a rate of one sample every 5 min for the first 270 min that will be used to infer the latent parameters. Analogous sensors are defined in the computational model at the same positions. Their observations will be compared during the inference procedure with those collected from the generative model. 
	
	\begin{table}[!h]
		{\caption{Sensor coordinates for the thermal application case.}}{
			\centering
			\begin{tabular}{|c|c|c|}
				\hline
				\textbf{Sensor} & \textbf{X} [m] & \textbf{Y} [m] \\
				\hline
				$T_1$ & 0.15 & 0.30 \\
				\hline
				$T_2$ & 0.30 & 0.30 \\
				\hline
				$T_3$ & 0.15 & 0.12 \\
				\hline
				$T_4$ & 0.30 & 0.12 \\
				\hline
		\end{tabular}}
		\label{tab:thermal_sensors}
	\end{table}
	
\begin{figure}[!h]
		\centering
		{
			\includegraphics[width=0.9\textwidth]{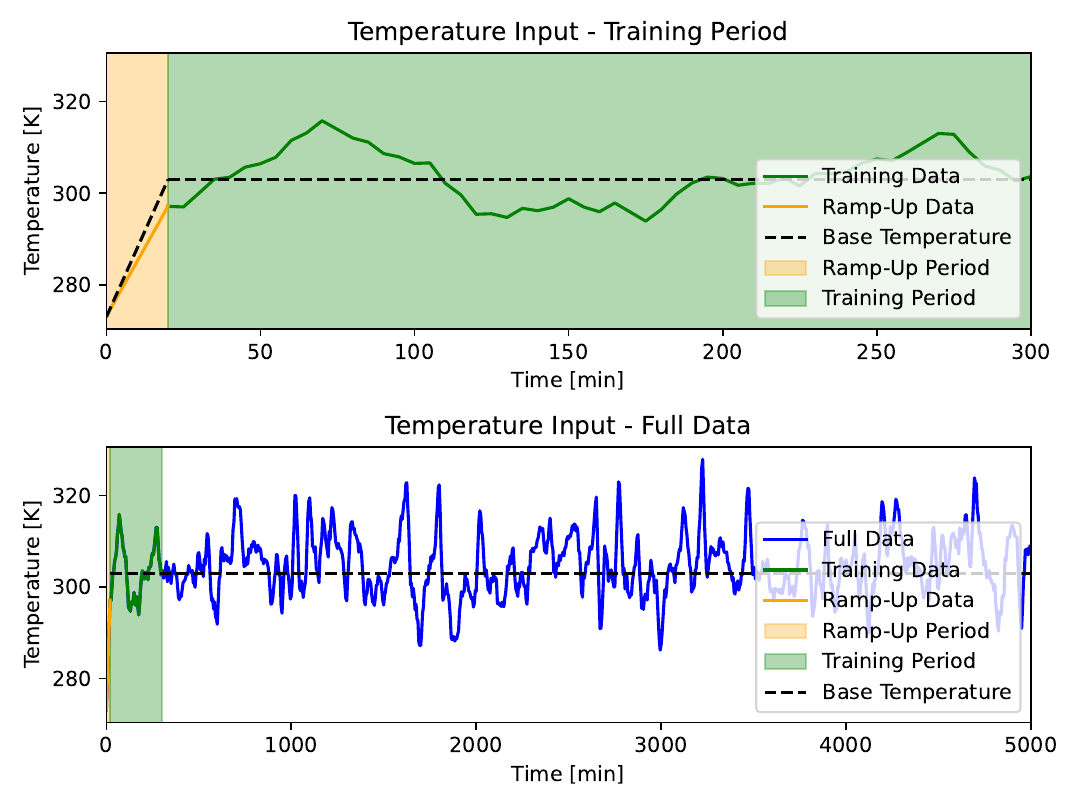}}{\caption{External temperature series for the reinforced concrete thermal example. Training series (top) and full temperature series (bottom) for QoI evaluation}
			\label{fig:thermal_temperature_series}}
	\end{figure}

\begin{figure}[!h]
		\centering
		{
			\includegraphics[width=0.9\textwidth]{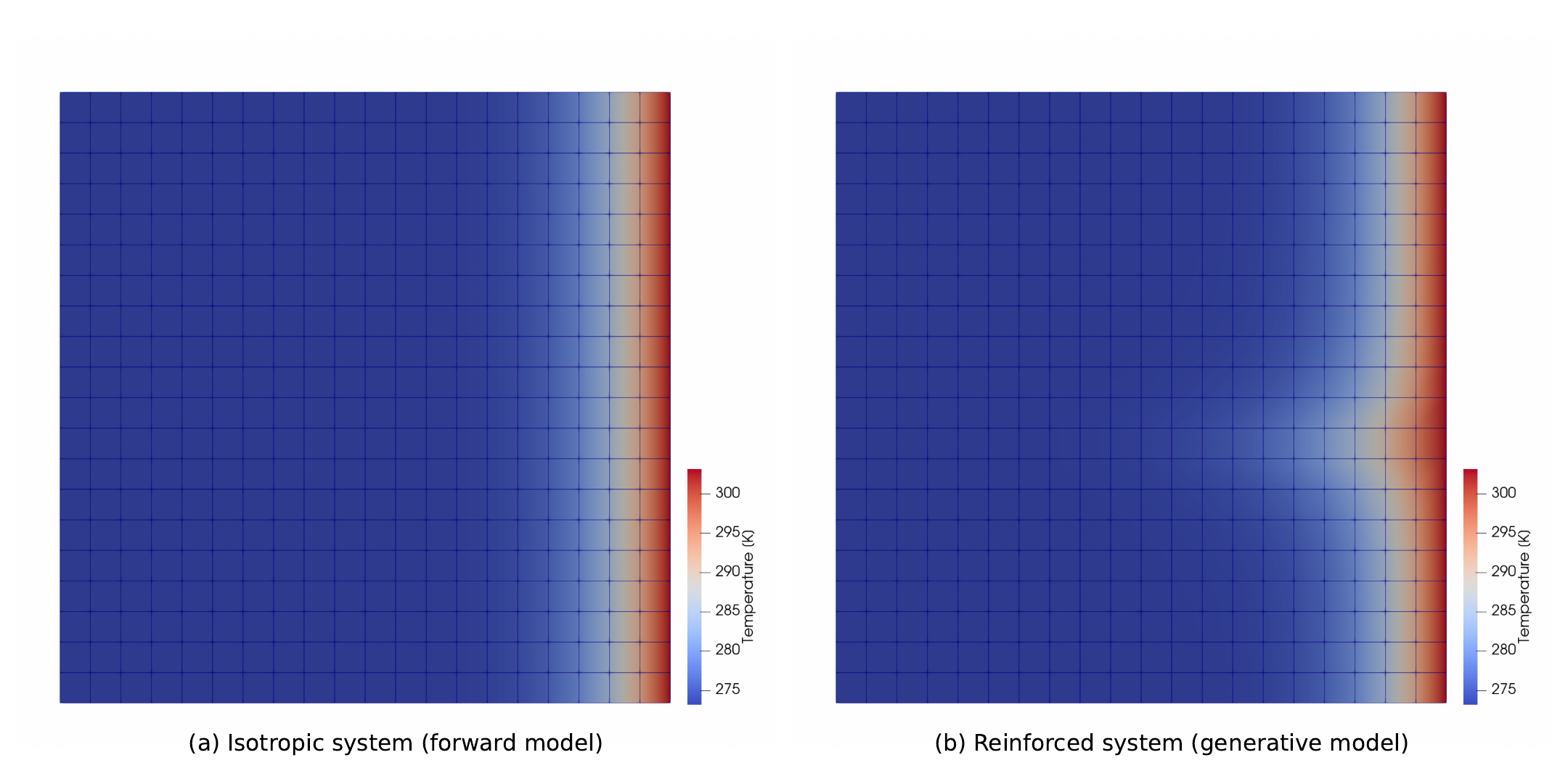}}{\caption{Resolved temperature field a $t=20$ min. (a) System with isotropic material properties $\mathcal{M}_\text{con}$ and (b) system with a band with modified material properties $\mathcal{M}_\text{rein}$ that represent the appearance of a reinforcement bar in that region. The isotropic system presents a uniform temperature gradient from right ($T=303$ K) to left, while the reinforced system presents a faster development of the temperature front at the position of the reinforcement than in the rest of the system}
			\label{fig:thermal_temperature_fields}}
	\end{figure}

	Equation \ref{eq:heat_eq} is solved over the system to obtain temperature measurements at the sensors. To this end, a finite element (FE) solution is implemented. For simplicity, the geometry is discretized in a regular mesh of $20\times20$ first-order Lagrange elements. The time integration is performed through a backward Euler scheme with 5 min timesteps, which is adequate for the slow evolution of the temperature profile in concrete. The same implementation is used to derive the temperature field to calculate the heat flux through the system. The first 29 min will be used as a transitory regime where the external temperature linearly ramps up starting from 273 K (the same temperature as the system) to 303K. Afterwards, the external temperature is composed of a constant component of 303K and an additional component that simulates variable external conditions. This variable component is composed of a short-term noise sampled at every timestep from a distribution $\mathcal{N}(0\text{ K},1.0\text{ K})$ and long-term component that is sampled once every five timesteps from a distribution $\mathcal{N}(0\text{ K},10.0\text{ K})$ and interpolated for the timesteps in between for a smooth transition. Notice that these additional noise components in the temperature are only considered in the data generation, with the isotropic computational model assuming constant external temperature, creating an additional source of model inadequacy. The training and full temperature series are plotted in Figure \ref{fig:thermal_temperature_series}. A comparison of the resolved temperature field at  $t=20$ min is shown in Figure \ref{fig:thermal_temperature_fields}.
	
	The solution of the application consists of two parts: the inference of the material parameters from the temperature measurements and the prediction of the heat flux through the midline section over time. Only thanks to the embedding it is possible to quantify the uncertainty in the heat due to the model inadequacy that steams from the misspecification of the forward model compared to the generative process. 
	
	\subsubsection{Temperature inference}
	\label{sec:thermal_temperature}
	To infer the diffusivity $\alpha$ of $\mathcal{M}_\text{con}$, an MCMC algorithm is implemented. A training dataset will be used for the parameter estimation and an additional testing dataset will be used for validation. The training dataset $\bm{y}$ corresponds to the measurements of $T=[T_1~T_2~T_3~T_4]$ between t=20 min and t=220 min using the generative model with the reinforcement. The first 20 min have been rejected as they belong to the ramp-up phase and the interest is when the external temperature reaches 303 K. A white noise perturbation with $\sigma_N=0.2$ K is prescribed at the output. For the testing dataset, the simulation is repeated with a new random seed, gathering measurements between t=20 min and t=270 min.The additinola observations in the testing compared to the training are for validation purposes.
	
\begin{figure}[!h]
		\centering
		\includegraphics[width=\textwidth]{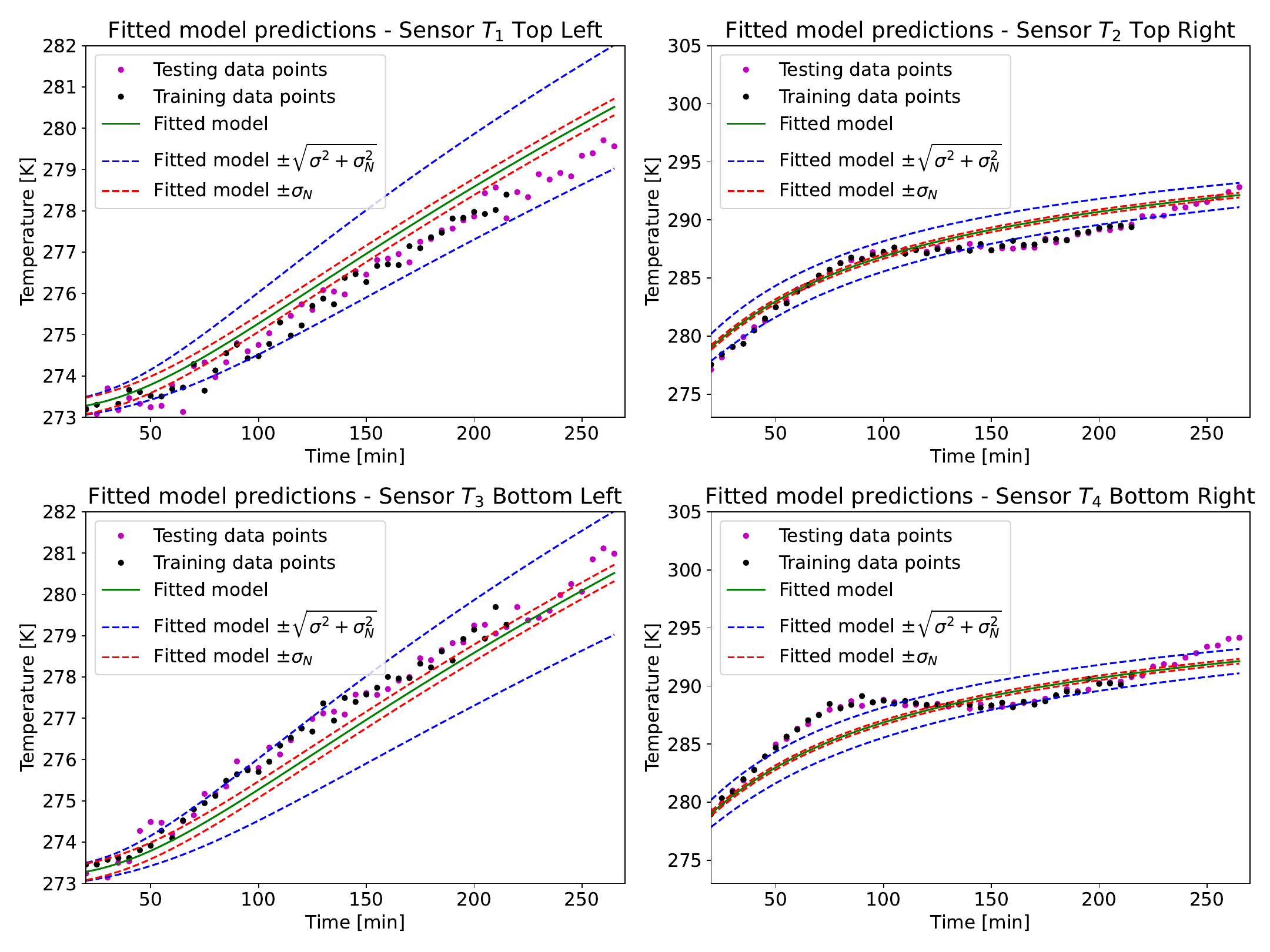}
		\caption{Temperature sensors predictions for ABC likelihood. Dashed lines indicate the interval of $\mu^h\pm\sigma^h$}
		\label{fig:thermal_temperature_abc}
	\end{figure}

	The embedding for the diffusivity is defined as
	\begin{equation}
		\alpha = \alpha_m + \delta_b, ~~~~\delta_b\sim\mathcal{N}(0,\sigma_b^2),
	\end{equation}
	where $\alpha_m$ and $\sigma_b$ are the latent parameters that must be estimated by solving the inverse problem. The prior distributions are $\pi(\alpha_m)\sim\mathcal{N}(10^{-6},(10^{-7})^2)$ and $\pi(\sigma_b)\sim\mathcal{LN}(-16,0.1)$ (which corresponds to a mean of approximately $1.1\text{e-}7$ and a standard deviation of approximately $1.1\text{e-}8$). The PCE for computing the forward model with the embedding consist of a Hermite polynomial expansion of degree 2. The noise $\sigma_N$ is properly specified with a value of 0.2, which is the same used for data generation. The MCMC algorithm is run until convergence with a threshold of $\widehat{ESS}=837$  after 250 steps of burn-in. The results for the ABC likelihood are presented in Figure \ref{fig:thermal_temperature_abc} and for the other likelihoods in \ref{ap:results_thermal}. The summarized inference results are presented in Table \ref{tab:thermal_results}.
	
	\begin{table}[h]
\centering
\caption{Posterior results for the thermal model and different likelihoods}
\label{tab:thermal_results}
\begin{tabular}{l|ccc|ccc}
 & \multicolumn{3}{c}{$\alpha_m$} & \multicolumn{3}{c}{$\sigma_b$} \\
Likelihood & Mean & Std & $\widehat{ESS}$ & Mean & Std & $\widehat{ESS}$ \\
ABC & 1.32e-06 & 4.42e-13 & 972 & 2.40e-07 & 5.24e-13 & 868 \\
GMM & 1.28e-06 & 2.31e-08 & 886 & 1.27e-07 & 1.31e-08 & 864 \\
RGMM & 1.27e-06 & 1.90e-08 & 1009 & 1.62e-07 & 9.66e-09 & 848 \\
IN & 1.27e-06 & 1.51e-08 & 852 & 1.87e-07 & 8.84e-09 & 893 \\
\end{tabular}
\end{table}

	The results do not present considerable variations between likelihood models for this case. As the noise is well prescribed, the data points with low sensitivity to changes in the parameters are not used for the training and the model can cover reasonably well the inadequacy in the model, all likelihood models produce roughly equivalent results. The posterior response is able to cover the test data points as well as the training ones thanks to the embedding being included in the latent variable itself. However, the values obtained for $\alpha_m$ with mean between $1.27\text{e-}6$ and $1.32\text{e-}6$ are not centered at $9.66\text{e-}7$, concrete's diffusivity value. This is expected, as there is a model inadequacy between the generative system and the computational model due to the reinforcement bars, that increase the diffusivity. The prescribed noise cannot explain the observed temperature measurements, therefore the embedded model inadequacy is justified. Inferring the noise scale $\sigma_N$ with an additive model would have allowed to explain the variance in the data points as a perturbation in the output, but would not necessarily be able to explain the variance in the testing range. Classical solutions to this case, such as non-additive or heteroscedastic could achieve better results for the prediction of temperature values, but they would not be transferable to other QoI such as the accumulated heat computations. An analogous situation occurs with the model inadequacy formulated following approaches based on Kennedy and O'Hagan's framework (\cite{Kennedy2001}), as shown in \cite{AndresArcones2023}.
	
	\subsubsection{Heat prediction}
\begin{figure}[!h]
		\centering
		{
			\includegraphics[width=0.7\textwidth]{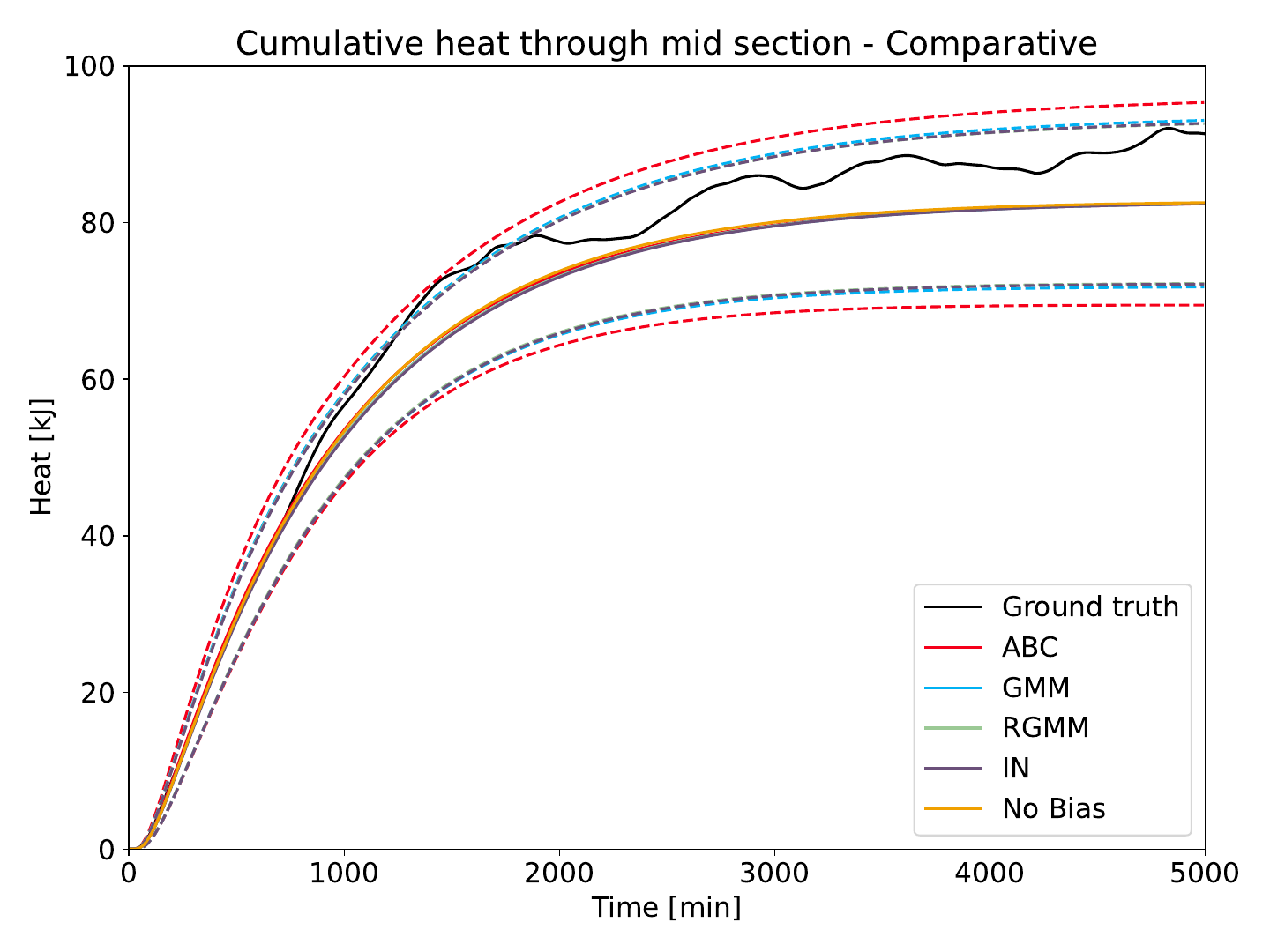}
		}{
			\caption{Temperature heat prediction for each likelihood. Dashed lines indicate the interval of $\mu^h\pm\sigma^h$}
			\label{fig:thermal_heat}
		}
		
	\end{figure}

\begin{figure}[!h]
		\centering
		{
			\includegraphics[width=\textwidth]{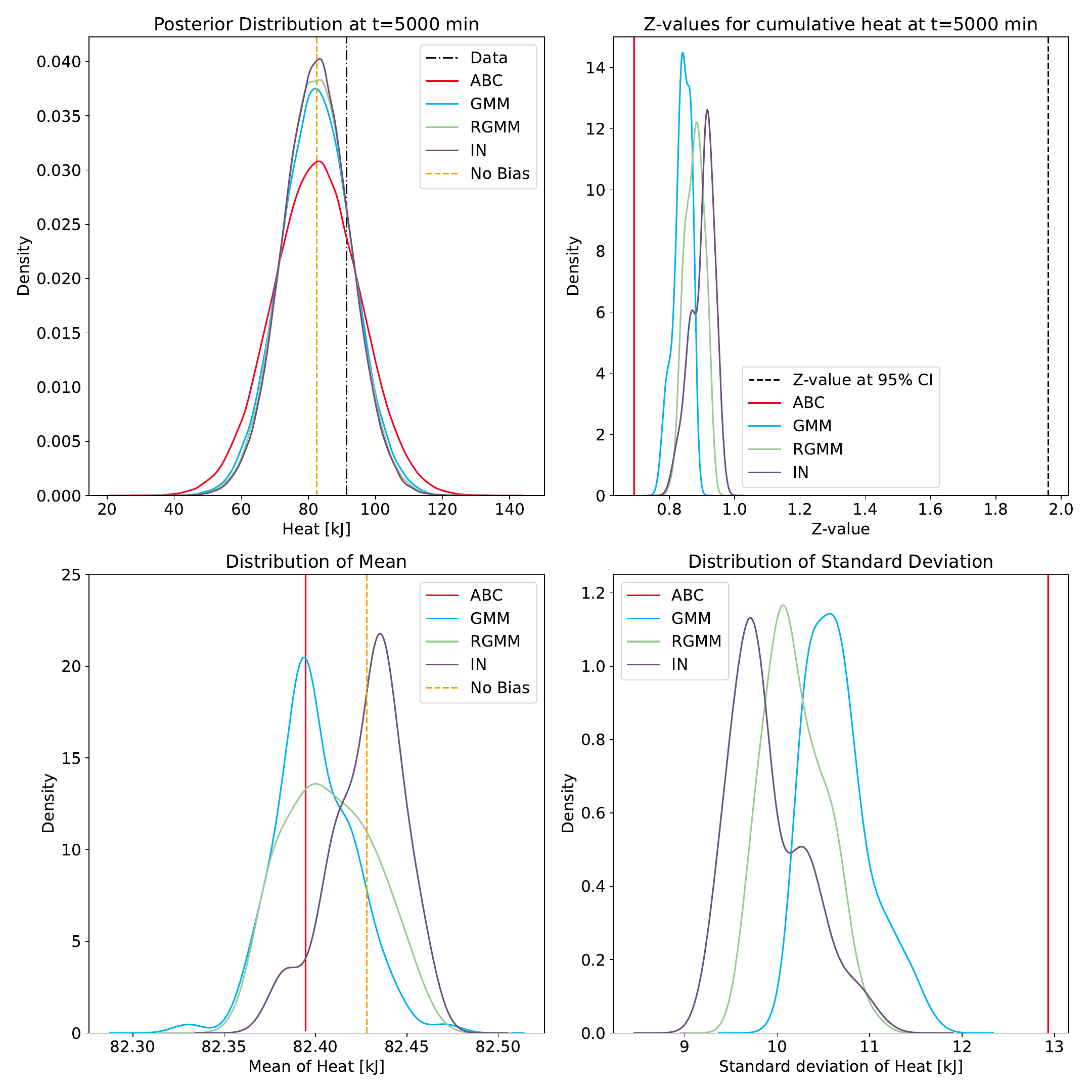}
		}{
			\caption{Analysis of Quantities of Interest by propagating the posterior distribution of $\alpha$ and $\sigma_b$ for the thermal example}
			\label{fig:thermal_qoi}
		}
	\end{figure}

	Once the posterior distribution of the latent variables $\alpha_m$ and $\sigma_b$ are estimated by solving the inverse problem, the updated system is used to calculate the cumulative heat through the cross-section. The simulation is run with the full external temperature series of 5000 min, which allows the development of the full thermal profile. Analogously with the simple example, four QoI based on the cumulative heat for $t=5000$ min are evaluated: the output $Q_\text{obj}(t=5000\text{ min})$, the mean of the predictive distribution $\mu_P^h$, the standard deviation $\sigma_P^h$ and z-value with a confidence of 95\% for $Q_\text{obj}(t=5000\text{ min})$. The predicted response taking the mean values of the posterior distribution and $\hat{\alpha}_m$ and $\hat{\sigma}_b$ solving Equation with them is presented in Figure \ref{fig:thermal_heat}. The propagated posterior distributions of the QoI are plotted in Figure \ref{fig:thermal_qoi}.
	As observed in Figure \ref{fig:thermal_heat}, the posterior distributions generated by the embeddings are able to reliably envelop the real cumulative heat value. The apparent offset is a product of a larger temperature input at the first 300 min in the especific realization of the dataset, as it can be observed in Figure \ref{fig:thermal_temperature_series}. Such is the case due to the similar sources of model inadequacy affecting the QoI calculation and the model updating, i.e. heterogeneity in the material and variations in the external temperature. If the heat simulation presented different sources of model inadequacy, such as larger temperature variations, the predicted distributions may not be able to reflect the true values. Nevertheless, they always provide a more informative assessment on the reliability of the model predictions. In comparison, the prediction when no model inadequacy term is included differs from the true value by more than 10 kJ despite not providing a significant confidence interval, leading to overconfident results.
	
	Analogous conclusions can be extracted from the analysis of the pushed-forward QoIs. While the model without inadequacy term produces a point distribution due to the lack of a noise model for the cumulative heat prediction, the embedded models produce distributions that could have generated the true value of 91338 J. This is reflected in the z-values, where all the sampled pairs of latent parameters generate distributions that produce z-values significantly smaller than 1.96. Due to its almost zero variance in the posterior distribution, a z-test would not have been reasonable for the model without inadequacy term unless an error model is prescribed. The choice of likelihood does not impact significantly the mean value of the predictions $\mu_P^h$, laying between 82300 and 82550 J. The main difference resides in the value for $\sigma_P^h$ and the spread of the distributions for each QoI. In general, ABC produces wider distributions of $Q_\text{obj}$, with more concentrated values for $\mu_P^h$ and $\sigma_P^h$. GMM, RGMM and IN provide comparable results with each other with a more concentrated $Q_\text{obj}$ but larger dispersion in the other QoIs. Only slight shifts and deformations in the distributions of the QoI can be observed, tending the global likelihoods to wider ones with more variance.
	
	\subsection{Discussion on likelihood selection}
	As it is clear from the formulation and applications, the likelihood models present substantial differences in their behaviour that influence their choice over the others depending on the characteristics of the dataset:
	\begin{itemize}
		\item \textbf{Number of parameters:} ABC requires prescribing $\gamma$ and $\epsilon$, while GMM and RGMM only require $\gamma$ and IN none. As already mentioned, the choice of $\epsilon$ is not trivial for datasets with prescribed noise and can largely impact the results. Additionally, the choice of $\gamma$ corresponds to previous beliefs on the expected predicted distribution. In particular, a value of $\gamma=\sqrt{\frac{\pi}{2}}$ is expected for ABC and $\gamma=1.0$ for the others. The difference in the value of choice for $\gamma$ steams from the formulation of the matching of the variance, which induces a half-normal distribution as discussed in Section \ref{sub:abc}, while the others do not.
		
		\item \textbf{Noise misspecification:} As shown in Appendix \ref{ap:behaviour_noise}, ABC is very sensitive to the specification of the noise amplitude, which can mislead the parameter updating procedure to values of $\bm{\theta}$ further from the expected ones. The misspecification of the prescribed noise generally leads to larger acceptance ratios and wider posterior distribution of $\bm{\theta}$ for GMM, RGMM and IN.
		\item \textbf{Assumptions of normality:} The four likelihoods considered suppose different distributions of the residuals while still keeping the assumption of normality. ABC fits exactly the moments under the consideration of Gaussian noise. GMM impose the condition that all the residuals are characterized by the same mixture model, which may not be reasonable. RGMM assume that the relative errors between observations and predicted distribution follow such a common mixture model instead. Finally, IN assumes independent normal distributions for each observation. Depending on the application case, some of these assumptions may not hold.
		\item \textbf{Inferred posterior distribution:} By construction, ABC aims to match the statistical moments exactly up to a tolerance $\epsilon$, which will lead to narrow inferred posterior distributions of the parameters $\bm{\theta}$ for acceptable values of $\epsilon$ compared to GMM, RGMM and IN. In particular, larger predicted variances imply more samples needed for the convergence of the chain (\cite{Schaelte2020}). Additionally, the formulation of $\mathcal{L}_2(\bm{\theta})$ for GMM and RGMM produces a flatter likelihood function in the direction of the variance-governing parameters, which further hinders the convergence for those parameters.
		\item \textbf{Observations out of the predictive range:} If the observations cannot be replicated by the computational model independently of the choice in the parameters, then embedded approaches lead to suboptimal results. RGMM and IN are expected to be the most impacted if those values are at points with low predictive variance. However, the regularizing effect of considering the whole dataset in RGMM and GMM for the goodness of fit instead of fitting every point, as in IN or ABC, should reduce the impact of these observations if they are not predominant in the dataset.
		\item \textbf{Fit to exact observations:} As pointed out in \cite{Sargsyan2015}, marginalizing formulations such as IN would lead to wrong results if $\sigma_i^h$ is significantly smaller for a given observation $i$ than for the others. In those cases, such observation will be overfitted by reducing the residual to the minimum to the detriment of the other observation points with larger variances. The presence of a prescribed noise reduces this effect in IN, GMM and RGMM likelihoods, as all observations will have at least variance $\sigma_N$. For an analysis on this phenomena, refer to \cite{Sargsyan2015}.
	\end{itemize}
	
	\section{Conclusions}
	\label{sec:conclusion}
	This paper has introduced several significant advances in the implementation of embedded model inadequacy approaches for quantifying model uncertainties. First, a more interpretable embedding representation was developed as an alternative to the one presented by \cite{Sargsyan2019}. Although this approach requires specifying a distribution for the model form uncertainty, it reduces the dimensionality of the Polynomial Chaos Expansion (PCE) needed to describe the embedding and improves the separation between fitting the embedding’s PCE and inferring the model parameters. Next, it was demonstrated that the Approximate Bayesian Computation (ABC) likelihood formulation, commonly used to update model parameters with an embedded model inadequacy, is highly sensitive to the prescribed measurement noise and requires careful selection of the parameters $\gamma$ and $\epsilon$. One of the main contributions of this work is the explicit treatment and analysis of measurement noise, which had previously been mentioned in past studies but not thoroughly examined. To address this, the ABC likelihood was extended to properly account for measurement noise and accurately quantify predictive variance by imposing $\gamma = \sqrt{\pi/2}$. Additionally, two alternative likelihood formulations —Global Moment-Matching (GMM) and Relative Global Moment-Matching (RGMM)— were proposed to mitigate sensitivity issues. These formulations were analyzed and compared against the modified ABC likelihood and the independent normal noise formulation using a simple linear and a complex transient thermal example. Finally, a key contribution of this paper is the analysis of how the uncertainties quantified through the embedding approach propagate to Quantities of Interest (QoI) calculated using the updated model and parameters. 
	
	The inclusion of the embedded model inadequacy term enables the quantification of uncertainty arising during parameter updating due to modeling assumptions, potentially capturing the full range of uncertainties in the measured observations. By introducing the embedding model inadequacy term as a stochastic extension of the latent parameters, the proposed formulation allows for straightforward propagation of these uncertainties to other predicted QoIs that depend on the same model parameters. However, as shown in this work, the prescribed measurement noise critically influences the selection of the appropriate likelihood formulation and affects the convergence of the Bayesian updating process when an embedding is used. In particular, ABC likelihoods have been identified as the most sensitive to poor noise prescriptions. Additionally, the embedded model inadequacy formulation limits predictions to the range of responses that the model is capable of producing. As a result, discrepancies that cannot be explained by changes in the model parameters, such as offsets or outlier values, can skew the inferred parameters. ABC methods tend to be more sensitive to outliers, while RGMM and IN are more impacted by offsets or outliers near values with low variance. GMM and RGMM, on the other hand, are reliable only if the assumption of the existence of a common mixture model in the residuals is valid. Identifying which of these effects are present during model updating is not always possible and often depends on data analysis or practitioner experience. However, selecting the right likelihood function is crucial in certain cases. Regardless of the chosen likelihood formulation, the inclusion of embedded model inadequacy improves predictive posterior distributions, allowing for more informed decision-making, albeit at the cost of increased computational effort during the update phase. While hierarchical Bayesian approaches can produce comparable results when the conditional distributions of the hierarchical formulation are known, a thorough comparison between these methodologies is left for future work.
	
	One of the key strengths of this methodology is its ability to propagate the uncertainty captured by the embedded model inadequacy. This paper has demonstrated how propagating the full posterior distribution of the latent parameters through QoI calculations provides insights that are not possible with point estimates or model inadequacy-free formulations. Generating distributions for the statistical moments of the predicted QoI enables the use of inference techniques for a more robust analysis of the results. Moreover, the interpretable embedding formulation proposed here allows the propagation of uncertainties related only to specific parameters that influence the QoI being calculated. Nevertheless, when multiple parameters are inferred alongside an associated model form uncertainty, it is likely that their inadequacy terms will be correlated. Implementing and propagating such correlation structures to QoIs where only certain parameters are relevant remains as future work. 
	
	In conclusion, the advances presented in this paper represent a significant step toward more effective quantification and propagation of uncertainties in model parameters through embedded model inadequacy approaches. One of the main contributions of this work is the explicit consideration and analysis of measurement noise, which had been previously acknowledged in past methods but never thoroughly examined in terms of its impact on the model updating process. Future work should focus on further validating this methodology by applying it to real measurement data from sensors to assess its practical applicability. Additionally, extending the current approach to hierarchical Bayesian formulations and exploring the propagation of correlated inadequacy terms to various QoIs are important avenues for continued research. These developments promise to enhance the robustness and reliability of uncertainty quantification in complex models.
	
	\appendix             
	
	\section{Monte Carlo-Markov Chain stopping criteria}
	\label{ap:ess_threshold}
	In this paper, the MCMC sampler based on ensembles of chains implemented in \texttt{emcee} (\cite{ForemanMackey2013}) is used for the Bayesian updating framework. In the chosen implementation, a stretch move as defined in \cite{Goodman2010} proposes the new samples for the next iteration of the MCMC algorithm based on the current values of the ensemble of chains. The chains in a given ensemble are therefore not independent. This correlation invalidates the use of those convergence criteria that are base on the independence of the chains such as Gelman-Rubin's $\hat{R}$ metric (\cite{Gelman1992, Vats2021}). Alternatively, the integrated autocorrelation time $\tau_\infty$ (\cite{Sokal1997}) quantifies how long it takes for the values in a chain or ensemble to become effectively uncorrelated with one another and is in practice a metric of the quality of the ensemble that does not require independence of the chains. A stopping criteria based on the effective sample size $ESS$ obtained from $\tau_\infty$ is imposed on the MCMC algorithm to determine the number of iterations that must be evaluated.
	
	First, an objective $ESS$ threshold must be define to determine when the chain is converged. Following \cite{Vats2019}, an estimated $\widehat{ESS}$ such that
	\begin{equation}
		\widehat{ESS}\geq W_{p,\alpha,\epsilon}
	\end{equation}
	\noindent
	is calculated a priori. The value $W_{p,\alpha,\epsilon}$ depends only on the dimensionality of the problem $p$, the level of confidence $\alpha$ of the considered regions and the desired precision $\epsilon$. For large samples, i.e. sufficiently long chains, \cite{Vats2019} prove that a good approximation for $W_{p,\alpha,\epsilon}$ is
	\begin{equation}
		\widehat{ESS}\geq W_{p,\alpha,\epsilon}=\frac{2^{2/p}\pi}{\left(p\Gamma(p/2)\right)^{2/p}}\frac{\chi^2_{1-\alpha,p}}{\epsilon^2}.
	\end{equation}
	\noindent
	Additionally, a maximum number of iterations steps $n_\text{max}$ is established. Therefore, the MCMC algorithm will be continued while, for a given iteration $i$,
	\begin{equation}
		ESS_i<\frac{2^{2/p}\pi}{\left(p\Gamma(p/2)\right)^{2/p}}\frac{\chi^2_{1-\alpha,p}}{\epsilon^2}\text{ and }i<n_\text{max}.
		\label{eq:ess_threshold}
	\end{equation}
	
	The effective sample size $ESS_i$ is directly related with $\tau_\infty$ at iteration $i$ as
	\begin{equation}
		ESS_i=\frac{n_im}{\tau_\infty},
	\end{equation}
	where $n_i$ is the number of steps at iteration $i$ and $m$ is the number of walkers, i.e. chains in the ensemble. In practice, $\tau_\infty$ is approximated by an estimation following the implementation described in \cite{Sokal1997}. To avoid excessive computations, $ESS_i$ is not calculated after every sample of the MCMC algorithm but after a batch of iterations with predefined size have been evaluated. The MCMC will be stopped when the thresholds of Equation \ref{eq:ess_threshold} are exceeded for the chains of each latent parameter $\theta$ that is being sampled.
	
\begin{figure}[!h]
		\centering
		{
			\includegraphics[width=\textwidth]{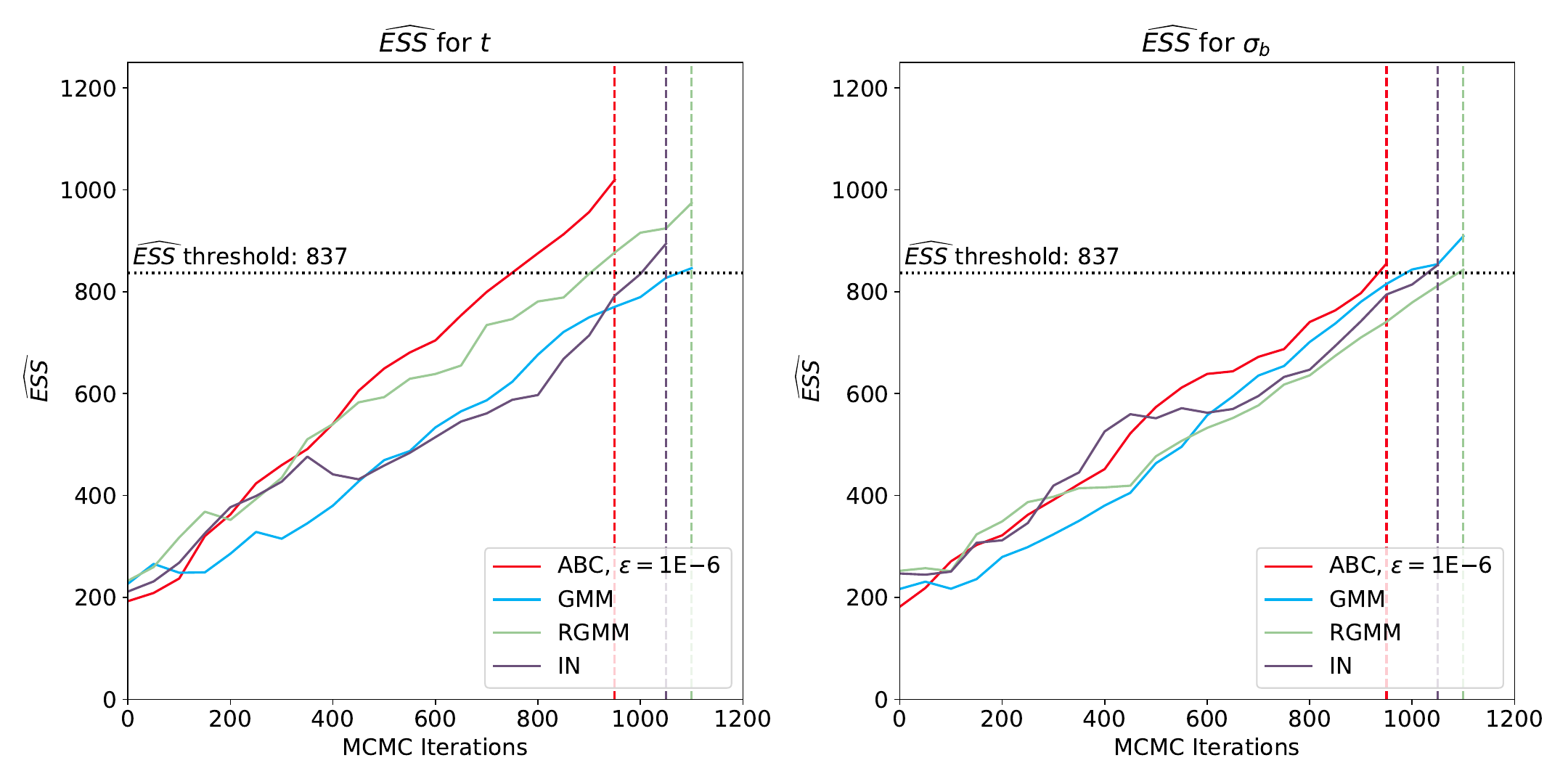}
		}{
			\caption{Estimated sample size (ESS) for the simple example over the MCMC iterations for simple linear application case}
			\label{fig:simple_convergence_ess}
		}
	\end{figure}

	Convergence plots for the simple case of Section \ref{sec:linear} are presented in Figure \ref{fig:simple_convergence_ess}, where the stopping criteria has been chosen following the procedure previously described.
	
	\section{Construction of moment-matching likelihoods with measurement noise}
	\label{ap:convergence_noise}
	In this section we build the general formulation of the moment-matching likelihoods with  measurement noise used in this article and that Markov chains that use them in the algorithm converge to the target posterior distribution that they define for a given prior.
	We will start by constructing likelihood functions from the noise $\varepsilon_\text{noise}$ as an error model. First, we will prove that for observations $y$, real system response $z$ and  latent parameters $\theta$
	\begin{equation}
		\pi(\theta, z|y)\propto \pi_\varepsilon(\rho(y,z))\pi(\theta).
	\end{equation}
	Then, if $t(\cdot)$ represents a sufficient statistic, it will be proven that 
	\begin{equation}
		\pi(\theta, z|y)\propto \pi_\varepsilon(\rho(t(y),t(z)))\pi(\theta).
	\end{equation}
	Finally, its convergence in a Markov chain context will be shown.
	
	From Equation \ref{eq:noise_model}, such noise model can be expressed as
	\begin{equation}
		\varepsilon_\text{noise}=\rho(y,z)
		\label{eq:noise_definition}
	\end{equation}
	\noindent
	where $\rho:\mathbb{R}^2\to \mathbb{R}$ defines a residual function between observations  $y$ and real values $z$ that depends on the prescribed structure of the noise, often defining a signed distance. For example, for additive noise, $\varepsilon_\text{noise}=y-z$ and follows a zero-mean normal distribution with prescribed variance $\varepsilon_\text{noise}\sim\mathcal{N}(0,\sigma_N)$. In case the model response is considered unbiased, i.e. $\varepsilon_\text{model}\overset{!}{=}0$, then $\varepsilon_\text{noise}=\rho(y,f(\theta))$. This assumption corresponds to the belief that the computational model can represent the system response in its totality.
	
	We define the random variable $Y$ from which the observations $y$ are realizations. Its probability given a model sample is
	\begin{equation}
		\pi(Y=y|f(\theta)=z,\theta) = \pi_\varepsilon(\rho(y,z))
		\label{eq:prob_noise}
	\end{equation}
	\noindent
	where $\pi_\varepsilon(\rho(y,z))$ is the probability associated with a Gaussian likelihood according to the distribution of $\varepsilon_\text{noise}$, which follows from Equation \ref{eq:noise_definition}. Assuming summary statistics $T(\cdot)$ and $t(\cdot)$, \emph{sufficiency} in a Bayesian context (\cite{Bernardo1994}) is formulated, for almost every $x$, as
	\begin{equation}
		\pi(\theta|X=x)=\pi(\theta|T(X)=t(x)),
	\end{equation}
	which directly allows to formulate the noise error model from Equation \ref{eq:prob_noise} with sufficient statistics as
	\begin{equation}
		\pi(T(Y)=t(y)|T(f(\theta))=t(z),\theta) = \pi_\varepsilon(\rho(t(y),t(z))).
		\label{eq:prob_noise_stats}
	\end{equation}
	\noindent
	To simplify notation, the convergence will be proven using the original variables $y$ and $z=f(\theta)$ but can be directly extended with the statistical moments. To define the posterior distribution of $\theta$, Bayes' theorem is applied using the predictions $z$ as auxiliary variable:
	\begin{align}
		\pi(\theta, z|y)& = \frac{\pi(y|\theta,z)\pi(\theta,z)}{\pi(y)} && \text{Bayes' theorem} \label{eq:bayes_app}\\
		& \propto \pi(y|\theta,z)\pi(\theta,z)\\
		& \propto \pi(y|\theta,z)\pi(z|\theta)\pi(\theta) && \text{conditioning on $\theta$}\\
		& \propto \pi(Y=y|f(\theta)=z,\theta)\pi(z|\theta)\pi(\theta)\\
		& \propto \pi_\varepsilon(\rho(y,z))\pi(z|\theta)\pi(\theta) && \text{from Equation \ref{eq:prob_noise}} \label{eq:noise_post_app}\\
		& \propto \pi_\varepsilon(\rho(y,z))\pi(\theta) && \text{if $f(\theta)$ is deterministic.}
	\end{align}
	\noindent
	In this paper, $\pi_\varepsilon(\cdot)$ is from an exponential family, denoting Gaussian noise. When summary statistics are involved, composed error models are possible. For example, a valid formulation would be
	\begin{equation}
		\pi_\varepsilon(\rho(t(y),t(z)))=\pi_{t_1}(t_1(y)-t_1(z))\pi_{t_2}\left(\frac{t_2(y)}{t_2(z)}\right)
		\label{eq:composed_noise}
	\end{equation}
	where $\pi_{t_1}$ and $\pi_{t_2}$ are the respective noise models for the statistics $t_1$ and $t_2$. 
	
	Markov chains converge to the aforementioned posterior distribution when using the noisy likelihood (\cite{Wilkinson2013}). In effect, given a transition kernel $q(\theta,\theta')$ defined in the chosen MCMC approach, it must be proven that the \textit{detailed balance equations} are satisfied. Considering $\pi(\cdot)$ the target stationary distribution and $p(\cdot,\cdot)$ the transition kernel of the chain, which also depends on $q(\theta,\theta')$, the detailed balance equations are formulated as
	\begin{equation}
		\pi(u)p(u,v)=\pi(v)p(v,u)~\forall u,v.
		\label{eq:balance_equation}
	\end{equation}
	\noindent
	The transition kernel of the chain is the product between the kernel defined by the MCMC approach and the acceptance rate imposed by the ABC approach. The acceptance ratio for a given proposal represents the probability of accepting a given sample and is defined as
	\begin{equation}
		r((\theta_t,z_t),(\theta',z'))=\min\left(1,\frac{\pi_\varepsilon(\rho(y,z'))q(\theta',\theta_t)\pi(\theta')}{\pi_\varepsilon(\rho(y,z_t))q(\theta_t,\theta')\pi(\theta_t)} \right) 
	\end{equation}
	\noindent
	Therefore, the transition kernel is
	\begin{equation}
		p((\theta_t,z_t),(\theta',z'))=q(\theta_t,\theta')\pi(z'|\theta')\min\left(1,\frac{\pi_\varepsilon(\rho(y,z'))q(\theta',\theta_t)\pi(\theta')}{\pi_\varepsilon(\rho(y,z_t))q(\theta_t,\theta')\pi(\theta_t)} \right) 
		\label{eq:transition_kernel}
	\end{equation}
	\noindent
	where $(\theta_t,z_t)$ define the current state of the chain and $(\theta', z')$, the new sample. The stationary distribution follows from Bayes' theorem and Equations \ref{eq:bayes_app}-\ref{eq:noise_post_app} as
	\begin{equation}
		\pi(\theta,z)=\pi(y)\pi(\theta, z|y) = \pi_\varepsilon(\rho(y,z))\pi(z|\theta)\pi(\theta).
	\end{equation}
	\noindent
	Combining the transition kernel and the stationary distribution, the detailed balance equations are satisfied directly. The current state of the chain $u=(\theta_t,z_t)$ is expressed as
	\begin{equation}
		\pi(u) = \pi(\theta_t,z_t) = \pi(y)\pi_\varepsilon(\rho(y,z_t))\pi(z_t|\theta_t)\pi(\theta_t),
	\end{equation}
	and the new sample $v=(\theta',z')$ as
	\begin{equation}
		\pi(v) = \pi(\theta',z') = \pi(y)\pi_\varepsilon(\rho(y,z'))\pi(z'|\theta')\pi(\theta').
	\end{equation}
	From the transition kernel defined in Equation \ref{eq:transition_kernel}, $p(u,v)$ is directly the transition probability $p((\theta_t,z_t),(\theta',z'))$, and $p(v,u)$ is its conjugated
	\begin{equation}
		p(v,u)=p((\theta',z'),(\theta_t,z_t))=q(\theta',\theta_t)\pi(z_t|\theta_t)\min\left(1,\frac{\pi_\varepsilon(\rho(y,z_t))q(\theta_t,\theta')\pi(\theta_t)}{\pi_\varepsilon(\rho(y,z'))q(\theta',\theta_t)\pi(\theta')} \right). 
		\label{eq:transition_kernel_conjugated}
	\end{equation}
	\noindent
	Notice that when $p(u,v)=q(\theta_t,\theta')\pi(z'|\theta')$, then $p(v,u)=q(\theta',\theta_t)\pi(z_t|\theta_t)\frac{\pi_\varepsilon(\rho(y,z_t))q(\theta_t,\theta')\pi(\theta_t)}{q(\theta',\theta_t)\pi_\varepsilon(\rho(yz'))\pi(\theta')}$ and vice-versa, as the non-constant part of the minimization function of one transition kernel is the inverse of the other. The balance equation from Equation \ref{eq:balance_equation} must be evaluated in the two possible cases of the minimization function:
	\begin{equation}
		\begin{split}
			\pi(u)p(u,v)&=\pi(y)\pi_\varepsilon(\rho(y,z_t))\pi(z_t|\theta_t)\pi(\theta_t)q(\theta_t,\theta')\pi(z'|\theta')= \\ &=\pi(y)\pi(z'|\theta')\pi(z_t|\theta_t)\pi_\varepsilon(\rho(y,z_t))q(\theta_t,\theta')\pi(\theta_t)=\pi(v)p(v,u),
		\end{split}
	\end{equation}
	\noindent
	and
	\begin{equation}
		\begin{split}
			\pi(u)p(u,v)&=\pi(y)\pi(z_t|\theta_t)\pi(z'|\theta')\pi_\varepsilon(\rho(y,z'))q(\theta',\theta_t)\pi(\theta')= \\ &=\pi(y)\pi_\varepsilon(\rho(y,z'))\pi(z'|\theta')\pi(\theta')q(\theta',\theta_t)\pi(z_t|\theta_t)=\pi(v)p(v,u),
		\end{split}
	\end{equation}
	which implies the fulfilment of the balance equations and the convergence of the MCMC chain for transition kernels $q(\theta,\theta')$ that are well defined. This derivation holds for the MCMC framework used in this paper, which is based on ensemble samplers with a stretch move.
	
	\section{Moment-matching likelihood behaviour with measurement noise}
	\label{ap:behaviour_noise}
	Let $A_i$ be the residual at a given point $i$ as $A_i=|\mu_i^h-y_i|$ and $B_i$ the total standard deviation of the model as $B_i=\sqrt{\sigma_N^2+{\sigma_i^h}^2}$. $A_i$ and $B_i$ are independent of each other, as $A_i$ depends on the value of the original latent parameters and $B_i$ on the value of the auxiliary ones introduced in the embedded model inadequacy. Then, the noisy moment-matching likelihood from Equation \ref{eq:noisy_moment_matching} in its logarithmic form rends
	\begin{equation}
		\mathcal{\log L}(\bm{\theta})={-\frac{n_y}{2}}\log\left( 2\pi\epsilon^2\sigma_N^2\right)- \sum_{i=1}^{n_y}\exp\left(\frac{A_i^2}{2\sigma_N^2}+\frac{\left( B_i-\gamma A_i\right) ^2}{2\epsilon^2} \right).
	\end{equation}
	\noindent
	The goal is to analyze the behaviour of the likelihood function for the different values of $A$ and $B$, especially for the cases that lead to the optimal values. Obtaining the cases for which increasing or decreasing $A$ or $B$ leads to higher likelihood allows discerning tendencies in the optimization procedure and potential optimums. To do so, the derivatives of the likelihood with respect to $A$ and $B$ are required. Deriving the likelihood with respect to the residuals $A_i$, we obtain
	\begin{align}
		\frac{\partial\log\mathcal{L}}{\partial A}&=-\sum_{i=1}^{n_y}\left( \frac{A_i}{\sigma_N^2}-\frac{\gamma(B_i-\gamma A_i)}{\epsilon^2}\right)\\
		&=-\sum_{i=1}^{n_y}\left( \left(\frac{1}{\sigma_N^2}+\frac{\gamma^2}{\epsilon^2}\right)A_i-\frac{\gamma}{\epsilon^2}B_i\right).
	\end{align}
	\noindent
	Positive values of these derivatives imply that an increase of the residuum would produce a larger likelihood, and consequently, a ``worse'' fit of the mean values is favoured. Analogously, negative values of the derivative with respect to $A$ favour a better fit of the mean values of the prediction with the observations. Alternatively, deriving with respect to the total deviation $B_i$ we obtain
	\begin{align}
		\frac{\partial\log\mathcal{L}}{\partial B}&=-\sum_{i=1}^{n_y}\frac{\gamma(B_i-\gamma A_i)}{\epsilon^2}\\
		&=-\sum_{i=1}^{n_y}\left(-\frac{\gamma}{\epsilon^2}A_i+\frac{1}{\epsilon^2}B_i\right).
	\end{align}
	\noindent
	In this case, positive values of the derivative with respect to  $B$ favour increasing the predicted model form uncertainty, while negative values favour reducing it. For an analysis of the optimality, the second derivatives have to be computed. The entries of the Hessian matrix $\bm{H}(A,B)$ are calculated as
	\begin{align}
		\frac{\partial^2\log\mathcal{L}}{\partial A^2}&=-n_y\left(\frac{1}{\sigma_N^2}+\frac{\gamma^2}{\epsilon^2}\right)\\
		\frac{\partial^2\log\mathcal{L}}{\partial B^2}&=-\frac{n_y}{\epsilon^2}\\
		\frac{\partial^2\log\mathcal{L}}{\partial A\partial B}=\frac{\partial^2\log\mathcal{L}}{\partial B\partial A}&=-\frac{\gamma n_y}{\epsilon^2}
	\end{align}
	\noindent
	and its determinant returns
	\begin{equation}
		\det(\bm{H}(A,B))=\frac{\partial^2\log\mathcal{L}}{\partial A^2}\frac{\partial^2\log\mathcal{L}}{\partial B^2}-\frac{\partial^2\log\mathcal{L}}{\partial A\partial B}\frac{\partial^2\log\mathcal{L}}{\partial B\partial A}=\frac{n_y^2}{\epsilon^2}\left(\frac{1}{\sigma_N^2}+\frac{\gamma^2}{\epsilon^2}\right)-\frac{\gamma^2 n_y^2}{\epsilon^4}=\frac{n_y^2}{\epsilon^2\sigma_N^2}>0.
	\end{equation}
	\noindent
	As $\frac{\partial^2\log\mathcal{L}}{\partial A^2}<0$ for any allowed value of the parameters, all critical points of $\log\mathcal{L}$ are local maximum. Therefore, the log-likelihood defined is concave as expected and $\frac{\partial\log\mathcal{L}}{\partial A},\frac{\partial\log\mathcal{L}}{\partial B}>0$ lead closer to the optimum. Taking means over $n_y$, it can be observed that
	\begin{equation}
		\frac{\partial\log\mathcal{L}}{\partial A}>0\iff\frac{1}{n_y}\frac{\partial\log\mathcal{L}}{\partial A}>0
	\end{equation}
	\noindent
	therefore
	\begin{equation}
		\frac{\partial\log\mathcal{L}}{\partial A}>0\iff -\left(\frac{1}{\sigma_N^2}+\frac{\gamma^2}{\epsilon^2}\right)\bar{A}+\frac{\gamma}{\epsilon^2}\bar{B}>0
	\end{equation}
	\noindent
	where $\bar{A}$ and $\bar{B}$ denote, respectively, the mean value of $A$ and $B$ over the observations. Using the mean value instead of the sum simplifies the analysis and allows for a more direct comparison between cases than using the sum of individual terms. Analogously,
	\begin{equation}
		\frac{\partial\log\mathcal{L}}{\partial B}>0\iff \frac{\gamma}{\epsilon^2}\bar{A}-\frac{1}{\epsilon^2}\bar{B}>0.
	\end{equation}
	\noindent
	Based on this information, the cases indicated in Table \ref{tab:cases_likelihood} are considered. Note that for the likelihood presented in Equation \ref{eq:likelihood_sargsyan}, the results will be equivalent by simply taking $\sigma_N^2=\epsilon^2$.
	\begin{table}[!b]
		\centering
		{\caption{Cases for noisy moment-matching log-likelihood behaviour}}
		{ \centering\begin{tabular}{|cc|c|c|}
					\hline
					&\textbf{Cases} & $\frac{\partial\log \mathcal{L}}{\partial A}$ & $\frac{\partial\log \mathcal{L}}{\partial B}$ \\
					\hline
					(I)& $\bar{B}<\gamma\bar{A}$ & Negative & Positive \\
					(II) &$\bar{B}=\gamma\bar{A}$ & Negative & Zero \\
					(III)& $\bar{B}\in\left(\gamma\bar{A},\left(\gamma+\frac{\epsilon^2}{\sigma_N^2}\right)\bar{A}\right)$ & Negative & Negative \\
					(IV)& $\bar{B}=\left(\gamma+\frac{\epsilon^2}{\sigma_N^2}\right)\bar{A}$  & Zero & Negative \\
					(V)& $\bar{B}>\left(\gamma+\frac{\epsilon^2}{\sigma_N^2}\right)\bar{A}$ & Positive & Negative \\
					(VI) & $\bar{A}=0,~\bar{B}>0$\footnote{$\bar{A}>0$ for all cases except (VI) and (VII).}& Positive & Negative \\
					(VII) & $\bar{A}=0,~\bar{B}=0$\footnote{$\bar{B}>0$ for all cases except (VII).}& Zero & Zero \\
					\hline
				\end{tabular}
		}
		\label{tab:cases_likelihood}
	\end{table}
	These cases allow for the following interpretations:
	\begin{enumerate}
		\item Case (I) corresponds to the situation where the variance predicted by the model cannot cover the residuals. The resulting tendency assuming an optimal path would be an increase in the predicted variance and a reduction of the residual. This case occurs when the predictions for a given sampled parameter vector are inaccurate or if the variance in the observations is under-represented.
		\item Case (II) corresponds to a model that predicts perfectly the variance (on average) as prescribed by the approximated model, while the residuals are not well represented. 
		\item Case (III) is produced by a model that predicts a larger variance than the one needed to obtain the observations and where the residuum can be reduced. In cases with prescribed noise, further reducing the variance may not be possible.
		\item Case (IV) represents the case where the fitting of the residuals is considered optimum. However $\bar{B}=\gamma\bar{A}$ as prescribed only if $\epsilon=0$. Otherwise, the predicted variation will have to be reduced to achieve the optimum.
		\item Case (V) is symmetrical to Case (I), where the model over-represents the variance in the observations. Therefore, the residuals are increased to present a larger variance while the predicted deviation is to be reduced.
		\item Case (VI) is a special case of (V) where the model can exactly represent on average the observations. However, $B>0$ by construction, unless the case without noise or predicted variance is allowed. The residuals will tend to be increased at least up to the minimum possible of the predicted variance.
		\item Case (VII) is the actual maximum, which can be achieved only if the predictions and their variance can be predicted exactly by the model. It is a special case of (II) for $\bar{A}=0$.
	\end{enumerate}
	Cases (VI) and (VII) are generally not possible in the presence of model inadequacy and noise. The minimum $\bar{A}$ possible is limited by the model inadequacy and the minimum $\bar{B}$ possible is limited by the noise term. Superior limits can only be prescribed manually. If $\min \bar{A} > \min \bar{B}$, then the residual can be covered by the predicted model form uncertainty, and the optimum will be at Case (II) for $\bar{B} = \gamma \min \bar{A}$. However, if $\min \bar{A} < \min \bar{B}$, the optimum will be at Case (IV) with $\bar{A}=\left(\gamma+\frac{\epsilon^2}{\sigma_N^2}\right)^{-1}\min\bar{B}$. In such a case, the residuals are minimized in order to fit better the predicted variance, which is generally undesirable for a good fit. This situation can be avoided with a proper quantification of the noise and the embedding of the right parameters such that $B$ covers the residuals of $A$ in the most probable range.
	
	\section{Results thermal example}
	\label{ap:results_thermal}
	The results for Section \ref{sec:thermal_temperature} for GMM, RGMM and IN likelihoods are presented here.
	\begin{figure}[!h]
		{\includegraphics[width=\textwidth]{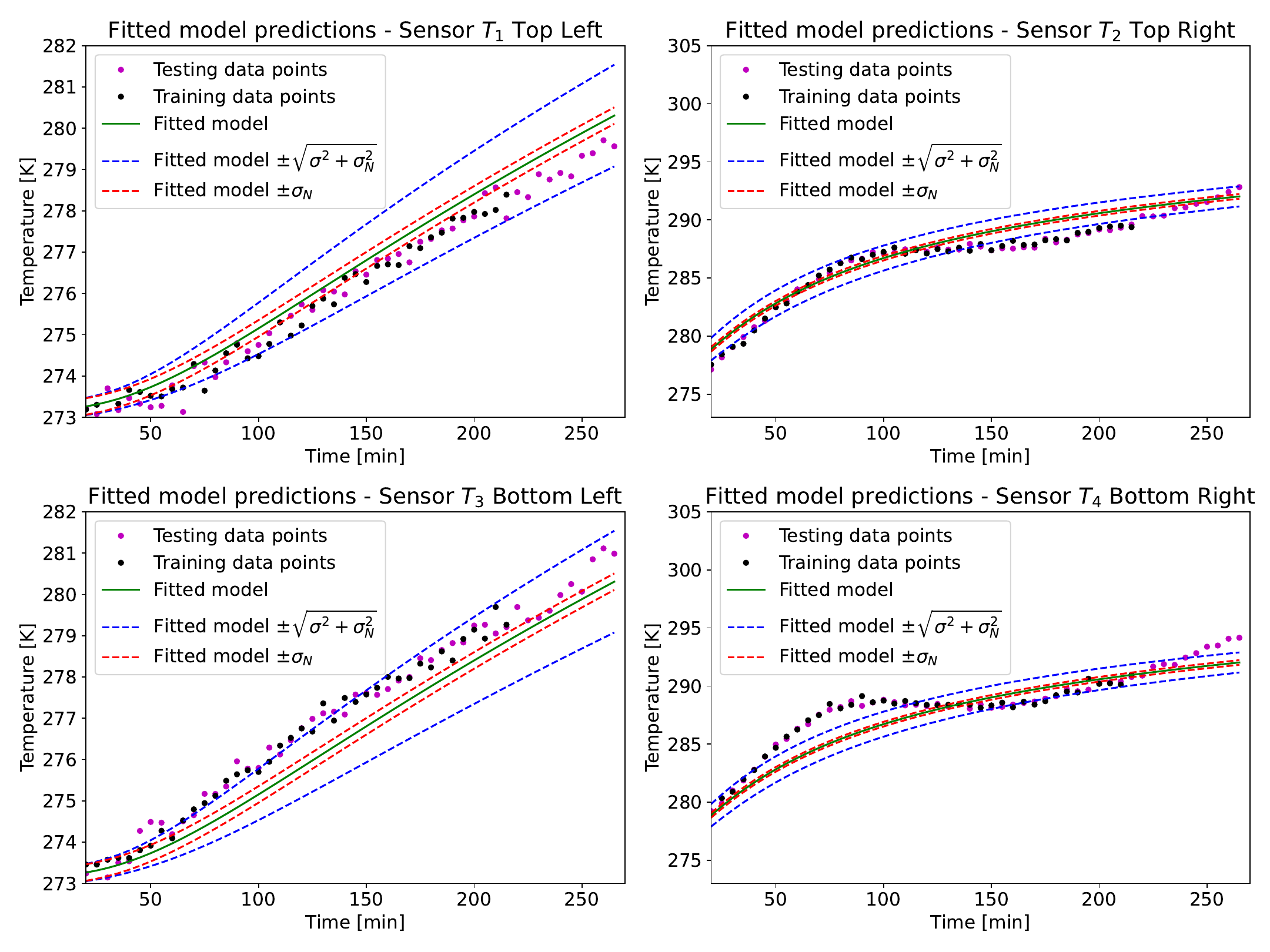}}{\caption{Temperature sensors predictions for GMM likelihood}}
		\label{fig:thermal_temperature_gmm}
	\end{figure}
	\begin{figure}[!h]
		{\includegraphics[width=\textwidth]{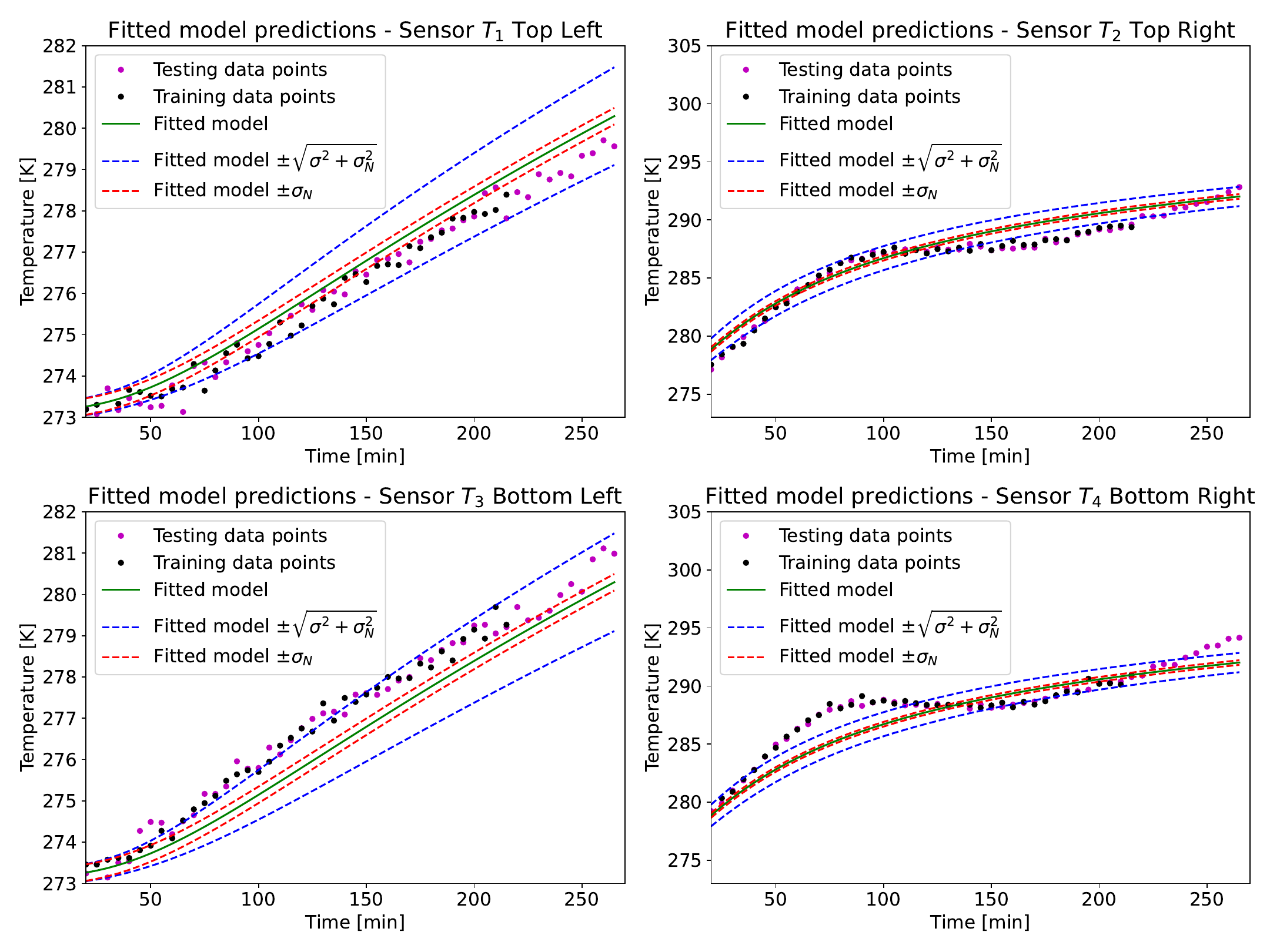}}{\caption{Temperature sensors predictions for RGMM likelihood}}
		\label{fig:thermal_temperature_rgmm}
	\end{figure}
	\begin{figure}[!h]
		{\includegraphics[width=\textwidth]{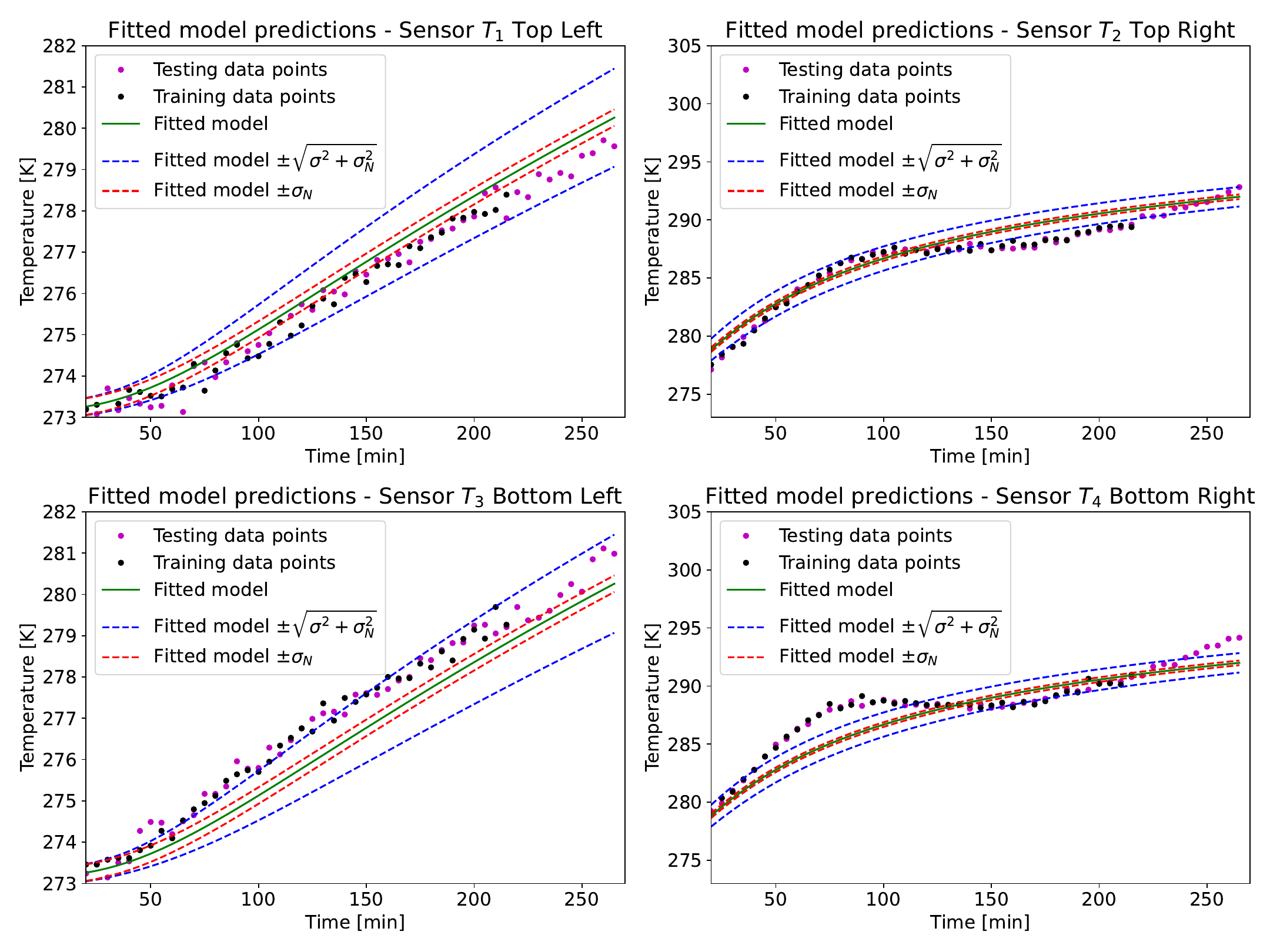}}{\caption{Temperature sensors predictions for IN likelihood}}
		\label{fig:thermal_temperature_in}
	\end{figure}

	\bibliographystyle{apalike}
	\bibliography{mybibliography}

\end{document}